\documentclass[11pt]{article}
\usepackage{graphicx, caption, subfigure}
\makeatletter
\newcommand{\singlespacing}{\let\CS=\@currsize\renewcommand{\baselinestretch}{1.0}\tiny\CS}
\newcommand{\doublespacing}{\let\CS=\@currsize\renewcommand{\baselinestretch}{1.5}\tiny\CS}

\begin{document}
\title{Understanding the Characteristics of Multiple Production of Light Hadrons in $Cu+Cu$ Interactions at Various
RHIC Energies: A Model-based Analysis}
\author{P.
Guptaroy$^1$\thanks{e-mail: gpradeepta@rediffmail.com}, Goutam
Sau$^2$\thanks{e-mail: gautamsau@yahoo.co.in}, S. K. Biswas$^3$ \&
S. Bhattacharyya$^4$\thanks{e-mail: bsubrata@isical.ac.in
(Communicating Author).}\\
{\small $^1$ Department of Physics, Raghunathpur College,}\\
 {\small P.O.: Raghunathpur 723133,  Dist.: Purulia(WB), India.}\\
{\small $^2$ Beramara Ram Chandrapur High School,}\\
 {\small South 24-Parganas, 743609(WB), India.}\\
 {\small $^3$ West Kodalia Adarsha Siksha Sadan, New Barrackpore,}\\
 {\small Kolkata-700131, India.}\\
 {\small $^4$ Physics
and Applied Mathematics Unit(PAMU),}\\
 {\small Indian Statistical Institute,}\\
 {\small 203 B. T. Road, Kolkata - 700108, India.}}
\date{}
\maketitle

\bigskip
\bigskip
\begin{abstract}
Experiments involving copper-copper collisions at the RHIC-BNL (USA)
at energies $\sqrt {s_{NN}}$ = 22.5, 62 and 200 GeV have produced a
vast amount of high-precision data which are to be analysed in the
light of various competing models in the domain of multiparicle
production scenario. We have chosen to analyse here the measured
data on the $p_T$-spectra of various light and non-strange
secondaries at various energies mentioned above, some of their very
important ratio-behaviours at the various centralities of the
collisions and the nuclear modification factors, $R_{AA}$ and
$R_{CP}$, in the light of a version of the Sequential Chain Model
(SCM). The agreements between the measured data and model-based
results are generally found to be modestly satisfactory. Besides,
our obtained results have also been compared with results based upon
two approaches with strong standard model flavour, of which one is
purely pQCD-oriented and is forwarded by Vitev.
\end{abstract}


\newpage
\doublespacing
\section{Introduction}
In a previous paper \cite{pgr07} we dealt with the properties of
$Au+Au$ interactions at RHIC-energies at $\sqrt{s_{NN}}$=200 GeV in
great detail with emphasis on the nature of the most important
observables. The gold-gold interactions constituted a relatively
heavy system. Compared to it, the $Cu+Cu$ system represents a
lighter and smaller system-size, with mass-number for copper being
much less than gold. Besides, in Ref. \cite{pgr07} we also gave the
details about the essential points of the basic multiparticle
production model applied therein.
\par
So, in order to avoid repetitions, we will just give a very brief
sketch of the main/major observables that would here be dealt with
and on the same model that we make use of here.
\par
Our purpose is to check (i) whether the same model could explain the
data measured for a comparatively lighter system; (ii) whether [and
to what extent (if at all)] there is any prominent system size
effect on the results obtained by the experimental measurements.
\par
Apart from the nature of $p_T$-spectra and some particle-production
ratios, we will dwell upon here properties of nuclear modification
factors $R_{AA}$ and $R_{CP}$ for $Cu+Cu$ collisions and the
centrality dependence of the results. But production of the heavies
and the heavy strange hadrons are left out in the present work.
\par
As our approach to the analysis of $Cu+Cu$ collisions (or for that
matter, of $Au+Au$ and even $p+p$ reactions) is somewhat different
from the `standard' varieties, our phrases and vocabularies are not
identical with them. We have inducted here the physics of partonic
multiple scatterings for both nucleus-nucleus interactions, and
$p+p$ reactions at large-$p_T$ through some mechanism as given in
the text in the appropriate place. Even the effects of rescatterings
have also been incorporated here by the same technique. But, the
associated hadronic collectivity factor (gauged by the elliptic flow
represented normally by `$v_2$') has not been touched upon here;
this flow-behaviour would be dwelt-upon in some detail in a future
work.
\par
In order to dispel any misunderstanding of any interested reader,
let us emphasise here, before closing up the introductory remarks,
that our intention is not to make any tall claim  that the model
tried out and tested here is much better than the existing models.
But, our objective is just to familiarise the fact that modest
description of data is quite possible even with some non-standard
approaches, like the present one, which is being applied and
validated to explain some observables in high energy collisions
since the mid-seventies.
\par
Finally, the organization of this work is as follows. In section 2
we sum up the main physics-aspects entailed in this work. A glance
and a glimpse into it would reveal the non-standard nature of the
basic approach. In the next section (section 3) the results arrived
at have been presented with tables and figures. The section 4
provides some comparison of our work with both data and the results
obtained by other model-based approaches for a very few observables
related to the production of some very select secondaries (for which
theory-based studies on $Cu+Cu$ collisions are available). And in
the last section (section 5) we offer the final comments and
conclusions.
\section{The Theoretical Framework: A Brief Outlook}
The description of the model-based features would be subdivided into
some parts. The first part(subsection 2.1) gives a brief overview of
the production mechanism of the secondary hadrons in nucleon-nucleon
($p+p$) interaction in the context of the Sequential Chain Model
(SCM). Then in the subsection 2.2 we present a brief outline of the
main and major achievements of the model; these points, in essence,
also highlight the important characteristics of the model.
Thereafter, the relevant transitions for different observables from
$p+p$ to $A+B$ interactions will be discussed in the subsection 2.3.
\subsection{Basic Model for Particle Production in $PP$ Scatterings: An Outline}
 According to
this Sequential Chain Model (SCM),  high energy hadronic
interactions boil down, essentially, to the pion-pion interactions;
as the protons and neutrons are conceived in this model as
$p$~=~($\pi^+$$\pi^0$$\vartheta$) and $n$~=~($1/\sqrt{\chi_1^2 +
\chi_2^2}$
[$\chi_1$($\pi^0$$\pi^0$$\vartheta$)+$\chi_2$($\pi^-$$\pi^+$$\vartheta$)])
repectively, where $\vartheta$ is a spectator particle needed for
the dynamical generation of quantum numbers of the nucleons and
$\chi_1$, $\chi_2$ are the weightage factors \cite{pgr07}-\cite{bhat882}. Our focus and concern would now be
confined and concentrated to the structure of protons alone. The
production of pions in the present scheme occurs as follows: the
incident energetic $\pi$-mesons in the structure of the projectile
proton(nucleon) emits a rho($\varrho$)-meson in the interacting
field of the pion lying in the structure of the target proton, the
$\varrho$-meson then emits a $\pi$-meson and is changed into an
omega($\omega$)-meson, the $\omega$-meson then again emits a
$\pi$-meson and is transformed once again into a $\varrho$-meson and
thus the process of production of pion-secondaries continue in the
sequential chain of $\varrho$-$\omega$-$\pi$ mesons. The two ends of
the diagram contain the baryons exclusively \cite{pgr07}-\cite{bhat882}.
\par
For $K^+$($K^-$)or  $K^0{\bar{K^0}}$ production the model proposes
the following mechanism. One of the interacting $\pi$-mesons emits a
$\varrho$-mesons; the $\varrho$-mesons in its turn emits a
$\phi^0$-meson and a $\pi$-meson. The $\pi$-meson so produced then
again emits $\varrho$ and $\phi^0$ mesons and the process continues.
The $\phi^0$ mesons so produced now decays into either $K^+K^-$ or
$K^0{\bar{K^0}}$ pairs. The $\varrho$-$\pi$ chain proceeds in any
Fenymann diagram in a line with alternate positions, pushing the
$\phi^0$ mesons  (as producers of $K^+K^-$ or $K^0{\bar{K^0}}$
pairs) on the sides. This may appear paradoxical as the $\phi^0$
production cross-section is generally smaller than the $K{\bar K}$
production cross-section; still the situation arises due to the fact
that the $\phi^0$ resonances produced in the collision processes
will quickly decay into $K{\bar K}$ pairs, for which the number of
$\phi^0$ will be lower than that of the $K{\bar K}$ pairs. Besides,
as long as $\phi^0$ mesons remain in the virtual state,
theoretically there is no problem, for $\phi^0 K^+ K^-$ ( or $\phi^0
K^0 {\bar {K^0}}$) is an observed and allowed decay mode, wherein
the strangeness conservation is maintained with the
strange-antistrange coupled production. Moreover, $\phi^0 K^+ K^-$ (
or $\phi^0 K^0 {\bar {K^0}}$) coupling constant is well known and is
measured by experiments with a modest degree of reliability. And we
have made use of this measured coupling strength for our
calculational purposes, whenever necessary. It is assumed that the
$K^+K^-$ and $K^0{\bar{K^0}}$ pairs are produced in equal
proportions \cite{pgr07}-\cite{bhat882}. The entire production
process of kaon-antikaons is controlled jointly by the coupling
constants, involving $\varrho$-$\pi$-$\phi$ and $\phi^0$-$K^+K^-$ or
$\phi^0$-$K^0{\bar {K^0}}$.
\par
Now we describe here the baryon-antibaryon production. According to
the SCM mechanism, the decay of the pion secondaries produces
baryon-antibaryon pairs in a sequential chain as before. The pions
producing baryons-antibaryons pairs are obviously turned into the
virtual states. And the proton-antiproton pairs are just a part of
these secondary baryon-antibaryon pairs. In the case of
baryon-antibaryon pairs it is postulated that protons-antiprotons
and neutrons-antineutrons constitute the major bulk, Production of
the strange baryons-antibaryons are far less
 due to the much smaller values of the coupling constants and due to their being much heavier.
\par
The field theoretical calculations for the average multiplicities of
the $\pi$, $K$ and $\bar p$-secondaries and for the inclusive
cross-sections of those secondary particles deliver some expressions
which we would pick up from \cite{pgr07}-\cite{bhat882}.
\par
The inclusive cross-section of the $\pi^-$-meson produced in the
$p+p$ collisions given by
\begin{equation}\displaystyle E\frac{d^3\sigma}{dp^3}|_{pp \rightarrow
\pi^- x}  \cong \Gamma_{\pi^-} \exp(- 2.38 <n_{\pi^-}>_{pp}
x)\frac{1}{p_T^{(N_R^{\pi^-})}} \exp(\frac{-2.68
p_T^2}{<n_{\pi^-}>_{pp}(1-x)})   ~ ,
\end{equation}
with \begin{equation}\displaystyle {<n_{\pi^+}>_{pp} ~ \cong  ~
<n_{\pi^-}>_{pp} ~ \cong
 ~ <n_{\pi^0}>_{pp}  ~ \cong  ~ 1.1s^{1/5} ~,}
 \end{equation}
 where $\Gamma_{\pi^-}$ is the
normalisation factor which will increase as the inelastic
cross-section increases and it is different for different energy
region and for various collisions, for example, $|\Gamma_{\pi^-}| \cong
90$ for Intersecting Storage Ring(ISR) energy region. The terms
$p_T$, $x$ in equation (1) represent the transverse momentum,
Feynman Scaling variable respectively. Moreover, by definition, $x ~
= ~ 2p_L/{\sqrt s}$ where $p_L$ is the longitudinal momentum of the
particle. The $s$ in equation (2) is the square of the c.m. energy.
\par
$1/p_T^{N_R^{\pi^-}}$ of the expression (1) is the `constituent
rearrangement term' arising out of the partons inside the proton
which essentially provides a damping term in terms of a power-law in
$p_T$ with an exponent of varying values depending on both the
collision process and the specific $p_T$-range. The choice of
${N_R}$ would depend on the following factors: (i) the specificities
of the interacting projectile and target, (ii) the particularities
of the secondaries emitted from a specific hadronic or nuclear
interaction and (iii) the magnitudes of the momentum transfers and
of a phase factor (with a maximum value of unity) in the
rearrangement process in any collision. And this is a factor for
which we shall have to parameterize alongwith some physics-based
points indicated earlier. The parametrization is to be done for two
physical points, viz., the amount of momentum transfer and the
contributions from a phase factor arising out of the rearrangement
of the constituent partons. Collecting and combining all these, we
proposed the relation to be given by \cite{pgr08}
\begin{equation}\displaystyle
N_R=4<N_{part}>^{1/3}\theta,
\end{equation}
where $<N_{part}>$ denotes the average number of participating
nucleons and $\theta$ values are to be obtained phenomenologically
from the fits to the data-points. In this context, the only
additional physical information obtained from the observations made
here is: with increase in the peripherality of the collisions the
values of $\theta$ gradually grow less and less, and vise versa.
\par
Similarly, for kaons of any specific variety ( $K^+$, $K^-$, $K^0$
or $\bar{K^0}$ ) we have

\begin{equation}
\displaystyle E \frac{d^3\sigma}{dp^3}|_{pp \rightarrow K^- x}
 ~ \cong  ~ \Gamma_{K^-}\exp(  -  6.55 <n_{K^-}>_{pp}
 x) ~ \frac{1}{p_T^{(N_R^{K^-})}}\exp(\frac{-  1.33
  p_T^2}{<n_{K^-}>^{3/2}_{pp}}) ~ ~ ,
\end{equation}
 with $|\Gamma_{K^-}| \cong 11.22$ for ISR energies and with

\begin{equation}
\displaystyle <n_{K^+}>_{pp}   \cong  <n_{K^-}>_{pp}  \cong
  <n_{K^0}>_{pp}  \cong  <n_{\bar{K^0}}>_{pp} \cong
5\times10^{-2}  s^{1/4}  .
\end{equation}
And for the antiproton production in $pp$ scattering at high
energies, the derived expression for inclusive cross-section is

\begin{equation}
\displaystyle E\frac{d^3\sigma}{dp^3}|_{pp\rightarrow{\bar p}x}
 ~ \cong \Gamma_{\bar p} \exp(-25.4 <n_{\bar{p}}>_{pp} x)\frac{1}{p_T^{({N_R}^{\bar p})}}\exp(\frac{-0.66 ((p_T^2)_{\bar
p}+ {m_{\bar p}}^2)}{<n_{\bar p}>^{3/2}_{pp} (1-x)})
 ~  ,
\end{equation}
with $|\Gamma_{\bar p}| ~  \cong  ~ 1.87\times10^3$ and $m_{\bar p}$
is the mass of the antiprotons. For ultrahigh energies

\begin{equation}
\displaystyle{ <n_{\bar p}>_{pp}  ~ \cong <n_p>_{pp}  ~ \cong
 ~ 2\times10^{-2} ~ s^{1/4} ~ ,}
 \end{equation}
\subsection{Some Cardinal Characteristics and Triumphs of the Model}
We agree that the word `non-standard' used in the preceding section
is a $\it{port~ manteau}$ adjective, with many layers of meaning
hidden within it. In this particular case, (i) the model is based on
some new ideas about the structure of hadrons and the nature of
hadronic interactions; (ii) the proposed mechanism underlying this
work does not admit of any airtight compartmentalisation of the
`soft' (low-$p_T$) and `hard'(large $p_T$, $p_T\geq2$ GeV/c)
production; (iii) rather, the model presents a unified approach to
the production of particle-secondaries; (iv) besides, the
fundamental expressions for final (analytical) calculations are
derived here on the basis of field-theoretic considerations and the
use of Feynman diagram techniques (albeit with some simplifying high
energy approximation and assumptions) with the infinite momentum
frame tools and under impulse approximation method; (v) this
approach establishes the $<universality>$ aspect of the multiplicity
of high energy interaction  called `globality property'; (vi) the
model explains the `jet'-structure for emanation of the secondary
particle as the $<<two-sided sprays>>$ of hadrons, (vii) it
reproduces the behaviour of average multiplicity, nature of
(invariant) inclusive cross-section and the properties of average
transverse momenta of various secondaries; (viii) the model could
also account for the very slow rising nature of the total
cross-sections. (ix) Besides, this model can/does explain the
majority of the characteristics of what are known or believed to be
the `quark gluon plasma'-diagnostics. And that could be obviously
done by an alternative approach and outlook. (x) Furthermore, the
application of the model can also accommodate a large amount of very
important Cosmic Ray Physics issues and problems that came to the
fore in the very recent times.
\par
The above-mentioned features are essentially ingrained in the
entirety of this totally non-standard mechanism emerging from an
alternative philosophy and outlook about the particle-constituents
and their interaction mode. On the whole, this is purely an
analytical approach with a reasonable number of valid assumptions
and approximations which are commonly used by all High Energy
Physicists. Uptil now, we have confined ourselves mostly to the
non-simulational calculations. The calculations for `soft'and `hard'
have been superposed here by virtue of the simple factorisation
property. One of the very strong points about this model is the fact
that the various coupling strengths used in this model are not only
known but also reliably well-measured by several experiments. This
factors helps to reduce considerably the speculative components in
the results.
\subsection{Results for $AA$ Collisions from $PP$ Reactions: The Connecting Bridge}
In order to study a nuclear interaction of the type $A+B\rightarrow
C^-+ x$, where $A$ and $B$ are projectile and target nucleus
respectively, and $C^-$ is the detected particle which, in the
present case, would be $\pi^-$, $K^-$ and $\bar p$, the SCM has been
adapted, on the basis of the suggested Wong \cite{wong} work to the
Glauber techniques by using Wood-Saxon distributions
\cite{eskola}-\cite{gorenstein}. The details of calculations and the
features of the SCM have been given in our previous paper
\cite{pgr07}.
\par
The general form of our SCM-based transverse-momentum distributions
for $A+B\rightarrow C^-+X$-type reactions can be written in the
following notation:
\begin{equation}\displaystyle{
\frac{1}{2\pi p_T}}{\frac{d^2N}{dp_T dy}|_{A+B\rightarrow C^-+
x}=\alpha_{C^-}\frac{1}{p_T^{N_R^{C^-}}}\exp(-\beta_{C^-} \times
p_T^2).}
\end{equation}
\par
 The set of relations to be used for estimating the
parameter $\alpha_{C^-}$ is given below \cite{pgr07}.
\begin{equation}\displaystyle{
\alpha_{C^-}={\frac{(A \sigma_B + B
\sigma_A)}{\sigma_{AB}}} {\frac{1}{1+a(A^{1/3}+B^{1/3})}}\Gamma_{C^-}\exp(- \eta
<n_{C^-}>_{pp} x)}
\end{equation}
Here, in the above equation [eqn.(9)], $\Gamma_{C^-}$, as stated
above, is the normalization constant which is different for the
different secondaries and the collider energies. It also depends on
the centrality of the collisions. The first factor in eqn.(9) gives
a measure of the number of wounded nucleons i.e. of the probable
number of participants, wherein $A\sigma_B$ gives the probability
cross-section of collision with `$B$' nucleus (target), had all the
nucleons of $A$ suffered collisions with $B$-target. And $B\sigma_A$
has just the same physical meaning, with $A$ and $B$ replaced.
Furthermore, $\sigma_A$ is the nucleon(proton)-nucleus(A)
interaction cross-section, $\sigma_B$ is the inelastic
nucleon(proton)-nucleus(B) reaction cross-section and $\sigma_{AB}$
is the inelastic $AB$ cross-section for the collision of nucleus $A$
and nucleus $B$. The values of $\sigma_{AB}$, $\sigma_{A}$,
$\sigma_{B}$ are worked here out in a somewhat heuristic manner by
the following formula \cite{na5002}
\begin{equation}
\displaystyle{ \sigma^{inel}_{AB} ~ = ~ \sigma_{0} ~
(A^{1/3}_{projectile} + A^{1/3}_{target} - \delta)^2}
\end{equation}
with $\sigma_{0} = 68.8$ mb, $\delta= 1.32$.
\par
Besides, in expression (9), the second term is a physical factor
related with energy degradation of the secondaries due to multiple
collision effects. The parameter $a$ occurring in eqn.(9) above is a
measure of the fraction of the nucleons that suffer energy loss. The
maximum value of $a$ is unity, while all the nucleons suffer energy
loss. This $a$ parameter is usually to be chosen \cite{wong},
depending on the centrality of the collisions and the nature of the
secondaries.
\par
The values of $\eta$ in eqn. (9) are different for different
secondary produced; for example, $\eta$= 2.38 for pions, 6.35 for
kaons and 25.4 for protons, as were given in eqn.(1), eqn.(4) and
eqn.(6).
\par
$1/p_T^{N_R^{C^-}}$ of the expression (8) is the `constituent
rearrangement term' arising out of the partons inside the proton
which essentially provides a damping term in terms of a power-law in
$p_T$ with an exponent of varying values depending on both the
collision process and the specific $p_T$-range. We have already
mentioned the details earlier.
\par
The values of $\beta_{C^-}$ of the equation (8) for different
secondaries have been calculated with the help of eqn.(1), eqn.(2),
eqn.(4)-eqn.(7).
\section{Steps Towards Calculations}
At the very start let us present a summary (which might be a rehash
of what we have mentioned) of the key physical facts that would be
of paramount importance which proceeding towards calculations. There
are some foundational steps that enable us to arrive at the final
working formulae which deliver the results to be reported here for
for $Cu+Cu$ collisions. The procedural steps are as follows: (i)
Firstly, we have the basic model for $p+p$ scattering at high
energies and low-$p_T$ (`soft') interactions, so we have to convert
the mathematical expressions for nucleus+nuclus ($Cu+Cu$) collisions
by introducing the nuclear dependence factor on the results arrived
at for $p+p$ collisions. (ii) Secondly, the data-points on $Cu+Cu$
reaction at various high energies exceed the range of the low-$p_T$
boundary, $p_T>2$ GeV/c, for which the large-$p_T$ effect is to be
superposed on the expressions for soft-production of the
secondaries. The constituent(partonic) rearrangement factor
introduced here takes care of this physical feature arising out of
the `hard' (large-$p_T$) contributions. (iii) In a model-dependent
way the SCM has some special and specific means of excess production
of the positive secondaries, specifically the light secondaries.
\subsection{Production of Main Varieties of Negatively Charged Secondaries}
The general expressions of inclusive cross-sections for the production of $\pi^-$, $K^-$ and $\bar p$ for $p+p$ collisions were stated in the previous section by eqn.(1), eqn.(4) and eqn.(6) respectively.
\par
For the production of $\pi^-$-mesons in $p+p$ collisions, we use eqn.(1), eqn.(2) and eqn.(9) with $\alpha_{\pi^-}=\Gamma_{\pi^-}\exp(- 2.38
<n_{\pi^-}>_{pp} x)$. The values of $(\alpha_{\pi^-})_{pp}$, $(N_R^{\pi^-})_{pp}$ and $(\beta_{\pi^-})_{pp}$ are given in Table 1. The experimental data for the inclusive cross-sections versus $p_T[GeV/c]$ for $\pi^0$ production in $p+p$
 interactions at $\sqrt{s_{NN}}$ = 20 GeV are taken
from Ref. \cite{enterria}. And for the data for inclusive cross-sections for $\pi^+$ at energies $\sqrt{s_{NN}}$ = 63 GeV
and $\sqrt{s_{NN}}$ = 200 GeV we use
 references \cite{alper}, \cite{phenix2}, \cite{yang} respectively. They are plotted
in Figure 1(a), Figure 1(b) and Figure 1(c) respectively. The solid
lines in those figures depict the SCM-based plots.

\begin{table}
\singlespacing\caption{Values of $(\alpha_{\pi^-})_{pp}$, $(N_R^{\pi^-})_{pp}$ and $(\beta_{\pi^-})_{pp}$ for
$\pi^-$ productions in $p+p$ collisions at
$\sqrt{s_{NN}}$=20, 63 and 200 GeV}
\begin{center}
\begin{tabular}{cccc}
\hline
$\sqrt{s_{NN}}$&$(\alpha_{\pi^-})_{pp}$&($N_R^{\pi^-})_{pp}$&$(\beta_{\pi^-})_{pp}$\\
\hline
20 GeV& 0.281&4.086&0.703\\
\hline
63 GeV& 0.545&3.327&0.468\\
\hline
200 GeV&0.007&3.867&0.293\\
\hline
\end{tabular}
\end{center}
\end{table}
\subsubsection{Production of $\pi^-$-mesons in $Cu+Cu$ Collisions}
We now, at first, turn our
attention to $\pi^-$ production in $Cu+Cu$ collisions at energies
$\sqrt{s_{NN}}$ =22.5, 62.4 and 200 GeV.
\par
Using eqn.(8) and eqn.(9) the SCM-based expressions for transverse momentum
distribution of negative pions produced in the $Cu+Cu$ collisions at
$\sqrt{s_{NN}}$ = 22.5, 62.4 and 200 GeV at RHIC and for different
centralities can be obtained. The values of $\alpha_{\pi^-}$,
$N_R^{\pi^-}$ and $\beta_{\pi^-}$ for different centralities and for
different energies are given in Table 2. The values of $N_{part}$,
for calculating $N_R^{\pi^-}$ from eqn.(3), in this context, have
been taken from Ref. \cite{alver06}, \cite{phenix08}. The
experimental results for the production of $\pi^-$ at different
centralities for energies $\sqrt{s_{NN}}$ = 22.5, 62.4 GeV are taken
from the Ref. \cite{phenix3} and for energy $\sqrt{s_{NN}}$ = 200
GeV we have used Ref. \cite{phenix1}. The invariant yields for
$\pi^-$ against $p_T[GeV/c]$ for different energies are plotted in
Figures 2(a), 2(b) and 2(c) respectively. The solid lines in those
figures show the theoretical SCM results.
\subsubsection{$K^-$ Production in $Cu+Cu$ Collisions}
With the help of the eqn.(3)-eqn.
(5), eqn. (8) and eqn. (9) the values of $\alpha_{K^-}$, $N_R^{K^-}$
and $\beta_{K^-}$ for the transverse momentum distributions for
different centralities of $K^-$-particles in $Cu+Cu$ collisions at
$\sqrt{s_{NN}}$ = 22.5, 62.4 and 200 GeV at RHIC have been
calculated and they are given in Table 3. The experimental results
are taken from the PHENIX group\cite{phenix3} for RHIC energies
$\sqrt{s_{NN}}$ = 22.5 and 62.4 GeV and they are plotted in Figures
4(a) and 4(b), whereas for $\sqrt{s_{NN}}$ = 200 GeV, the Ref.
\cite{phenix1} has been used and data are plotted in Fig. 4(c). The lines
in those figures depict the theoretical outcomes.
\subsubsection{Production of $\bar p$ in $Cu+Cu$ Collisions at Different Energies}
 Using the eqn.(6)-eqn.(9), the transverse
momentum distributions for antiproton in $Cu+Cu$ collisions at
energies $\sqrt{s_{NN}}$ = 22.5, 62.4 and 200 GeV at RHIC have been
calculated the corresponding values of $\alpha_{\bar p}$, $N_R^{\bar
p}$ and $\beta_{\bar p}$ for different centralities and different
energies are given in Table 4. The experimental results  of
invariant yields for the production of antiproton productions for
energies $\sqrt{s_{NN}}$ = 22.5, 62.4 GeV  and for $\sqrt{s_{NN}}$ =
200 Gev are taken from the Refs. \cite{phenix3}, \cite{phenix1}
respectively. They are plotted against $p_T$ in Figures 6(a), 6(b)
and 6(c) for different centralities. The solid lines in those
figures show the theoretical SCM-based results.
\subsection{On Excess Production of Positive Particles}
True, on the average, the particles are produced in
charge-independent equal measure, for which roughly one-third of the
particles could be reckoned to be positively charged, one third are
negatively charged and the rest one third are neutral. But,
according to the present mechanism of particle production, there are
some specifically exclusive means to produce positive particles of
which $\pi^+$, $K^+$, and $p$ are the members. They are produced
from within the structure of protons (nucleons). These production
characteristics and the quantitative expressions for their special
production have been dwelt upon in detail in Ref. \cite{bhat882}.
Let us assort the relevant expressions therefrom as results to be
used here.
\subsubsection{Production of Positive Pions in $Cu+Cu$ Collisions}
 For production of positive pions
[$\pi^+$ mesons] the excess term could be laid down
by the following expressions \cite{bhat882}:

\begin{equation}\displaystyle
(B_{\pi^+})_{pp} = {\frac {4}{3}}{g^2_{p \pi
\pi}}{\frac{(P'+K)^2}{[(P'+K)^2-m_p^2]^2}} A(\nu,
q^2)_{\pi}\int{\frac{d^3k_{\pi}}{2k_0(2 \pi)^3}\exp(-ik_{\pi}x)},
\end{equation}
where the symbols have their contextual connotation with the
following hints to the physical reality of extraneous $\pi^+$, as
non-leading secondaries. The first parts of the above equations
(Eqn.(11)), contain the coupling strength parameters, the second
terms of the above equations are just the propagator for excited
nucleons. The third terms represent the common multiparticle
production amplitudes along with extraneous production modes and the
last terms indicate simply the phase space integration terms on the
probability of generation of a single $\pi^+$. These expressions are
to be calculated by the typical field-theoretical techniques and are
to be expressed -- if and when necessary -- in terms of the relevant
variable and/or measured observables.
\par
In order to arrive at the transverse momentum distribution of
$\pi^+$, one has to consider the Eqn. (1), eqn. (8) along with eqn. (11). For
excess $\pi^+$ production, a factor represented by $(1+\gamma^{\pi^+}
p_T^{\pi^+})$ is to be operated on $\frac{1}{2\pi
p_T}\frac{d^2N}{dp_Tdy}|_{Cu+Cu\rightarrow\pi^-+ X}$ as an
multiplier \cite{bhat882}. $\gamma^{\pi^+}\simeq (20\pi
g^2_{{\rho}{\pi}{\pi}}/<n_{\pi}>)/\sqrt{s}\simeq 0.44$ \cite{pgr07},
\cite{pgr03}. Taking $<p_T>_{\pi^+} \simeq 0.31$ GeV/c
\cite{phobos08}, the calculated values of $\alpha_{\pi^+}$,
$N_R^{\pi^+}$ and $\beta_{\pi^+}$ for different centralities and for
different energies are given in Table 2. In Figures 3(a), 3(b) and
3(c), we have plotted experimental versus theoretical results for
$\pi^+$ production in $Cu+Cu$ collisions at energies $\sqrt{s_{NN}}$
= 22.5, 62.4 and 200 GeV, respectively. Data are taken from the
Refs. \cite{phenix3} and \cite{phenix1}. The solid lines in those
Figures are the SCM-based plots.
 \subsubsection{Production of Positive K-mesons in
$Cu+Cu$ Collisions}
For the excess production of $K^+$-mesons we proceed in the same
path as we did earlier for the case of $\pi^+$ production. The
equation for the extraneous production of $K^+$ is given hereunder
\begin{equation}\displaystyle
(B_{K^+})_{pp} = {\frac{1}{2}}(4{\pi}g^2_{K N
\Lambda}+4{\pi}g^2_{\Sigma K N}){\frac{1} {[(P'+K)^2-m_p^2]^2}}
A(\nu, q^2)_K\int{\frac{d^3k_K}{2k_0(2 \pi)^3}\exp(-ik_K x)},
\end{equation}
\par
Adopting the above procedure, as we indicated for the production of
positive pions, we obtain for the transverse momentum distribution
of $K^+$ a multiplicative factor $\sim (1+\gamma^{K^+} p_T^{K^+})$to be
operated on $\frac{1}{2\pi
p_T}\frac{d^2N}{dp_Tdy}|_{Cu+Cu\rightarrow {K^-}+ X}$
\cite{bhat882}. For the production of $K^+$, the factor, calculated
from eqn.(12),
 $\gamma^{K^+}\simeq (4\pi g^2_{K N \Lambda} + 4\pi g^2_{\Sigma K N})/2\sqrt{s}\simeq 0.082$
\cite{pgr07}, \cite{pgr03}. We use the value of $<p_T>_{K^+} \simeq
0.36$ GeV/c \cite{phobos08}  and the corresponding
values of $\alpha_{K^+}$, $N_R^{K^+}$ and $\beta_{K^+}$ for
different energies are presented in the Table 3. The experimental
results \cite{phenix3}, \cite{phenix1} for the production of $K^+$
of different centralities and for different energies are plotted in
Figures 5(a), 5(b) and 5(c). The solid lines in those figures show
the theoretical plots.
\subsubsection{Production of Excess Protons}
Similarly, for the excess production of protons, the extraneous term
can be can be picked up from our previous work \cite{bhat882} in the
following form:
\begin{equation}\displaystyle
 (B_p)_{pp}= \frac{4 {\pi}{g^2_{N N
 \pi}}}{[(P'+K)^2-m_p^2]^2} A(\nu,q^2)_{p_s}
 \int{\frac{d^3k_p}{2(2 \pi)^3}\exp(-ik_p x)},
\end{equation}

 For the production of protons, we
obtain for the transverse momentum distribution of $p$ by operating
a multiplicative factor $\sim (1+\gamma^p p_T^{p})$, which is an
outcome of eqn. (13), on $\frac{1}{2\pi p_T}\frac{d^2N}{dp_Tdy}$.
The value of $\gamma^p\sim 0.32$ \cite{pgr07} and by taking $<p_T>_p
\simeq 0.50$ GeV/c \cite{phobos08}, we finally obtain the values of
$\alpha_{p}$, $N_R^{p}$ and $\beta_{p}$, which are given in the
Table 4. In Figures 7(a), 7(b) and 7(c), we have presented the
experimental values of invariant yields for proton-production versus
the theoretical SCM-based results for energies $\sqrt{s_{NN}}$ =
22.5, 62.4 and 200 GeV respectively. Data are taken from
\cite{phenix3} and \cite{phenix1}. The lines in the figures show the
theoretical outcomes.
\subsection{The Ratio Behaviours for Different Secondaries}
\subsubsection{The $\pi^-/\pi^+$ Ratios at $\sqrt{s_{NN}}$ = 62.4 and 200 GeV}
The  model-based $\pi^-/\pi^+$ ratios for different participating
nucleons $N_{part}$ at energies $\sqrt{s_{NN}}$ = 62.4 and 200 GeV
have been obtained from the expression (8) and Table 2. Data in
Figs. 8(a) and 8(b), shown by filled squares and blank circles
respectively, are taken from the PHOBOS group \cite{veres},
\cite{wosiek}. The theoretical values in this regard are plotted by
solid line in Figure 8(a) and by filled circles in 8(b).
\subsubsection{The $K^-/K^+$ Ratios at $\sqrt{s_{NN}}$ = 62.4 and 200 GeV}
In a similar way, the $N_{part}$ versus $K^-/K^+$ at $\sqrt{s_{NN}}$
= 62.4 and 200 GeV can be obtained from equation (8) and Table 3.
The calculated values of $K^-/K^+$ against the $N_{part}$ in the
light of SCM are shown by solid line in Fig. 9(a) at energy
$\sqrt{s_{NN}}$ = 62.4 GeV and by solid squares in Fig. 9(b) at
$\sqrt{s_{NN}}$ = 200 GeV. The data in those figures are taken from
PHOBOS \cite{veres}, \cite{wosiek}.
\subsubsection{Some Other Ratio-Behaviours at $\sqrt{s_{NN}}$ = 62.4 and 200 GeV}
Based on the SCM, the $\bar p/p$ ratios at different energies like
$\sqrt{s_{NN}}$ = 62.4 and 200 GeV and for different participating
nucleons $N_{part}$ are obtained with the help of equation (8) and
Table 4. The calculated values are plotted in Figs. 10(a) and 10(b)
at energies $\sqrt{s_{NN}}$ = 62.4 and 200 GeV by solid line and
filled circles respectively. Data in those figures are taken from
PHOBOS \cite{veres}, \cite{wosiek}.
\subsubsection{The $\bar p/\pi^-$ and $p/\pi^+$ Ratios at $\sqrt{s_{NN}}$ = 22.5, 62.4 and 200 GeV}
The expressions for $\bar p/\pi^-$ ratios against $p_T$ for central
reactions at energies $\sqrt{s_{NN}}$ = 22.5, 62.4 and 200 GeV can
be obtained from equation (1) and Table 2 and Table 4 and they are
given hereunder
\begin{equation}\displaystyle{
\frac{\bar
p}{\pi^-}=0.32p_T^{2.038}\exp(-0.15p_T^2)}~~~~~for~~\sqrt{s_{NN}}=22.5
~~GeV,
\end{equation}
\begin{equation}\displaystyle{
\frac{\bar
p}{\pi^-}=0.28p_T^{1.932}\exp(-0.15p_T^2)}~~~~~for~~\sqrt{s_{NN}}=62.4
~~GeV,
\end{equation}
\begin{equation}\displaystyle{
\frac{\bar
p}{\pi^-}=0.22p_T^{1.821}\exp(-0.13p_T^2)}~~~~~for~~\sqrt{s_{NN}}=200
~~GeV.
\end{equation}
And for $p/\pi^+$ ratios versus $p_T$ at energies $\sqrt{s_{NN}}$ =
22.5, 62.4 and 200 GeV the SCM-based equations are written as
\begin{equation}\displaystyle{
\frac{
p}{\pi^+}=0.85p_T^{2.038}\exp(-0.15p_T^2)}~~~~~for~~\sqrt{s_{NN}}=22.5
~~GeV,
\end{equation}
\begin{equation}\displaystyle{
\frac{
p}{\pi^+}=0.42p_T^{1.932}\exp(-0.15p_T^2)}~~~~~for~~\sqrt{s_{NN}}=62.4
~~GeV,
\end{equation}
\begin{equation}\displaystyle{
\frac{
p}{\pi^+}=0.35p_T^{1.821}\exp(-0.13p_T^2)}~~~~~for~~\sqrt{s_{NN}}=200
~~GeV.
\end{equation}
In Figures 11(a) and 11(b) we have plotted $p_T$ versus $\bar
p/\pi^-$ and $p/\pi^+$ respectively for central  $Cu+Cu$ collisions
at $\sqrt{s_{NN}}$ = 22.5, 62.4 and 200 GeV. Data of these figures
are taken from PHENIX \cite{chujo07}. Lines in Figures 11(a) and
11(b) represent eqn. (14)- eqn. (16) and eqn. (17) to eqn. (19)
respectively.
\subsection{Nuclear Modification Factors}
In this subsection we would dwell upon the Nuclear Modification
Factors of two types viz., $R_{AA}$ and $R_{CP}$. In 3.4.1 the
former ($R_{AA}$) would be defined and the results would be hinted,
though the figures for both would be shown in the next section. And
in 3.4.2 the second one ($R_{CP}$) would be treated in some detail.
\subsubsection{The Nuclear Modification Factor, $R_{AA}$ } The nuclear modification
factor (NMF), designated as $R_{AA}$, for any secondary $C$, is
defined by \cite{chujo07}
\begin{equation}\displaystyle{
R_{AA}^C=\frac{(1/N^{evt}_{AA})d^2N_{AA}^C/dp_Tdy}{<N_{coll}(b)>/\sigma_{pp}^{inel}\times
d^2\sigma_{pp}^C/dp_Tdy }.}
\end{equation}
Depending on this definition, the SCM-based results  on NMFs for
$Cu+Cu$ collisions at $\sqrt{s_{NN}}$ = 22.5, 62.4 and 200 GeV are
deduced on the basis of Eqn.(8), Table 1 and Table 2 and they are
given by the undernoted relations
\begin{equation}\displaystyle{
R_{AA}=0.730p_T^{0.621}~~~~~for~~\sqrt{s_{NN}}=22.5 ~~GeV,}
\end{equation}
\begin{equation}\displaystyle{
R_{AA}=0.680p_T^{0.311}~~~~~for~~\sqrt{s_{NN}}=62.4 ~~GeV,}
\end{equation}
\begin{equation}\displaystyle{
R_{AA}=0.315p_T^{0.234}~~~~~for~~\sqrt{s_{NN}}=200 ~~GeV.}
\end{equation}
wherein the values of $<N_{coll}(b)>$ to be used are $\approx$
140.7, 152.3 and 182.7 \cite{phenix08} for $Cu+Cu$ collisions at
$\sqrt{s_{NN}}$ = 22.5, 62.4 and 200 GeV respectively. For
$N^{evt}$, we use the values $\approx$ $5.8\times 10^6$, $192\times
10^6$ and $794\times 10^6$ \cite{phenix08} at three different
energies and $\sigma_{pp}$ to be used as 30 mb \cite{sb02}. Our
model-based plot is shown in a figure in the next section.
\subsubsection{The Nuclear Modification Factor, $R_{CP}$ } There is
yet another nuclear modification factor, $R_{CP}$ which reflects
precisely the hadron $p_T$ spectra in different centrality bins and
presents comparison of the $p_T$-spectra between a collision at a
specific centrality and the one at the relatively peripheral
collision. It is quantified as
\begin{equation}\displaystyle{
R_{CP}=\frac{[d^2N/(2\pi p_T dp_Tdy)/N_{bin}]^{central}}{[d^2N/(2\pi p_T dp_Tdy)/N_{bin}]^{peripheral}}.}
\end{equation}
  According to above definition, with the centrality set at
  $0-10\%$ and the peripherality at $60-94\%$, our model-based
  expression for $R_{CP}$ in case of neutral pions ($\pi^0$s) is
  given by
\begin{equation}\displaystyle{
R_{CP}^{\pi^0}=0.543p_T^{-0.054}.}
\end{equation}
wherein we have made use of the values shown in Table-2. The
$R_{CP}$ for neutral kaons ($K^0$s) with the same centrality and the
peripherality changed to $40-60\%$, our model-based result is
\begin{equation}\displaystyle{
R_{CP}^{K^0}=0.664p_T^{-0.235}.}
\end{equation}
wherein we have taken $K^0=1/2(K^+ +K^-)$, and used the necessary
values from Table-3.\\ Our model-dependent plots on these two
particular observables, $R_{AA}$ and $R_{CP}$, are shown in the
subsequent section on comparative studies.
\section{Data and Results on Some Select Observables: A Comparison between Models}
We strongly uphold the view that point-to-point or
secondary-to-secondary specific comparisons between our model-based
results with both data and the other model-based calculations would
be quite meaningful and physically significant. But the main and
major constraint in this attempt is the lack of availability of such
comprehensive calculations encompassing all the light secondaries.
In so far as $Cu+Cu$ collisions are concerned, we have, so far, come
across two model-based studies, of which one is the application of
the Quark Combination Model (QCM) made by Fei et al. \cite{fei} for
the production of neutral pion/kaon and the other is the
pQCD-oriented theoretical study done by Vitev \cite{vitev} for
production of only the neutral pion and for no other secondary. This
constitutes a gross limitation to the successful completion of the
comparison-aspects in the present study. However, we have tried here
to show some comparison(s) only for one or two observables related
to a few selected neutral secondaries, with whatever little other
model-dependent studies could be obtained uptil now.
\par
In spite of the difficulties mentioned in the above paragraph, in the adjoining Fig.12 we compare the data-versus-results
based on two models for production of (a) neutral pion and (b) neutral kaon in $Cu+Cu$ collision at $\sqrt{s_{NN}}$ = 200 GeV.
The standard variety of models that are reckoned with here for comparison with the present SCM are (i) The Quark Combination Model (QCM) and (ii) Perturbative QCD-inspired Vitev's Model. In the former (QCM), the main idea is to line up the quarks and antiquarks in a one-dimensional order in phase space, e.g. in rapidity and let them combine into initial hadrons one by one following a combination rule \cite{fei}. These initial hadrons through combination of constituent quarks are then allowed to decay into the final state hadrons through the decay program of PYTHIA 6.1. The calculational steps, in this model, proceed on the basis of two-prong assumptions of two-component model (based on `soft'-`hard' artifact) and the concept of parton-hadron duality.
\par
The latter model used by us for comparison is one of Vitev \cite{vitev}. This is essentially a pQCD-oriented model with an analytic model of jet-quenching which embraces medium-induced energy loss after hard partonic scattering. This approach reduces the jet cross-section in the presence of the medium but leaves the parton fragmentation function unaltered. In fact, this feature was conveniently implemented in the analytic model of Vitev \cite{vitev} for QGP-induced leading hadron suppression. Vitev actually made use of a Hagedornian form of power law expressions for invariant cross-section(s) and inducted also the radiative energy-loss formalism.
\par
In Fig. (12), the invariant yields versus $p_T$ (GeV/c) for (a) $\pi^0$ and (b) $K_S^0$ for different centralities in $Cu+Cu$ collision at $\sqrt{s_{NN}}$ = 200 GeV have been plotted. Data are taken from \cite{phenix07} and \cite{star07}. The solid lines in those Figures represent the SCM-based calculations wherein the dashed lines show the QCM-oriented results \cite{fei}. And the dotted line in Fig. 12(a) shows pQCD-inspired calculations done by Vitev \cite{vitev} for production of only the neutral pions in $Cu+Cu$ collision.
\subsection{The Nuclear Modification Factor, $R_{AA}$ }
 In Fig.
13, we plot $R_{AA}$ vs. $p_T$ at energies (a) $\sqrt{s_{NN}}$ =
22.5 GeV, (b) $\sqrt{s_{NN}}$ = 62.4 GeV and (c) $\sqrt{s_{NN}}$ =
200 GeV. The solid lines in the figures show the SCM-based results,
wherein the experimental results  are taken from Refs.
\cite{chujo07}, \cite{awes} and \cite{reygers}. The dotted lines in
the figures represent the pQCD-based calculations
\cite{phenix08},\cite{vitev}. Moreover, in Fig. 13(d) we have plotted
the $<R_{AA}>$ vs. $N_{part}$ between the range $2.5<p_T<3.5$ Gev/c for the same collision and at the stated energies.
Data-points are taken from the experiments by PHENIX Collaboration
\cite{phenix08}. The solid lines in the Fig.13(d) represent the our
calculationl results and the dashed lines show the corresponding
pQCD-oriented theoretical calculations made by Vitev  \cite{phenix08},\cite{vitev}.
\subsection{The Nuclear Modification Factor, $R_{CP}$ }
 In Fig. 14(a) we have plotted $R_{CP}$ for the centralities $0-10\%$
and $60-94\%$ against $p_T$ for the $\pi^0$.The solid line in that
Figure represents the SCM-predicted results arising out of the
eqn.(25) and the dotted line in that gives the prediction from Quark
Combination Model \cite{fei}. No data on $R_{CP}$ for production of
neutral pions at RHIC energies have yet been reported.
\par
Similarly, in Fig. 14 (b) we have plotted $R_{CP}$  for the
centralities $0-10\%$ and $40-60\%$ against $p_T$ for the $K^0$. The
solid line in the Figure 14(b) depicts the SCM-based plot against
the experimental result \cite{star07} while the dotted line in the
same figure represents the results attained by the Quark Combination
Model \cite{fei}.
\section{Concluding Remarks}
The model applied here gives fair description of the $p_T$-spectra
of all the light secondaries with the chosen values of the two
parameters. Besides, the centrality-dependence of the $p_T$-spectra
is also well-reproduced as is indicated by the figures. Slight
disagreements observed at very low-$p_T$ ($p_T<<1$ GeV/c) are due to
the fact that the model has turned effectively into a mixed one with
the entry of a power-law form due to the physics of partonic rearrangement factor.
This dominance of power-law form disturbs, to a considerable extent,
the agreement between data and model-based calculations for the
extremely `soft' (very low-$p_T$) values. Among the secondaries
produced, the particle-antiparticle ratios and the proton-pion
ratios are also in good agreement with the measured values. The
obtained nuclear modification factors represented by $R_{AA}$ and
$R_{CP}$ (central-to-peripheral)are also in accord with the
measurements. It is to be noted that we achieve all these with a new
mechanism and introduction of some simple and basic ansatz like, the
physics of large-$p_T$ nucleus-nucleus collisions and by introducing
the properties of factorization, scale-breaking, mixed models with the combine of
power-and-exponential laws, along with the principles of structural
rearrangement factors at large transverse momenta.
\par
Selected comparisons of our model-based results with two other
model-dependent calculations reveal neither sharp disagreement with
any of them, nor very splendid agreement with either of them which
are generically of standard model variety. Rather, on an overall
basis, our results are in better agreement with data than either of
them. This is a factor which is of some substance and importance to
us.
\par
From a careful scrutiny of the fit-parameters we discover the
following properties of them: (i) the structural rearrangement
factor is clearly centrality dependent; it increases very slowly
with gradual rise in the peripherality of the collisions. (ii)
Secondly, the coefficients of the $p_T^2$ in the exponential term
are clearly energy-dependent in nature; on the contrary they
manifest themselves to be independent of the centrality-measure of
the interactions.
\par
The so-called suppression of the cross-sections at large-$p_T$ in
heavy ion collisions is addressed here without resorting to the
ideas of the `jet-quenching' which is perceived to be to be one of
the main conceptual pillars of heavy ion physics.
\par
An interesting question crops up in this connection. Here,we dealt
with some aspects of $Cu+Cu$ collisions at RHIC energies. We have
chosen to remain silent about the physics of `quark-gluon plasma'
(QGP) formation. The perturbative quantum chromodynamics (pQCD)
predicted the formation of quark-gluon plasma (QGP). But the RHIC
experiments failed to detect any plasma state; rather they found a
``new kind of fluid state with very low viscosity''. So QCD
prediction faltered at the first pillar. Secondly, how perfect the
fluid observed at RHIC is cannot still be ascertained; the answer is
not yet without caveats. Thirdly, we do not consider this proposed
`plasma'-state to be any startling revelation, because when heated
to very high temperatures, caused by the thermal motion of the
molecules in the macroscopic matter,the solid substances melt down
and turn into a variety of liquid.  Almost in a similar manner, if
microscopic particle-constituent matter is raised to very high
temperatures attained by the extremely energetic collisions, the
microscopic matter might also be converted to a liquid of somewhat
unknown nature, and thus obviously to a `new' kind
\cite{majumder}-\cite{goutam09}. So we do not pay much attention to
the pQCD-based predictions on QGP and /or of suppression
phenomena.
\par
Finally, we sum up by stating that (i) the model under consideration
here explains and accommodates quite well the data on $Cu+Cu$
collisions at various energies; and (ii) Quite agreeably, the values
of $R_{CP}$ obtained by the present calculations are not in good
agreement with data. This could be attributed to our neglect of the
effects of final state re-scatterings and some other complex
physical factors.
\par
{\bf{Acknowledgements}}\\
The authors would like to express their thankful gratitude to the learned Referee for his/her valuable remarks and constructive suggestions in   improving the earlier draft of the manuscript.

\begin{table}
\singlespacing \caption{Values of $\alpha$, $N_R$ and $\beta$ for
$\pi^-$ and $\pi^+$ productions in $Cu+Cu$ collisions at
$\sqrt{s_{NN}}$=22.5, 62.4 and 200 GeV}
\begin{center}
\begin{tabular}{c|c}
\hline Centrality&$\sqrt{s_{NN}}$=22.5 GeV\\
 &\begin{tabular}{c|c} $\pi^-$&$\pi^+$\\
\end{tabular}\\
 &\begin{tabular}{c|c} \begin{tabular}{ccc}
$\alpha_{\pi^-}$&$N_R^{\pi^-}$&$\beta_{\pi^-}$\\
\end{tabular}
&\begin{tabular}{ccc}
$\alpha_{\pi^+}$&$N_R^{\pi^+}$&$\beta_{\pi^+}$\\
\end{tabular}\\
\end{tabular}\\
\hline 0-10$\%$&\begin{tabular}{c|c} \begin{tabular}{ccc}
0.901&3.454&0.703\\
\end{tabular}
&\begin{tabular}{ccc}
1.001&3.454&0.703\\
\end{tabular}\\
\end{tabular}\\
 10-30$\%$&\begin{tabular}{c|c} \begin{tabular}{ccc}
0.492&3.444&0.703\\
\end{tabular}
&\begin{tabular}{ccc}
0.572&3.444&0.703\\
\end{tabular}\\
\end{tabular}\\
 30-60$\%$&\begin{tabular}{c|c} \begin{tabular}{ccc}
0.239&3.413&0.703\\
\end{tabular}
&\begin{tabular}{ccc}
0.244&3.413&0.703\\
\end{tabular}\\
\end{tabular}\\
 60-100$\%$&\begin{tabular}{c|c} \begin{tabular}{ccc}
0.033&3.410&0.703\\
\end{tabular}
&\begin{tabular}{ccc}
0.034&3.410&0.703\\
\end{tabular}\\
\end{tabular}\\
 Minbias&\begin{tabular}{c|c} \begin{tabular}{ccc}
0.395&3.431&0.703\\
\end{tabular}
&\begin{tabular}{ccc}
0.468&3.431&0.703\\
\end{tabular}\\
\end{tabular}\\
\hline
Centrality&$\sqrt{s_{NN}}$=62.4 GeV\\
 &\begin{tabular}{c|c} $\pi^-$&$\pi^+$\\
\end{tabular}\\
 &\begin{tabular}{c|c} \begin{tabular}{ccc}
$\alpha_{\pi^-}$&$N_R^{\pi^-}$&$\beta_{\pi^-}$\\
\end{tabular}
&\begin{tabular}{ccc}
$\alpha_{\pi^+}$&$N_R^{\pi^+}$&$\beta_{\pi^+}$\\
\end{tabular}\\
\end{tabular}\\
\hline 0-10$\%$&\begin{tabular}{c|c} \begin{tabular}{ccc}
0.707&3.035&0.468\\
\end{tabular}
&\begin{tabular}{ccc}
0.786&3.035&0.468\\
\end{tabular}\\
\end{tabular}\\
 10-30$\%$&\begin{tabular}{c|c} \begin{tabular}{ccc}
0.464&3.030&0.468\\
\end{tabular}
&\begin{tabular}{ccc}
0.474&3.030&0.468\\
\end{tabular}\\
\end{tabular}\\
 30-60$\%$&\begin{tabular}{c|c} \begin{tabular}{ccc}
0.180&3.026&0.468\\
\end{tabular}
&\begin{tabular}{ccc}
0.228&3.026&0.468\\
\end{tabular}\\
\end{tabular}\\
 60-100$\%$&\begin{tabular}{c|c} \begin{tabular}{ccc}
0.038&3.016&0.468\\
\end{tabular}
&\begin{tabular}{ccc}
0.044&3.016&0.468\\
\end{tabular}\\
\end{tabular}\\
 Minbias&\begin{tabular}{c|c} \begin{tabular}{ccc}
0.204&3.142&0.468\\
\end{tabular}
&\begin{tabular}{ccc}
0.207&3.142&0.468\\
\end{tabular}\\
\end{tabular}\\
\hline
Centrality&$\sqrt{s_{NN}}$=200 GeV\\
 &\begin{tabular}{c|c} $\pi^-$&$\pi^+$\\
\end{tabular}\\
 &\begin{tabular}{c|c} \begin{tabular}{ccc}
$\alpha_{\pi^-}$&$N_R^{\pi^-}$&$\beta_{\pi^-}$\\
\end{tabular}
&\begin{tabular}{ccc}
$\alpha_{\pi^+}$&$N_R^{\pi^+}$&$\beta_{\pi^+}$\\
\end{tabular}\\
\end{tabular}\\
\hline 0-5$\%$&\begin{tabular}{c|c} \begin{tabular}{ccc}
1.098&3.597&0.293\\
\end{tabular}
&\begin{tabular}{ccc}
1.113&3.597&0.293\\
\end{tabular}\\
\end{tabular}\\
 5-10$\%$&\begin{tabular}{c|c} \begin{tabular}{ccc}
0.898&3.572&0.293\\
\end{tabular}
&\begin{tabular}{ccc}
0.910&3.572&0.293\\
\end{tabular}\\
\end{tabular}\\
 10-15$\%$&\begin{tabular}{c|c} \begin{tabular}{ccc}
0.750&3.567&0.293\\
\end{tabular}
&\begin{tabular}{ccc}
0.760&3.567&0.293\\
\end{tabular}\\
\end{tabular}\\
 15-20$\%$&\begin{tabular}{c|c} \begin{tabular}{ccc}
0.687&3.552&0.293\\
\end{tabular}
&\begin{tabular}{ccc}
0.696&3.552&0.293\\
\end{tabular}\\
\end{tabular}\\
 20-30$\%$&\begin{tabular}{c|c} \begin{tabular}{ccc}
0.635&3.544&0.293\\
\end{tabular}
&\begin{tabular}{ccc}
0.643&3.544&0.293\\
\end{tabular}\\
\end{tabular}\\
30-40$\%$&\begin{tabular}{c|c} \begin{tabular}{ccc}
0.523&3.525&0.293\\
\end{tabular}
&\begin{tabular}{ccc}
0.530&3.525&0.293\\
\end{tabular}\\
\end{tabular}\\
40-50$\%$&\begin{tabular}{c|c} \begin{tabular}{ccc}
0.353&3.518&0.293\\
\end{tabular}
&\begin{tabular}{ccc}
0.358&3.518&0.293\\
\end{tabular}\\
\end{tabular}\\
50-60$\%$&\begin{tabular}{c|c} \begin{tabular}{ccc}
0.283&3.508&0.293\\
\end{tabular}
&\begin{tabular}{ccc}
0.287&3.508&0.293\\
\end{tabular}\\
\end{tabular}\\
60-70$\%$&\begin{tabular}{c|c} \begin{tabular}{ccc}
0.113&3.491&0.293\\
\end{tabular}
&\begin{tabular}{ccc}
0.115&3.491&0.293\\
\end{tabular}\\
\end{tabular}\\
70-80$\%$&\begin{tabular}{c|c} \begin{tabular}{ccc}
0.085&3.476&0.293\\
\end{tabular}
&\begin{tabular}{ccc}
0.086&3.476&0.293\\
\end{tabular}\\
\end{tabular}\\
80-92$\%$&\begin{tabular}{c|c} \begin{tabular}{ccc}
0.004&3.453&0.293\\
\end{tabular}
&\begin{tabular}{ccc}
0.004&3.453&0.293\\
\end{tabular}\\
\end{tabular}\\
\hline
\end{tabular}
\end{center}
\end{table}
\begin{table}\singlespacing
\caption{Values of $\alpha$, $N_R$ and $\beta$ for $K^-$ and $K^+$
productions in $Cu+Cu$ collisions at $\sqrt{s_{NN}}$=22.5, 62.4 and
200 GeV}
\begin{center}
\begin{tabular}{c|c}
\hline Centrality&$\sqrt{s_{NN}}$=22.5 GeV\\
 &\begin{tabular}{c|c} $K^-$&$K^+$\\
\end{tabular}\\
 &\begin{tabular}{c|c} \begin{tabular}{ccc}
$\alpha_{K^-}$&$N_R^{K^-}$&$\beta_{K^-}$\\
\end{tabular}
&\begin{tabular}{ccc}
$\alpha_{K^+}$&$N_R^{K^+}$&$\beta_{K^+}$\\
\end{tabular}\\
\end{tabular}\\
\hline 0-10$\%$&\begin{tabular}{c|c} \begin{tabular}{ccc}
0.406&2.304&0.863\\
\end{tabular}
&\begin{tabular}{ccc}
0.469&2.304&0.863\\
\end{tabular}\\
\end{tabular}\\
 10-30$\%$&\begin{tabular}{c|c} \begin{tabular}{ccc}
0.185&2.285&0.863\\
\end{tabular}
&\begin{tabular}{ccc}
0.255&2.285&0.863\\
\end{tabular}\\
\end{tabular}\\
 30-60$\%$&\begin{tabular}{c|c} \begin{tabular}{ccc}
0.069&2.274&0.863\\
\end{tabular}
&\begin{tabular}{ccc}
0.114&2.274&0.863\\
\end{tabular}\\
\end{tabular}\\
 60-100$\%$&\begin{tabular}{c|c} \begin{tabular}{ccc}
0.012&2.270&0.863\\
\end{tabular}
&\begin{tabular}{ccc}
0.018&2.270&0.863\\
\end{tabular}\\
\end{tabular}\\
 Minbias&\begin{tabular}{c|c} \begin{tabular}{ccc}
0.118&2.264&0.863\\
\end{tabular}
&\begin{tabular}{ccc}
0.151&2.264&0.863\\
\end{tabular}\\
\end{tabular}\\
\hline
Centrality&$\sqrt{s_{NN}}$=62.4 GeV\\
 &\begin{tabular}{c|c} $K^-$&$K^+$\\
\end{tabular}\\
 &\begin{tabular}{c|c} \begin{tabular}{ccc}
$\alpha_{K^-}$&$N_R^{K^-}$&$\beta_{K^-}$\\
\end{tabular}
&\begin{tabular}{ccc}
$\alpha_{K^+}$&$N_R^{K^+}$&$\beta_{K^+}$\\
\end{tabular}\\
\end{tabular}\\
\hline 0-10$\%$&\begin{tabular}{c|c} \begin{tabular}{ccc}
0.675&2.714&0.571\\
\end{tabular}
&\begin{tabular}{ccc}
0.755&2.714&0.571\\
\end{tabular}\\
\end{tabular}\\
 10-30$\%$&\begin{tabular}{c|c} \begin{tabular}{ccc}
0.405&2.704&0.571\\
\end{tabular}
&\begin{tabular}{ccc}
0.524&2.704&0.571\\
\end{tabular}\\
\end{tabular}\\
 30-60$\%$&\begin{tabular}{c|c} \begin{tabular}{ccc}
0.162&2.688&0.571\\
\end{tabular}
&\begin{tabular}{ccc}
0.187&2.688&0.571\\
\end{tabular}\\
\end{tabular}\\
 60-100$\%$&\begin{tabular}{c|c} \begin{tabular}{ccc}
0.023&2.658&0.571\\
\end{tabular}
&\begin{tabular}{ccc}
0.028&2.658&0.571\\
\end{tabular}\\
\end{tabular}\\
 Minbias&\begin{tabular}{c|c} \begin{tabular}{ccc}
0.217&2.681&0.571\\
\end{tabular}
&\begin{tabular}{ccc}
0.251&2.681&0.571\\
\end{tabular}\\
\end{tabular}\\
\hline
Centrality&$\sqrt{s_{NN}}$=200 GeV\\
 &\begin{tabular}{c|c} $K^-$&$K^+$\\
\end{tabular}\\
 &\begin{tabular}{c|c} \begin{tabular}{ccc}
$\alpha_{K^-}$&$N_R^{K^-}$&$\beta_{K^-}$\\
\end{tabular}
&\begin{tabular}{ccc}
$\alpha_{K^+}$&$N_R^{K^+}$&$\beta_{K^+}$\\
\end{tabular}\\
\end{tabular}\\
\hline 0-5$\%$&\begin{tabular}{c|c} \begin{tabular}{ccc}
0.286&2.939&0.417\\
\end{tabular}
&\begin{tabular}{ccc}
0.294&2.939&0.417\\
\end{tabular}\\
\end{tabular}\\
 5-10$\%$&\begin{tabular}{c|c} \begin{tabular}{ccc}
0.250&2.839&0.417\\
\end{tabular}
&\begin{tabular}{ccc}
0.257&2.839&0.417\\
\end{tabular}\\
\end{tabular}\\
 10-15$\%$&\begin{tabular}{c|c} \begin{tabular}{ccc}
0.214&2.819&0.417\\
\end{tabular}
&\begin{tabular}{ccc}
0.220&2.819&0.417\\
\end{tabular}\\
\end{tabular}\\
 15-20$\%$&\begin{tabular}{c|c} \begin{tabular}{ccc}
0.210&2.805&0.417\\
\end{tabular}
&\begin{tabular}{ccc}
0.216&2.805&0.417\\
\end{tabular}\\
\end{tabular}\\
 20-30$\%$&\begin{tabular}{c|c} \begin{tabular}{ccc}
0.181&2.795&0.417\\
\end{tabular}
&\begin{tabular}{ccc}
0.186&2.795&0.417\\
\end{tabular}\\
\end{tabular}\\
30-40$\%$&\begin{tabular}{c|c} \begin{tabular}{ccc}
0.165&2.765&0.417\\
\end{tabular}
&\begin{tabular}{ccc}
0.170&2.765&0.417\\
\end{tabular}\\
\end{tabular}\\
40-50$\%$&\begin{tabular}{c|c} \begin{tabular}{ccc}
0.112&2.734&0.417\\
\end{tabular}
&\begin{tabular}{ccc}
0.115&2.734&0.417\\
\end{tabular}\\
\end{tabular}\\
50-60$\%$&\begin{tabular}{c|c} \begin{tabular}{ccc}
0.051&2.714&0.417\\
\end{tabular}
&\begin{tabular}{ccc}
0.053&2.714&0.417\\
\end{tabular}\\
\end{tabular}\\
60-70$\%$&\begin{tabular}{c|c} \begin{tabular}{ccc}
0.041&2.698&0.417\\
\end{tabular}
&\begin{tabular}{ccc}
0.042&2.698&0.417\\
\end{tabular}\\
\end{tabular}\\
70-80$\%$&\begin{tabular}{c|c} \begin{tabular}{ccc}
0.024&2.672&0.417\\
\end{tabular}
&\begin{tabular}{ccc}
0.026&2.672&0.417\\
\end{tabular}\\
\end{tabular}\\
80-92$\%$&\begin{tabular}{c|c} \begin{tabular}{ccc}
0.013&2.652&0.417\\
\end{tabular}
&\begin{tabular}{ccc}
0.014&2.652&0.417\\
\end{tabular}\\
\end{tabular}\\
\hline
\end{tabular}
\end{center}
\end{table}
\begin{table}\singlespacing
\caption{Values of $\alpha$, $N_R$ and $\beta$ for $\bar p$ and $p$
productions in $Cu+Cu$ collisions at $\sqrt{s_{NN}}$=22.5, 62.4 and
200 GeV}
\begin{center}
\begin{tabular}{c|c}
\hline Centrality&$\sqrt{s_{NN}}$=22.5 GeV\\
 &\begin{tabular}{c|c} $\bar p$&$p$\\
\end{tabular}\\
 &\begin{tabular}{c|c} \begin{tabular}{ccc}
$\alpha_{\bar p}$&$N_R^{\bar p}$&$\beta_{\bar p}$\\
\end{tabular}
&\begin{tabular}{ccc}
$\alpha_{p}$&$N_R^{p}$&$\beta_{p}$\\
\end{tabular}\\
\end{tabular}\\
\hline 0-10$\%$&\begin{tabular}{c|c} \begin{tabular}{ccc}
0.211&0.826&0.853\\
\end{tabular}
&\begin{tabular}{ccc}
0.344&0.826&0.853\\
\end{tabular}\\
\end{tabular}\\
 10-30$\%$&\begin{tabular}{c|c} \begin{tabular}{ccc}
0.111&0.780&0.853\\
\end{tabular}
&\begin{tabular}{ccc}
0.129&0.780&0.853\\
\end{tabular}\\
\end{tabular}\\
 30-60$\%$&\begin{tabular}{c|c} \begin{tabular}{ccc}
0.044&0.695&0.853\\
\end{tabular}
&\begin{tabular}{ccc}
0.051&0.695&0.853\\
\end{tabular}\\
\end{tabular}\\
 60-100$\%$&\begin{tabular}{c|c} \begin{tabular}{ccc}
0.008&0.691&0.853\\
\end{tabular}
&\begin{tabular}{ccc}
0.010&0.691&0.853\\
\end{tabular}\\
\end{tabular}\\
 Minbias&\begin{tabular}{c|c} \begin{tabular}{ccc}
0.070&0.798&0.853\\
\end{tabular}
&\begin{tabular}{ccc}
0.101&0.798&0.853\\
\end{tabular}\\
\end{tabular}\\
\hline
Centrality&$\sqrt{s_{NN}}$=62.4 GeV\\
 &\begin{tabular}{c|c} $\bar p$&$p$\\
\end{tabular}\\
 &\begin{tabular}{c|c} \begin{tabular}{ccc}
$\alpha_{\bar p}$&$N_R^{\bar p}$&$\beta_{\bar p}$\\
\end{tabular}
&\begin{tabular}{ccc}
$\alpha_{p}$&$N_R^{p}$&$\beta_{p}$\\
\end{tabular}\\
\end{tabular}\\
\hline 0-10$\%$&\begin{tabular}{c|c} \begin{tabular}{ccc}
0.468&1.118&0.618\\
\end{tabular}
&\begin{tabular}{ccc}
0.957&1.118&0.618\\
\end{tabular}\\
\end{tabular}\\
 10-30$\%$&\begin{tabular}{c|c} \begin{tabular}{ccc}
0.256&0.958&0.618\\
\end{tabular}
&\begin{tabular}{ccc}
0.513&0.958&0.618\\
\end{tabular}\\
\end{tabular}\\
 30-60$\%$&\begin{tabular}{c|c} \begin{tabular}{ccc}
0.108&0.945&0.618\\
\end{tabular}
&\begin{tabular}{ccc}
0.192&0.945&0.618\\
\end{tabular}\\
\end{tabular}\\
 60-100$\%$&\begin{tabular}{c|c} \begin{tabular}{ccc}
0.012&0.930&0.618\\
\end{tabular}
&\begin{tabular}{ccc}
0.023&0.930&0.618\\
\end{tabular}\\
\end{tabular}\\
 Minbias&\begin{tabular}{c|c} \begin{tabular}{ccc}
0.184&1.178&0.618\\
\end{tabular}
&\begin{tabular}{ccc}
0.298&1.178&0.618\\
\end{tabular}\\
\end{tabular}\\
\hline
Centrality&$\sqrt{s_{NN}}$=200 GeV\\
 &\begin{tabular}{c|c} $\bar p$&$p$\\
\end{tabular}\\
 &\begin{tabular}{c|c} \begin{tabular}{ccc}
$\alpha_{\bar p}$~&$N_R^{\bar p}$~&$\beta_{\bar p}$\\
\end{tabular}
&\begin{tabular}{ccc}
$\alpha_{p}$&$N_R^{p}$&$\beta_{p}$\\
\end{tabular}\\
\end{tabular}\\
\hline 0-5$\%$&\begin{tabular}{c|c} \begin{tabular}{ccc}
0.166&1.251&0.426\\
\end{tabular}
&\begin{tabular}{ccc}
0.185&1.251&0.426\\
\end{tabular}\\
\end{tabular}\\
 5-10$\%$&\begin{tabular}{c|c} \begin{tabular}{ccc}
0.156&1.231&0.426\\
\end{tabular}
&\begin{tabular}{ccc}
0.181&1.231&0.426\\
\end{tabular}\\
\end{tabular}\\
 10-15$\%$&\begin{tabular}{c|c} \begin{tabular}{ccc}
0.113&1.211&0.426\\
\end{tabular}
&\begin{tabular}{ccc}
0.131&1.211&0.426\\
\end{tabular}\\
\end{tabular}\\
 15-20$\%$&\begin{tabular}{c|c} \begin{tabular}{ccc}
0.105&1.192&0.426\\
\end{tabular}
&\begin{tabular}{ccc}
0.122&1.192&0.426\\
\end{tabular}\\
\end{tabular}\\
 20-30$\%$&\begin{tabular}{c|c} \begin{tabular}{ccc}
0.087&1.172&0.426\\
\end{tabular}
&\begin{tabular}{ccc}
0.101&1.172&0.426\\
\end{tabular}\\
\end{tabular}\\
30-40$\%$&\begin{tabular}{c|c} \begin{tabular}{ccc}
0.074&1.154&0.426\\
\end{tabular}
&\begin{tabular}{ccc}
0.086&1.154&0.426\\
\end{tabular}\\
\end{tabular}\\
40-50$\%$&\begin{tabular}{c|c} \begin{tabular}{ccc}
0.038&1.133&0.426\\
\end{tabular}
&\begin{tabular}{ccc}
0.044&1.133&0.426\\
\end{tabular}\\
\end{tabular}\\
50-60$\%$&\begin{tabular}{c|c} \begin{tabular}{ccc}
0.028&1.112&0.426\\
\end{tabular}
&\begin{tabular}{ccc}
0.032&1.112&0.426\\
\end{tabular}\\
\end{tabular}\\
60-70$\%$&\begin{tabular}{c|c} \begin{tabular}{ccc}
0.017&1.105&0.426\\
\end{tabular}
&\begin{tabular}{ccc}
0.020&1.105&0.426\\
\end{tabular}\\
\end{tabular}\\
70-80$\%$&\begin{tabular}{c|c} \begin{tabular}{ccc}
0.006&1.099&0.426\\
\end{tabular}
&\begin{tabular}{ccc}
0.007&1.099&0.426\\
\end{tabular}\\
\end{tabular}\\
80-92$\%$&\begin{tabular}{c|c} \begin{tabular}{ccc}
0.003&1.075&0.426\\
\end{tabular}
&\begin{tabular}{ccc}
0.004&1.075&0.426\\
\end{tabular}\\
\end{tabular}\\
\hline
\end{tabular}
\end{center}
\end{table}

\newpage
\singlespacing

\newpage
\begin{figure}
\subfigure[]{
\begin{minipage}{.5\textwidth}
\centering
\includegraphics[width=2.5in]{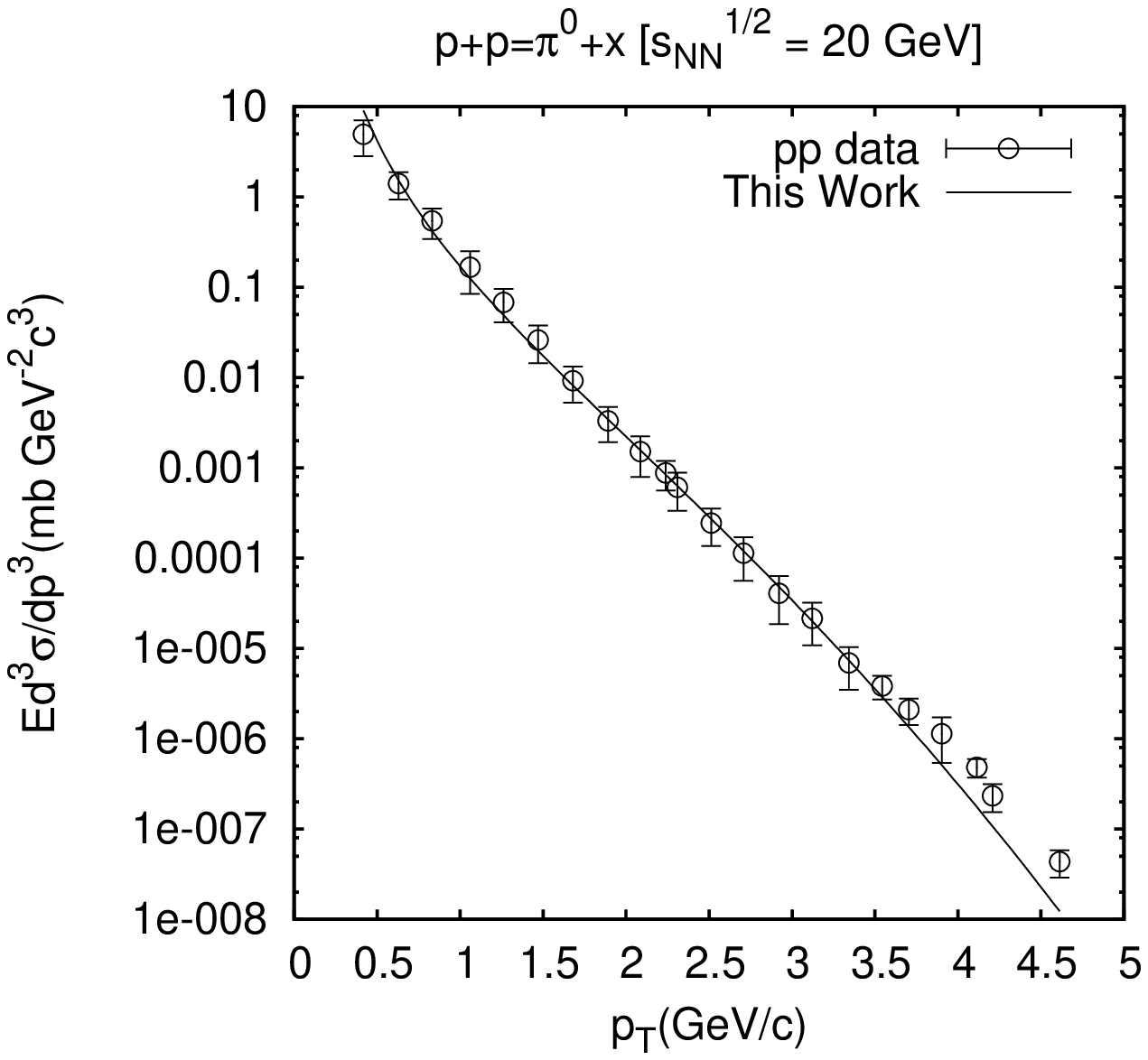}
\setcaptionwidth{2.6in}
\end{minipage}}%
\subfigure[]{
\begin{minipage}{0.5\textwidth}
\centering
 \includegraphics[width=2.5in]{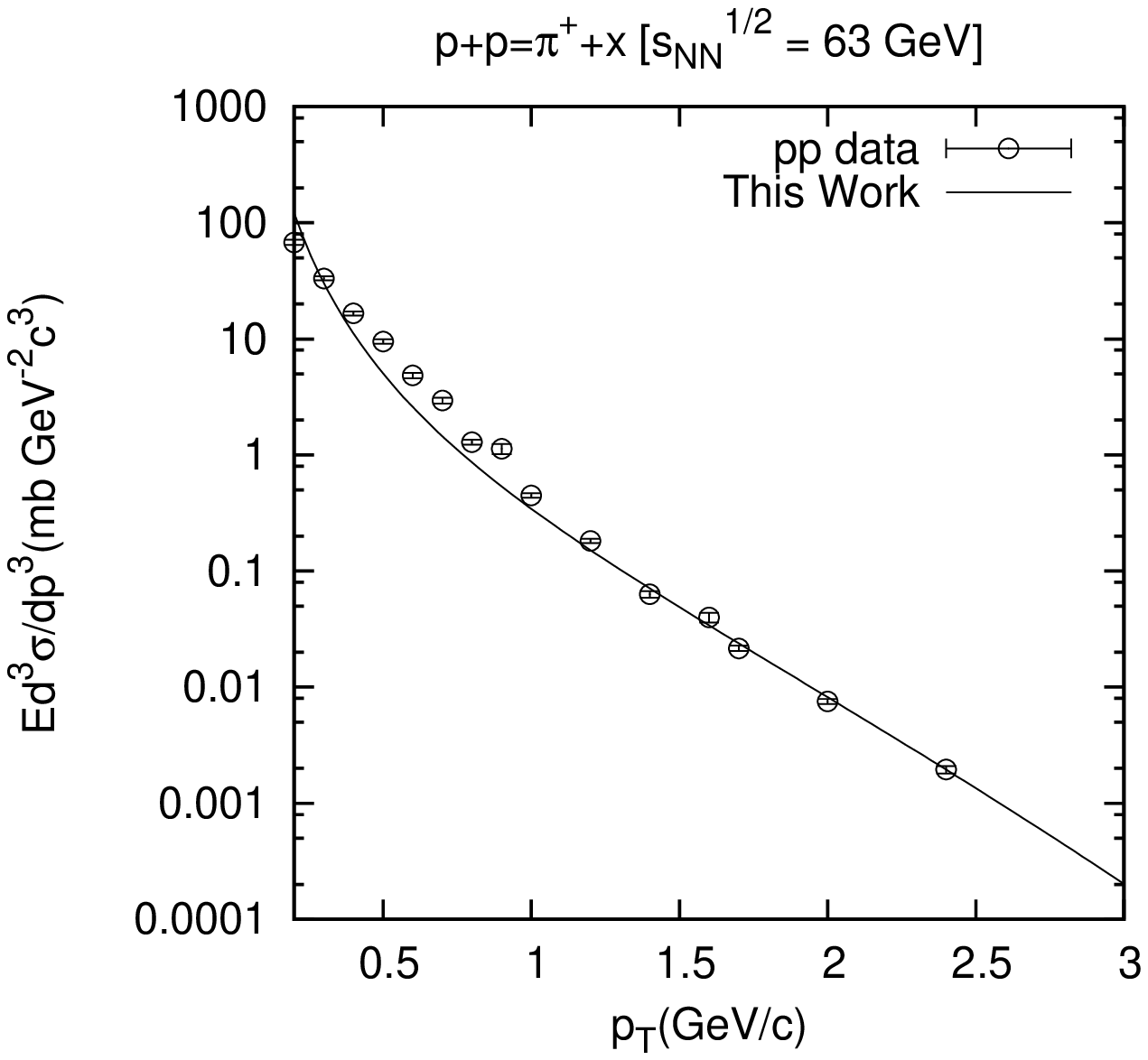}
  \end{minipage}}%
  \vspace{0.01in}
 \subfigure[]{
\centering
 \includegraphics[width=2.5in]{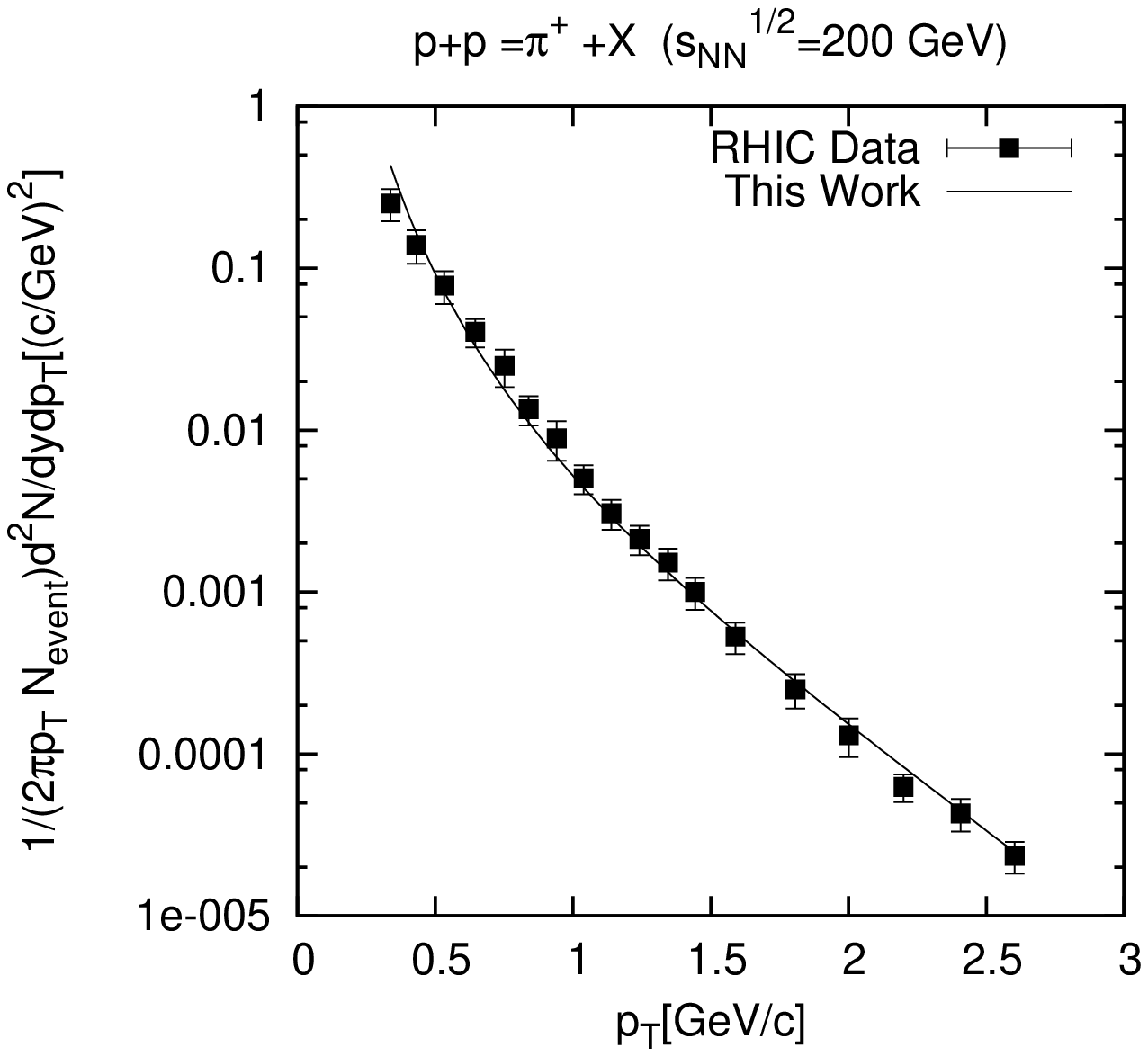}}
\caption{{\small Plots for $\pi$ production in $p+p$ collisions at
energies (a) $\sqrt{s_{NN}}$ = 20 GeV, (b)$\sqrt{s_{NN}}$ = 63 GeV
and (c) $\sqrt{s_{NN}}$ = 200 GeV. Data are taken (a) from Ref.
\cite{enterria}, (b) from Ref. \cite{phenix1} and (c) from Ref.s
\cite{phenix2}, \cite{yang}. Solid lines in the Figures show the SCM-based
 plots.} }
\end{figure}
\begin{figure}
\subfigure[]{
\begin{minipage}{.5\textwidth}
\centering
\includegraphics[width=2.5in]{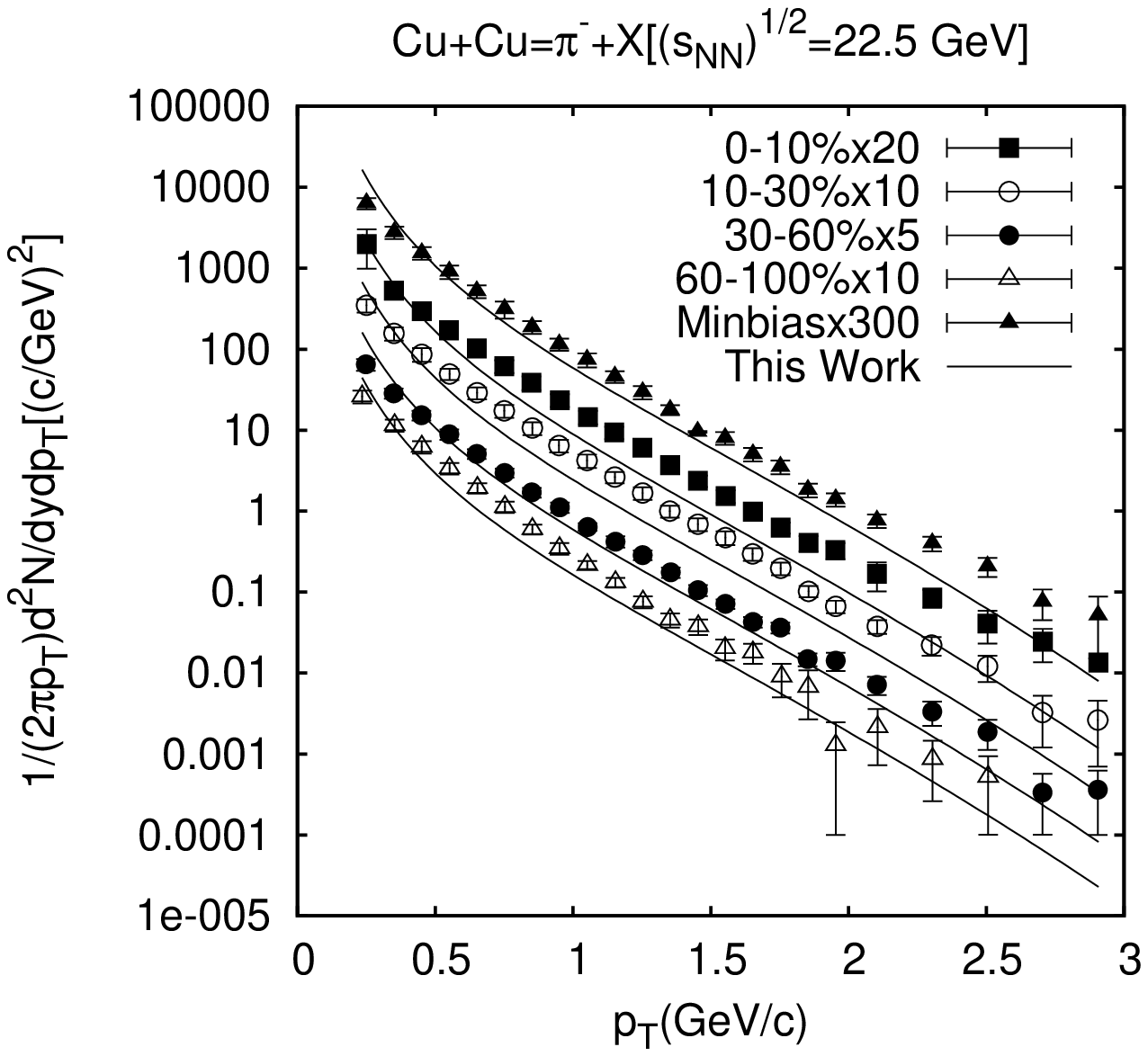}
\setcaptionwidth{2.6in}
\end{minipage}}%
\subfigure[]{
\begin{minipage}{0.5\textwidth}
\centering
 \includegraphics[width=2.5in]{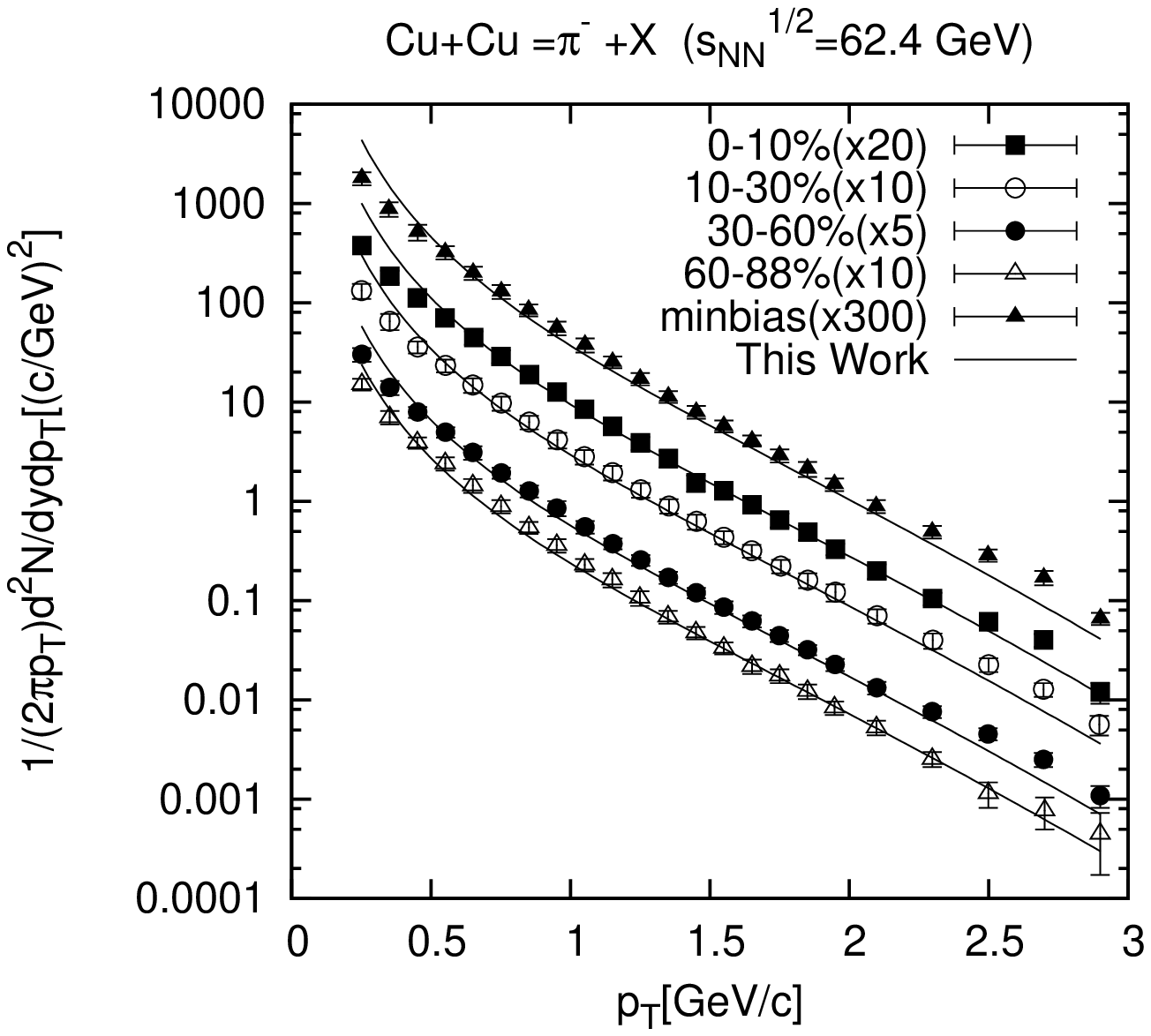}
  \end{minipage}}%
  \vspace{0.01in}
 \subfigure[]{
\centering
 \includegraphics[width=4.5in]{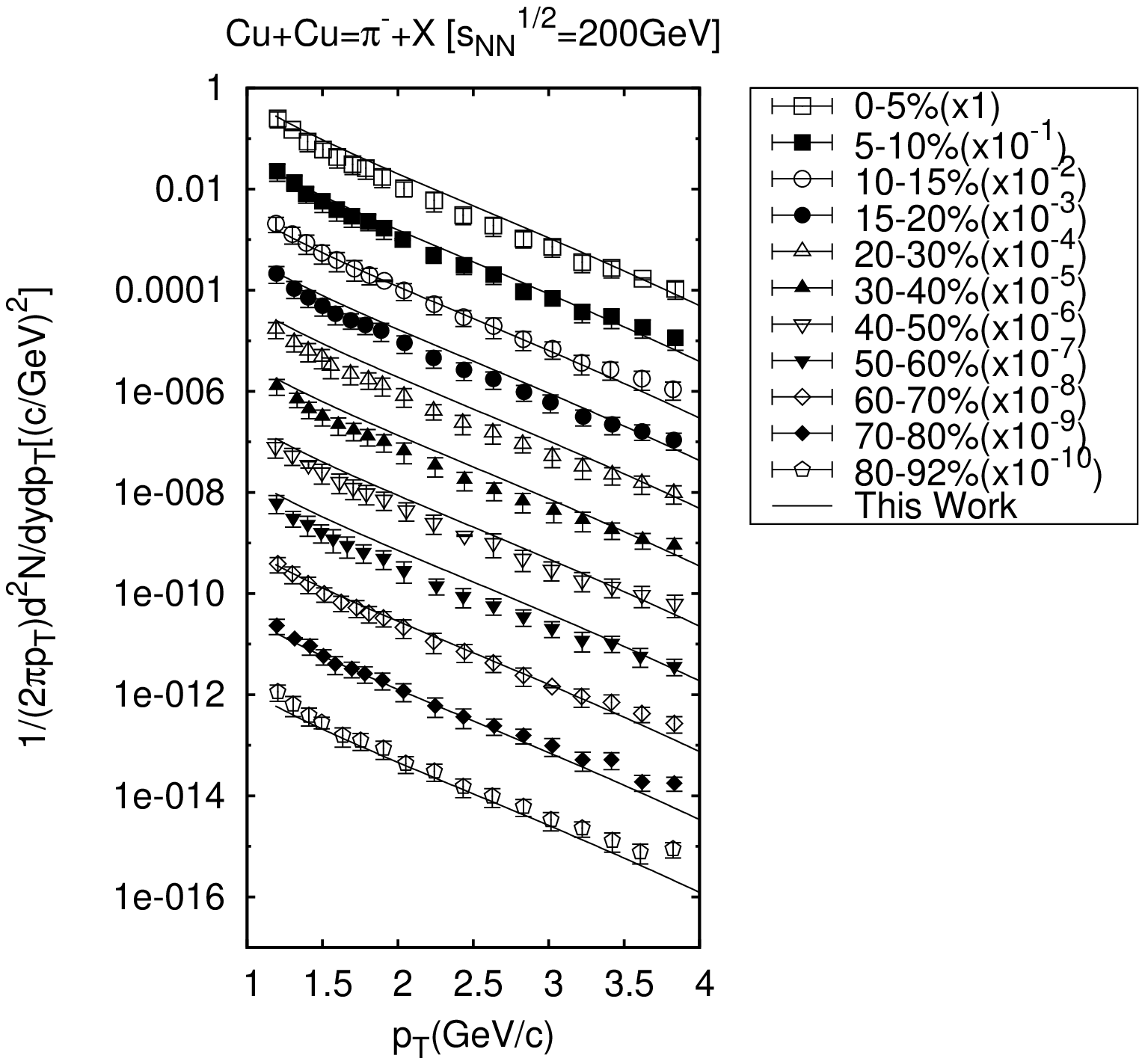}}
 \caption{{\small Centrality dependence of
the $p_T$ distribution for $\pi^-$ for different centralities and at
energies (a) 22.5 GeV \cite{phenix3}, (b) 62.4 GeV \cite{phenix3}
and (c) 200 GeV \cite{phenix1} in $Cu+Cu$ collisions. The solid lines in
the Figures 2(a), 2(b) and 2(c) show the SCM
calculations for different centralities.  } }
\end{figure}
\begin{figure}
\subfigure[]{
\begin{minipage}{.5\textwidth}
\centering
\includegraphics[width=2.5in]{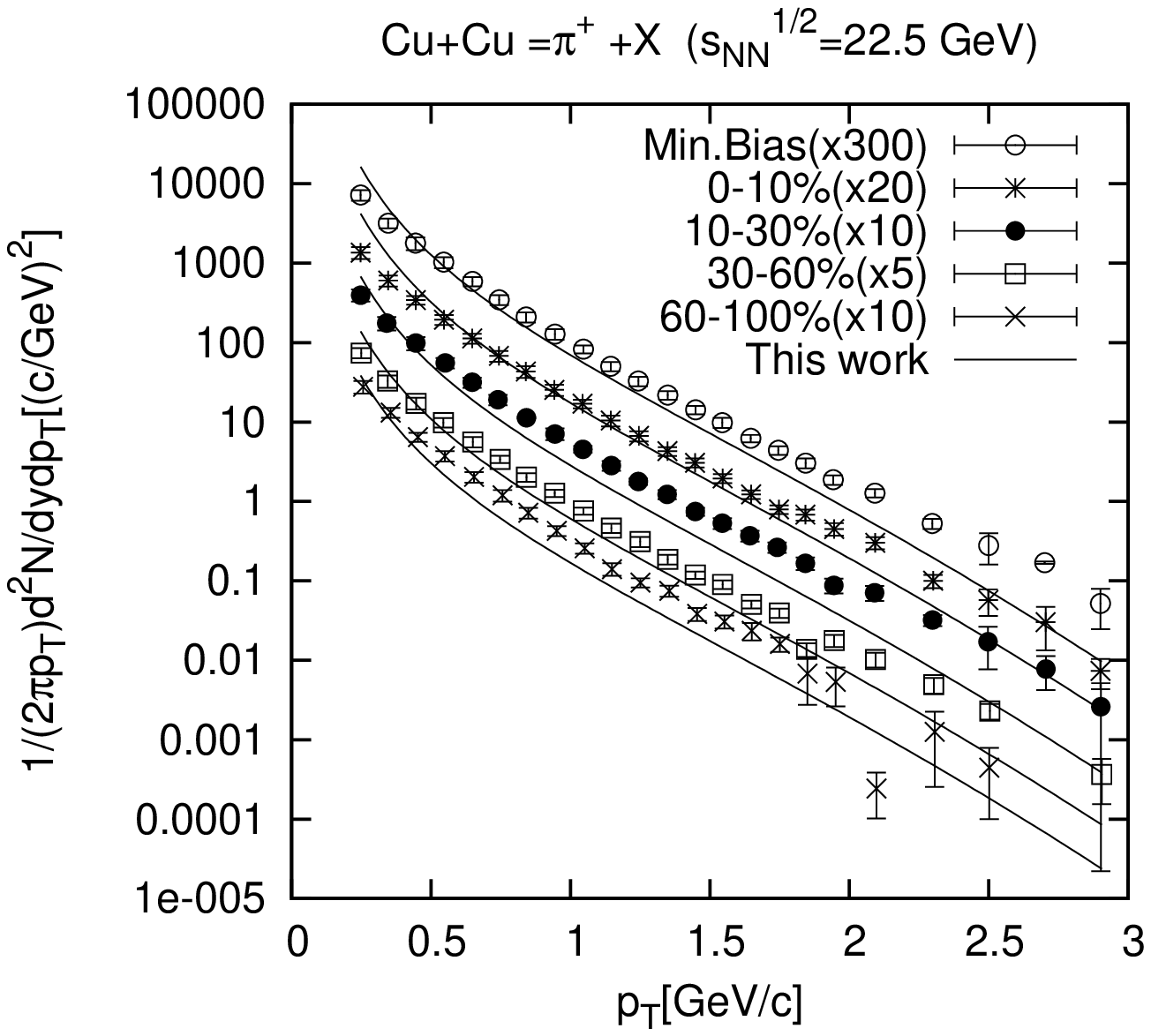}
\setcaptionwidth{2.6in}
\end{minipage}}%
\subfigure[]{
\begin{minipage}{0.5\textwidth}
\centering
 \includegraphics[width=2.5in]{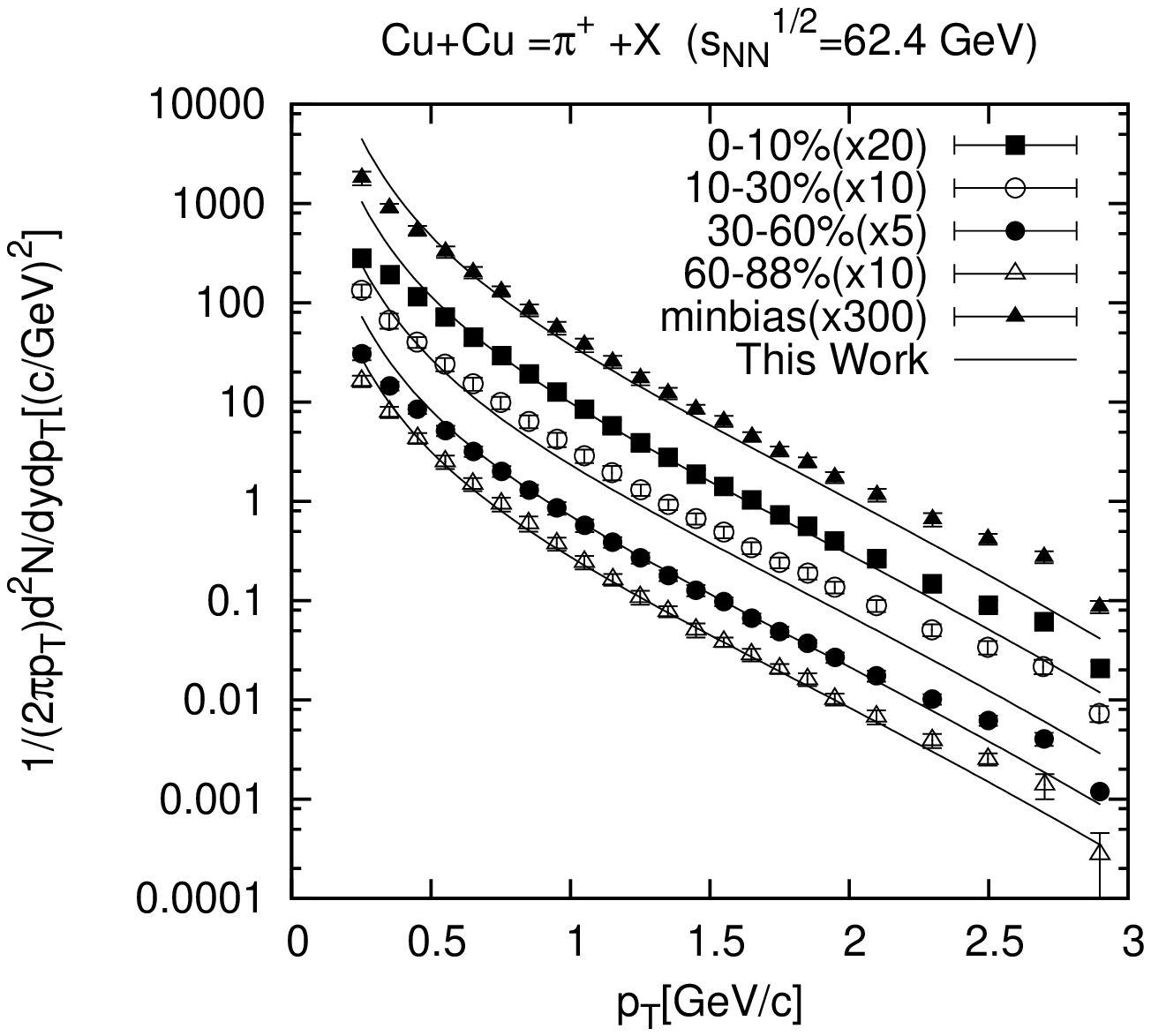}
  \end{minipage}}%
  \vspace{0.01in}
 \subfigure[]{
\centering
 \includegraphics[width=4.5in]{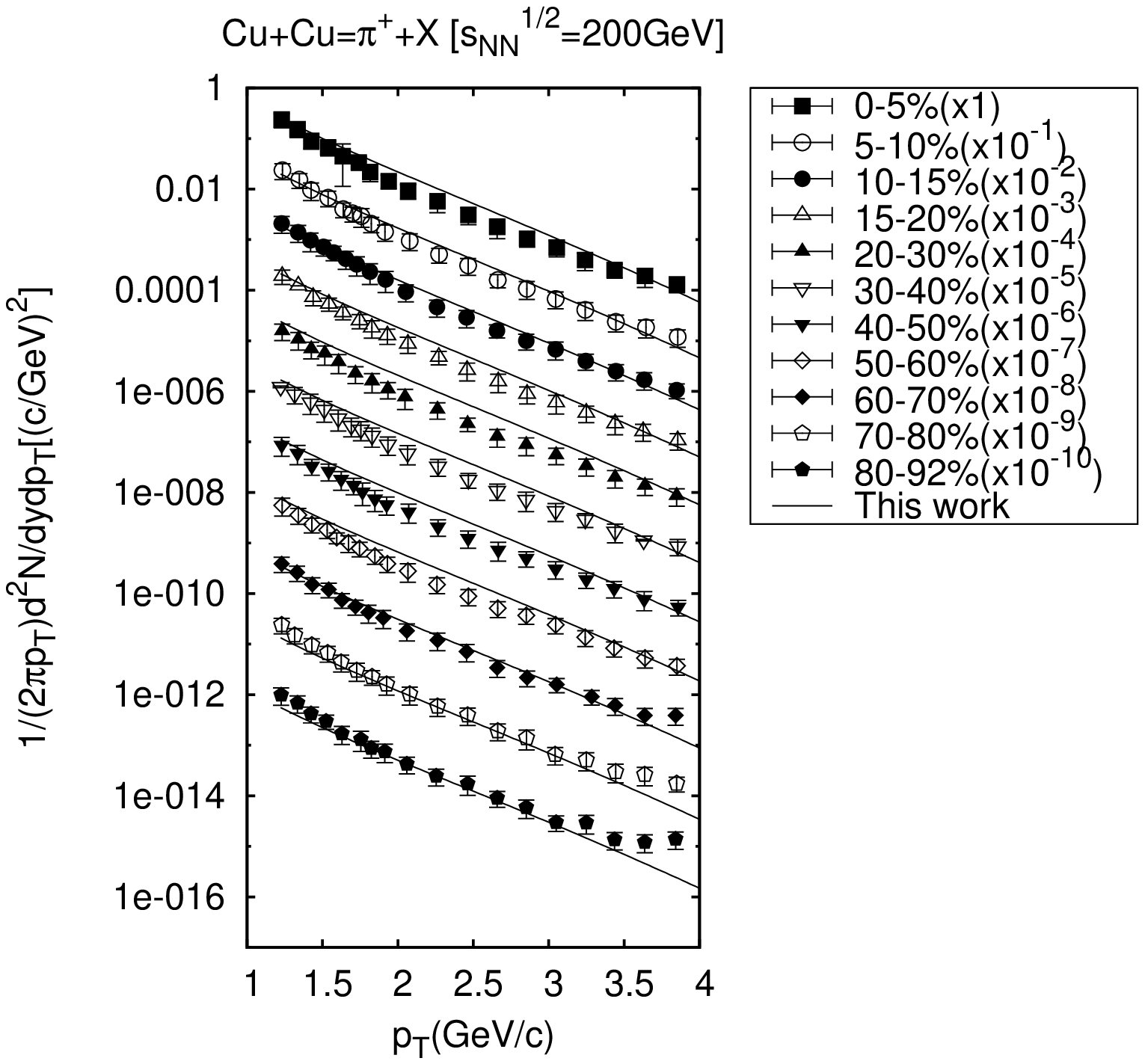}}
 \caption{{\small Centrality dependence of
the $p_T$ distribution for $\pi^+$ for different centralities and at
energies (a) 22.5 GeV \cite{phenix3}, (b) 62.4 GeV \cite{phenix3}
and (c) 200 GeV \cite{phenix1} in $Cu+Cu$ collisions. The solid lines in
the Figures 3(a), 3(b) and 3(c) show the SCM
calculations for different centralities.}  }
\end{figure}
\begin{figure}
\subfigure[]{
\begin{minipage}{.5\textwidth}
\centering
\includegraphics[width=3.2in]{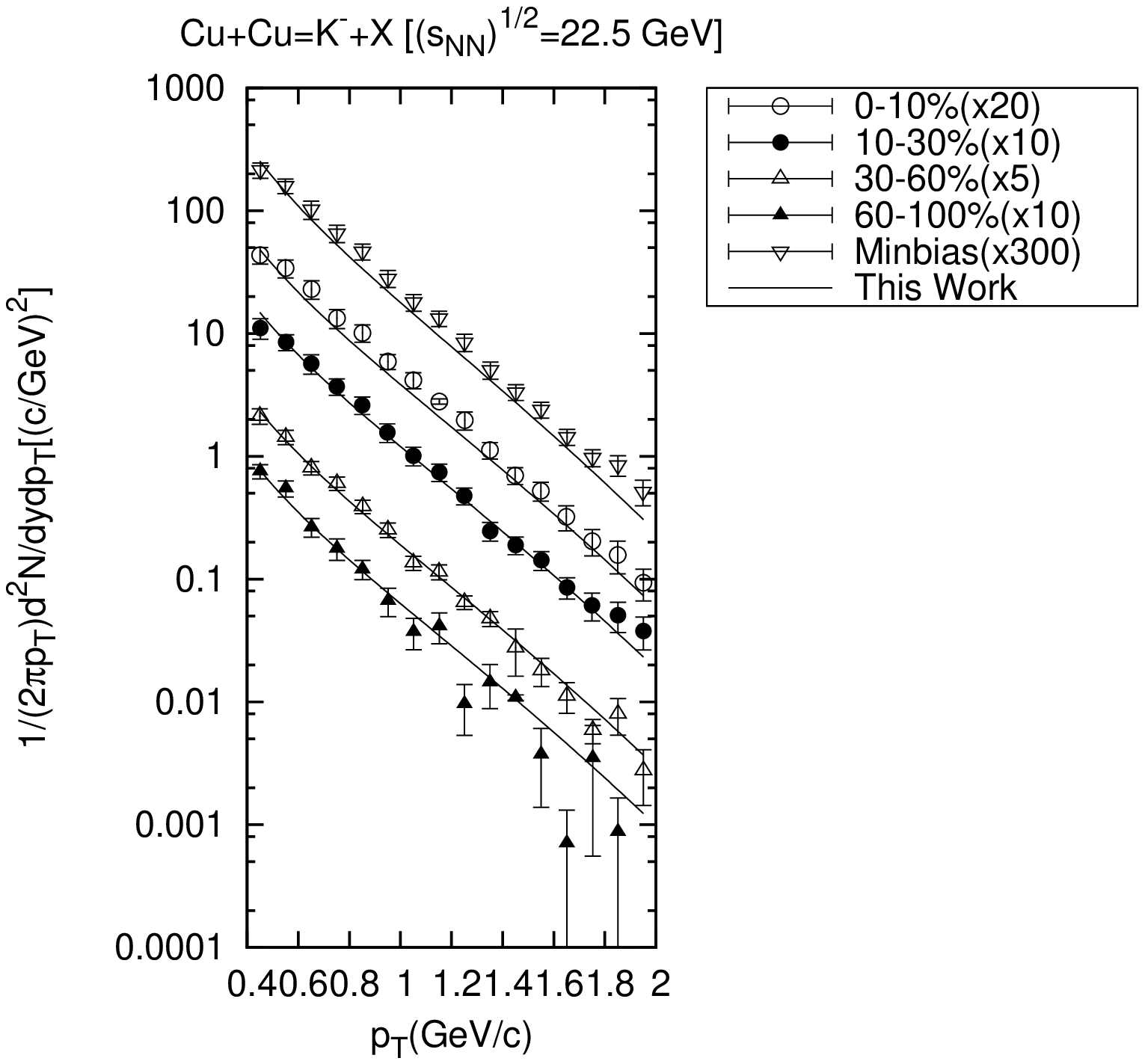}
\setcaptionwidth{2.6in}
\end{minipage}}%
\subfigure[]{
\begin{minipage}{0.5\textwidth}
\centering
 \includegraphics[width=3.2in]{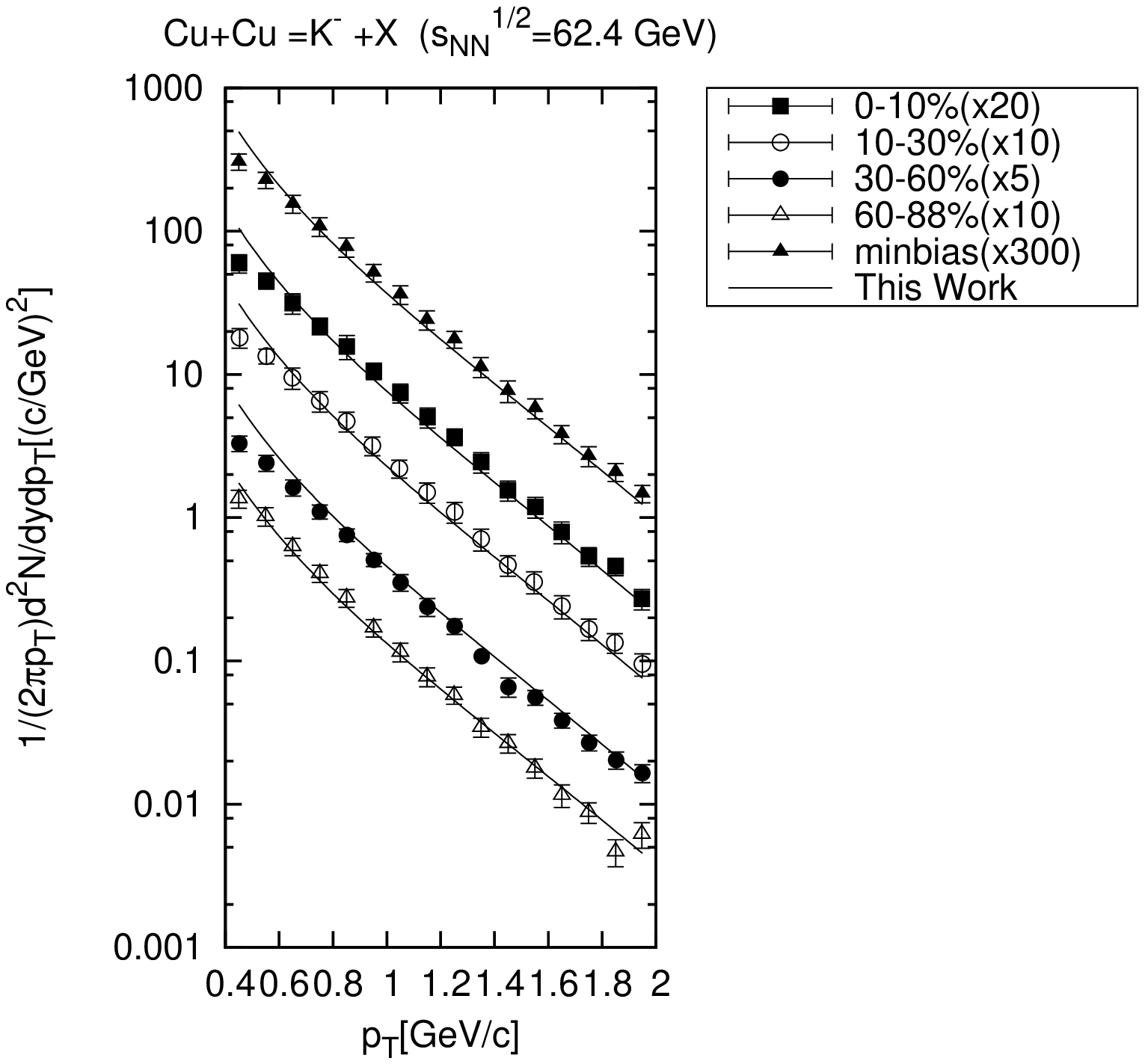}
  \end{minipage}}%
  \vspace{0.01in}
 \subfigure[]{
\centering
 \includegraphics[width=4.5in]{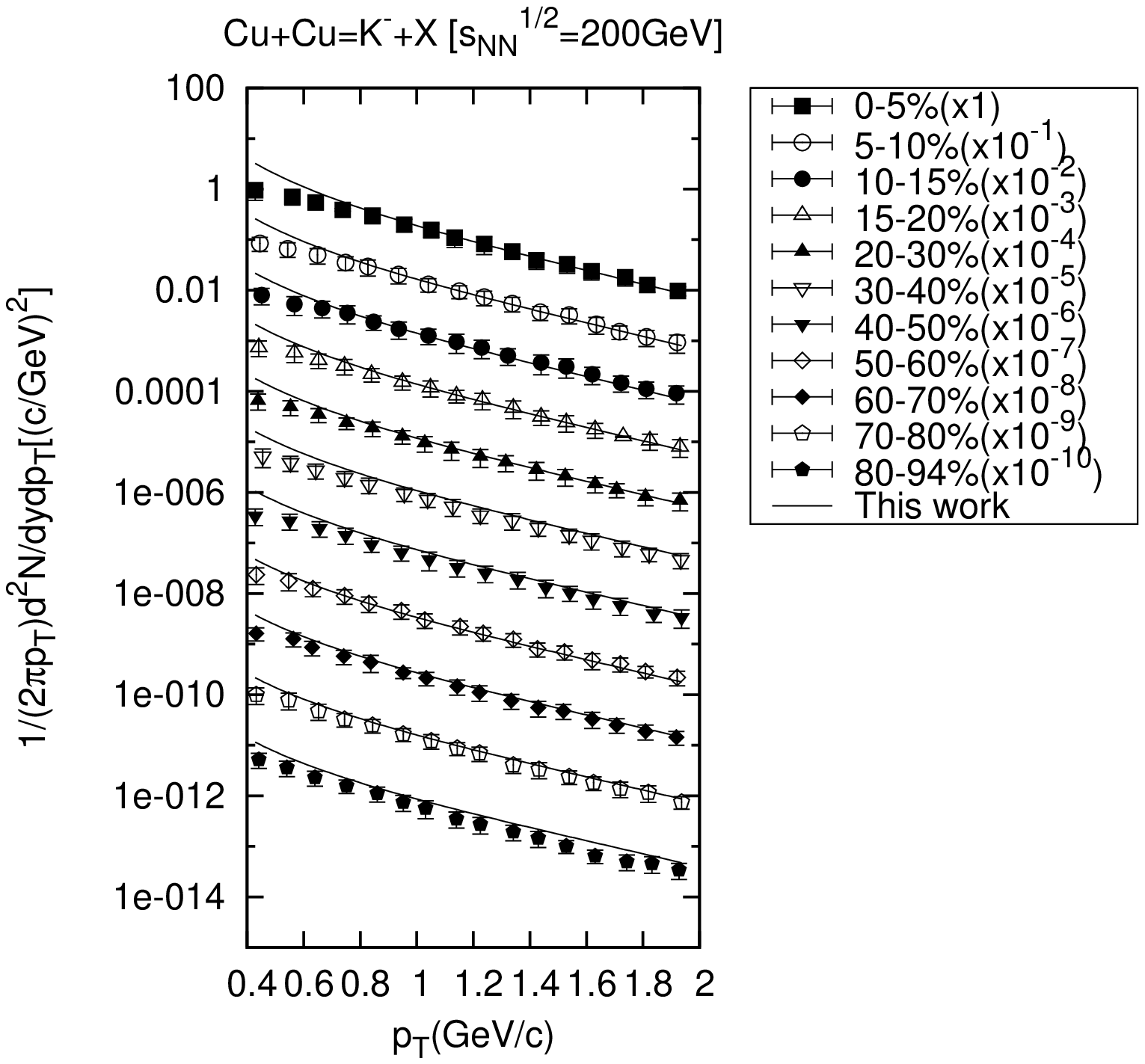}}
 \caption{{\small Invariant
spectra as function of $p_T$ for $K^-$ production in $Cu+Cu$
collisions at (a) $\sqrt {s_{NN}}$ =22.5 GeV \cite{phenix3},
(b)$\sqrt s_{NN}$ =62.4 GeV \cite{phenix3} and for (c) $\sqrt
s_{NN}$ =200 GeV \cite{phenix1}. The solid lines show the SCM-based
results.} }
\end{figure}
\begin{figure}
\subfigure[]{
\begin{minipage}{.5\textwidth}
\centering
\includegraphics[width=3.2in]{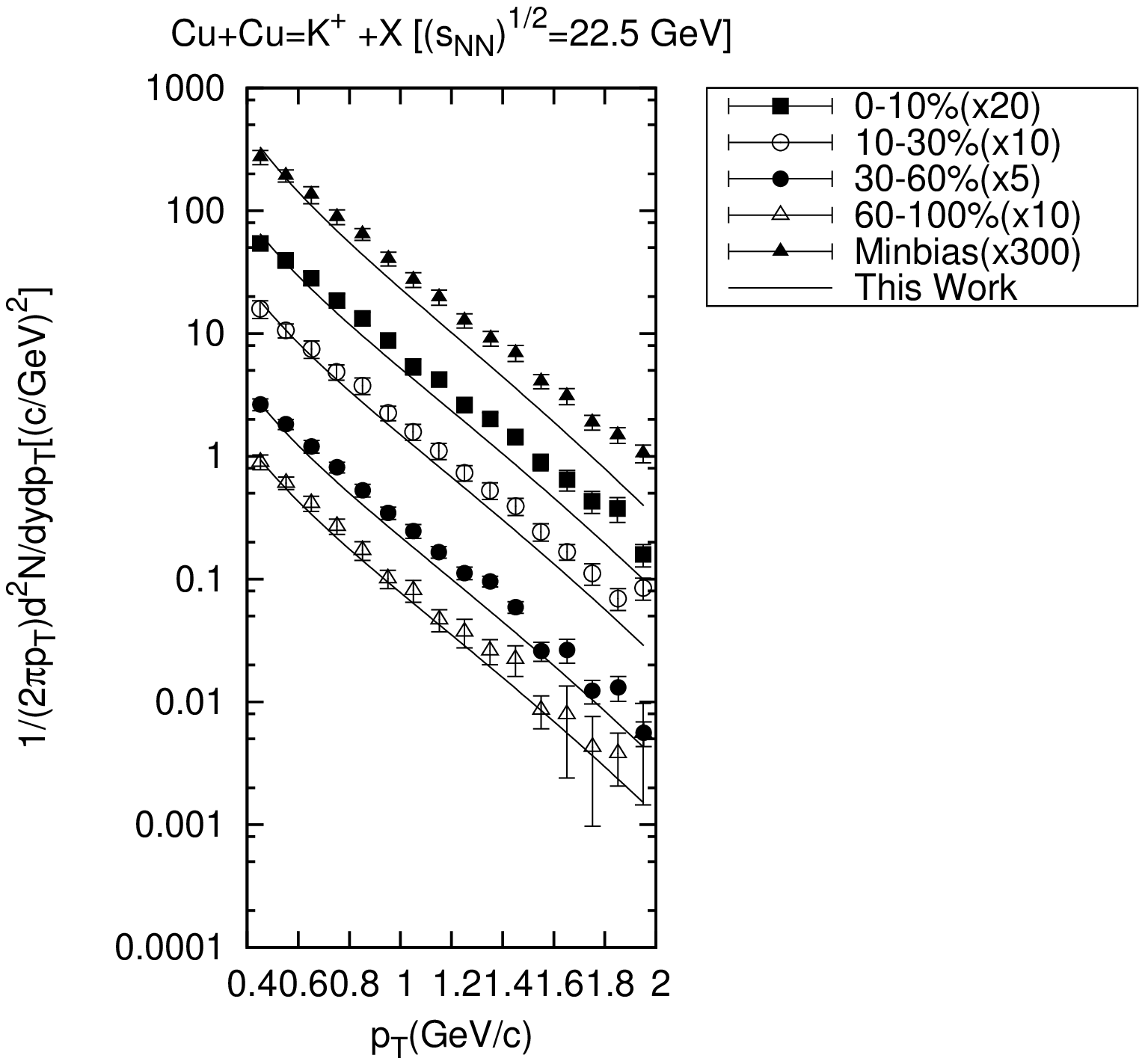}
\setcaptionwidth{2.6in}
\end{minipage}}%
\subfigure[]{
\begin{minipage}{0.5\textwidth}
\centering
 \includegraphics[width=3.2in]{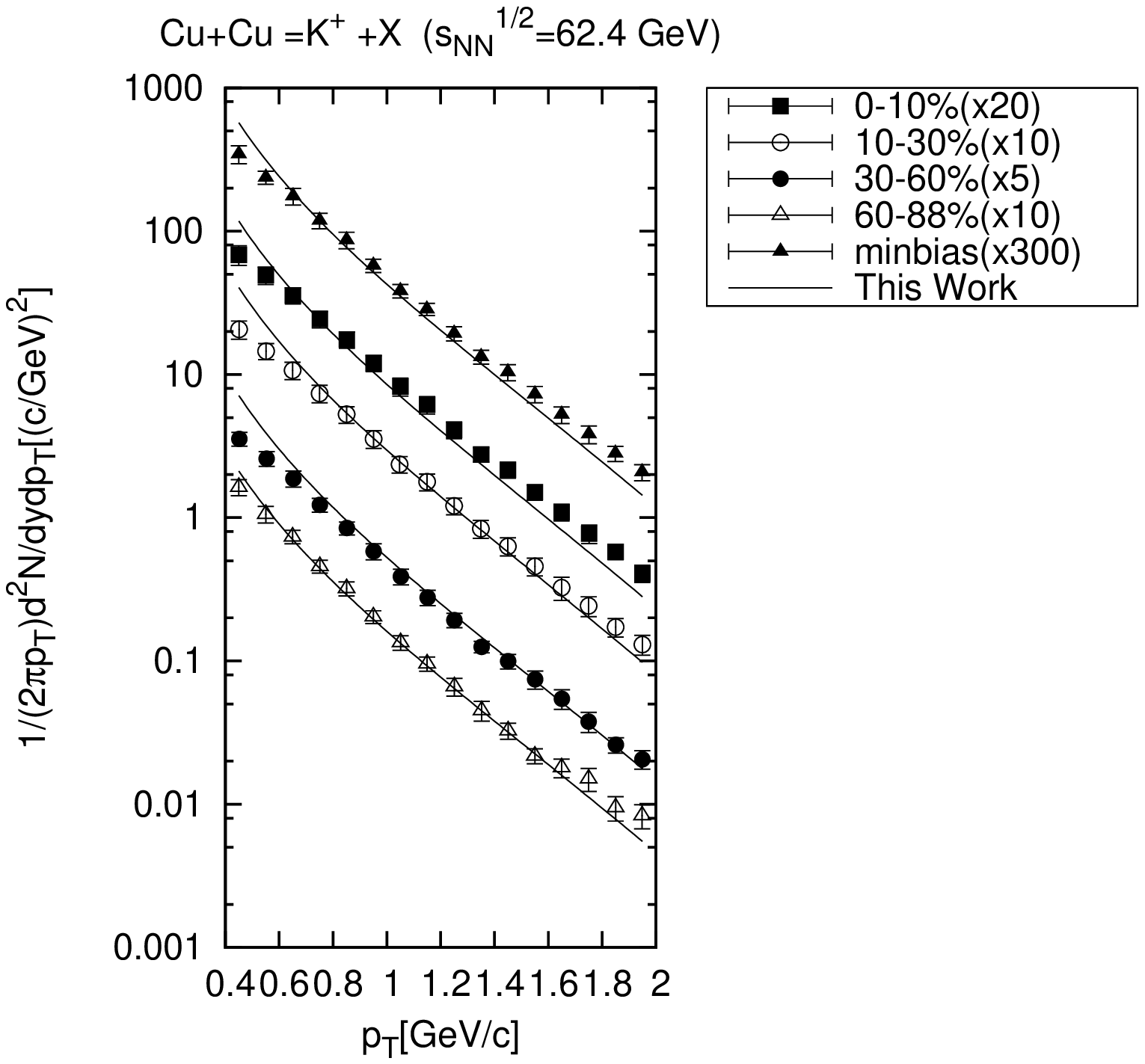}
  \end{minipage}}%
  \vspace{0.01in}
 \subfigure[]{
\centering
 \includegraphics[width=4.5in]{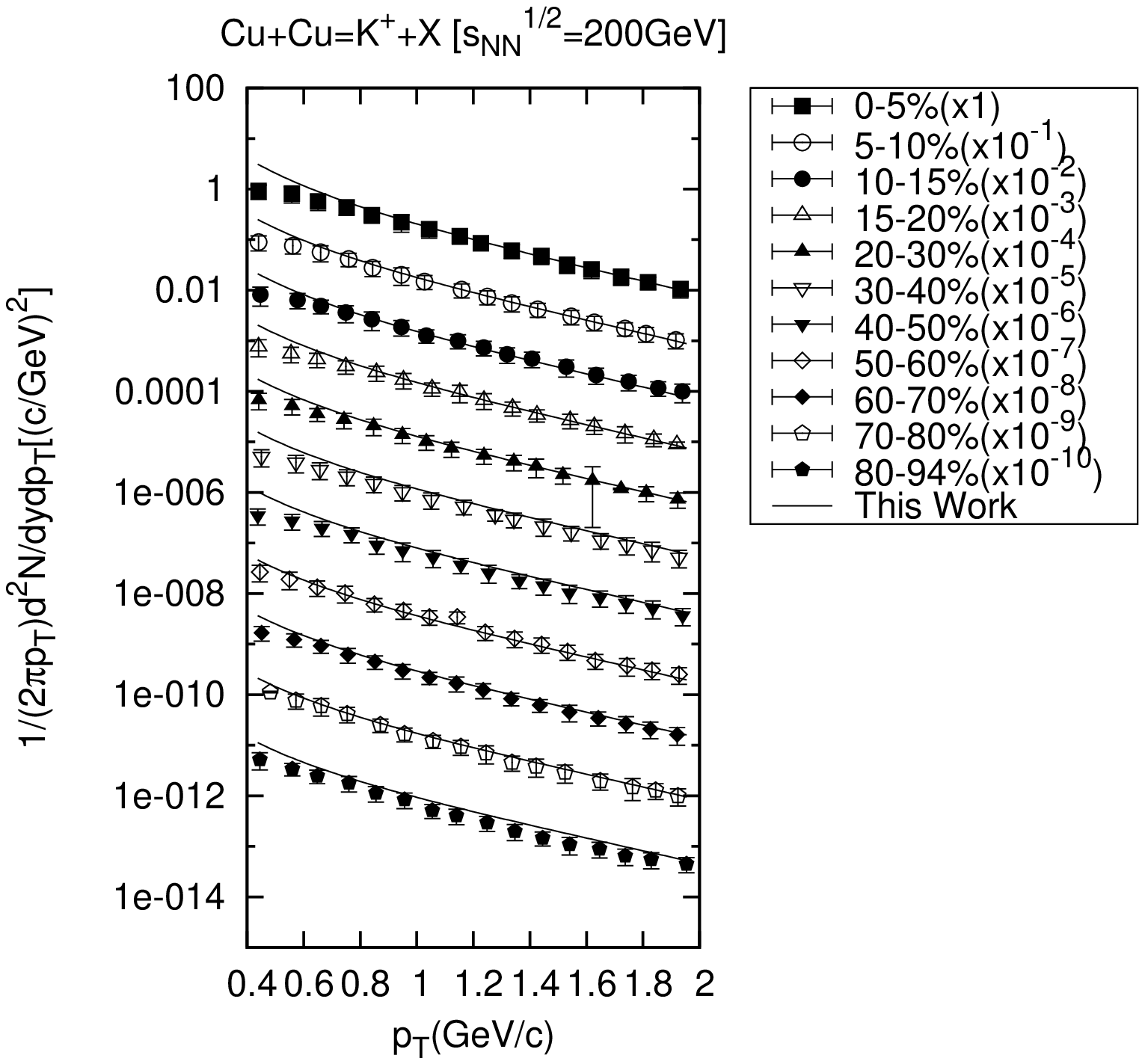}}
 \caption{{\small Invariant
spectra as function of $p_T$ for $K^+$ production in $Cu+Cu$
collisions at (a) $\sqrt {s_{NN}}$ =22.5 GeV \cite{phenix3},
(b)$\sqrt s_{NN}$ =62.4 GeV \cite{phenix3} and for (c) $\sqrt
s_{NN}$ =200 GeV \cite{phenix1}. The solid lines show the SCM-based
results.}  }
\end{figure}
\begin{figure}
\subfigure[]{
\begin{minipage}{.5\textwidth}
\centering
\includegraphics[width=3.2in]{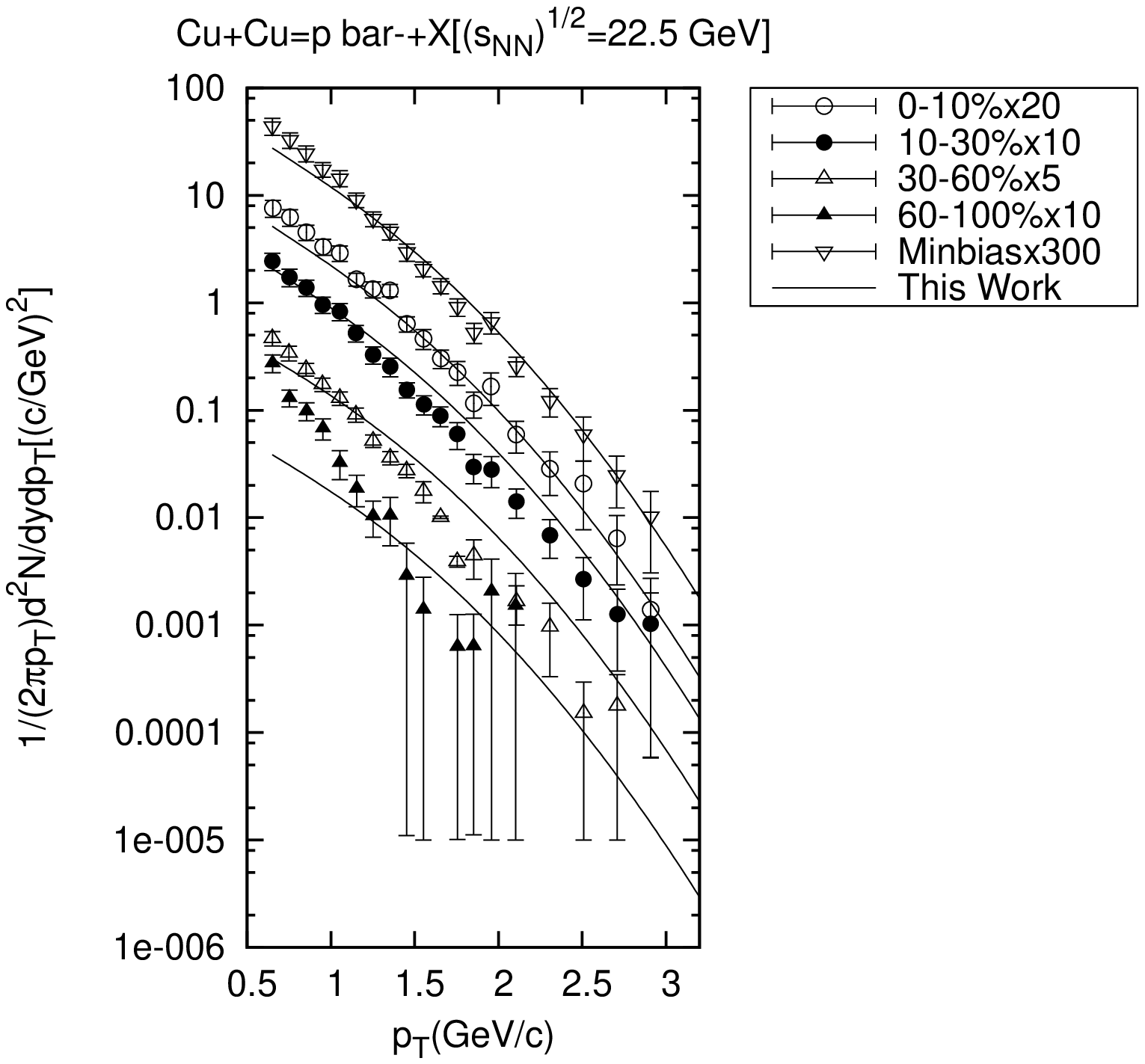}
\setcaptionwidth{2.6in}
\end{minipage}}%
\subfigure[]{
\begin{minipage}{0.5\textwidth}
\centering
 \includegraphics[width=3.2in]{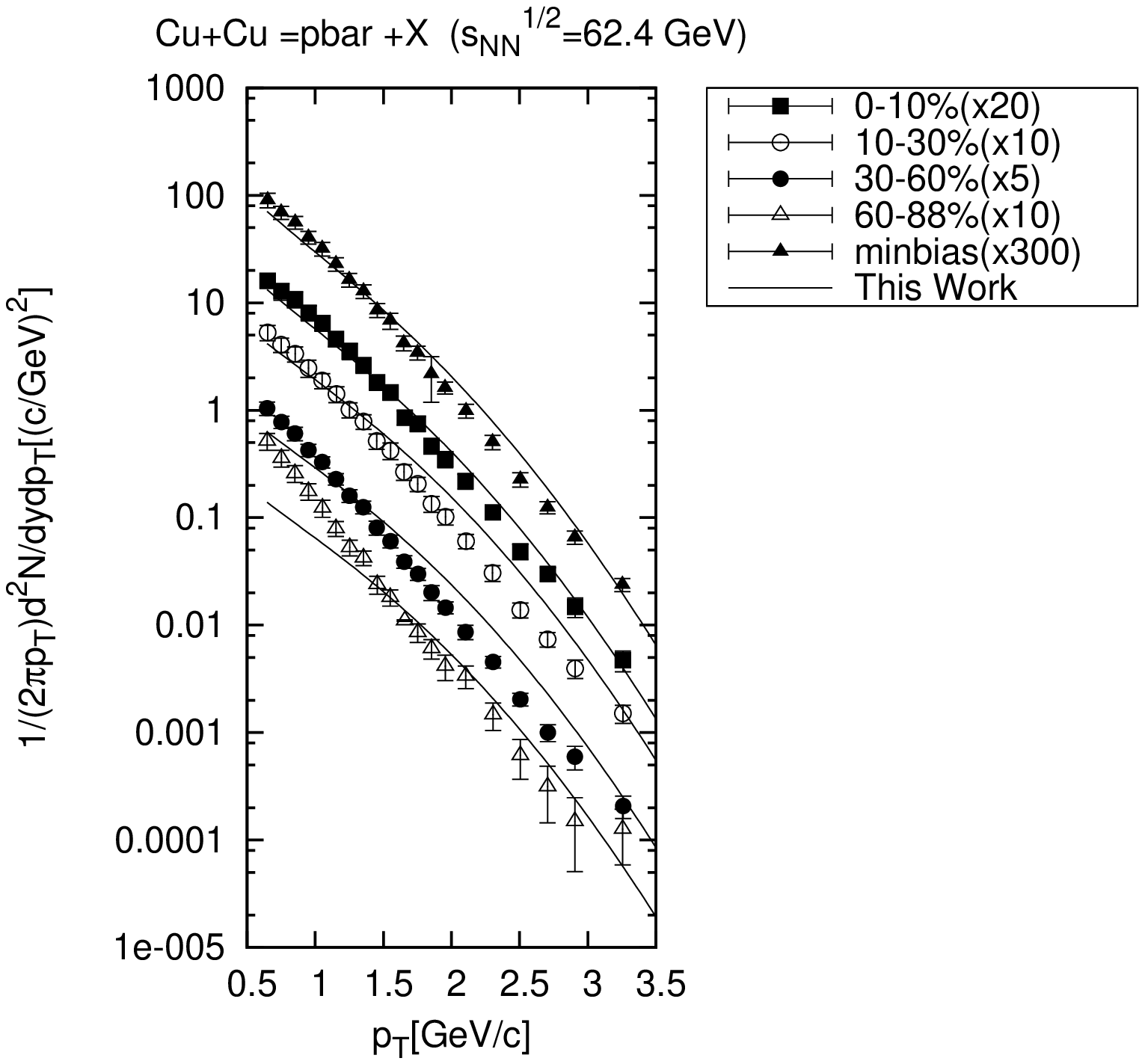}
  \end{minipage}}%
  \vspace{0.01in}
 \subfigure[]{
\centering
 \includegraphics[width=4.5in]{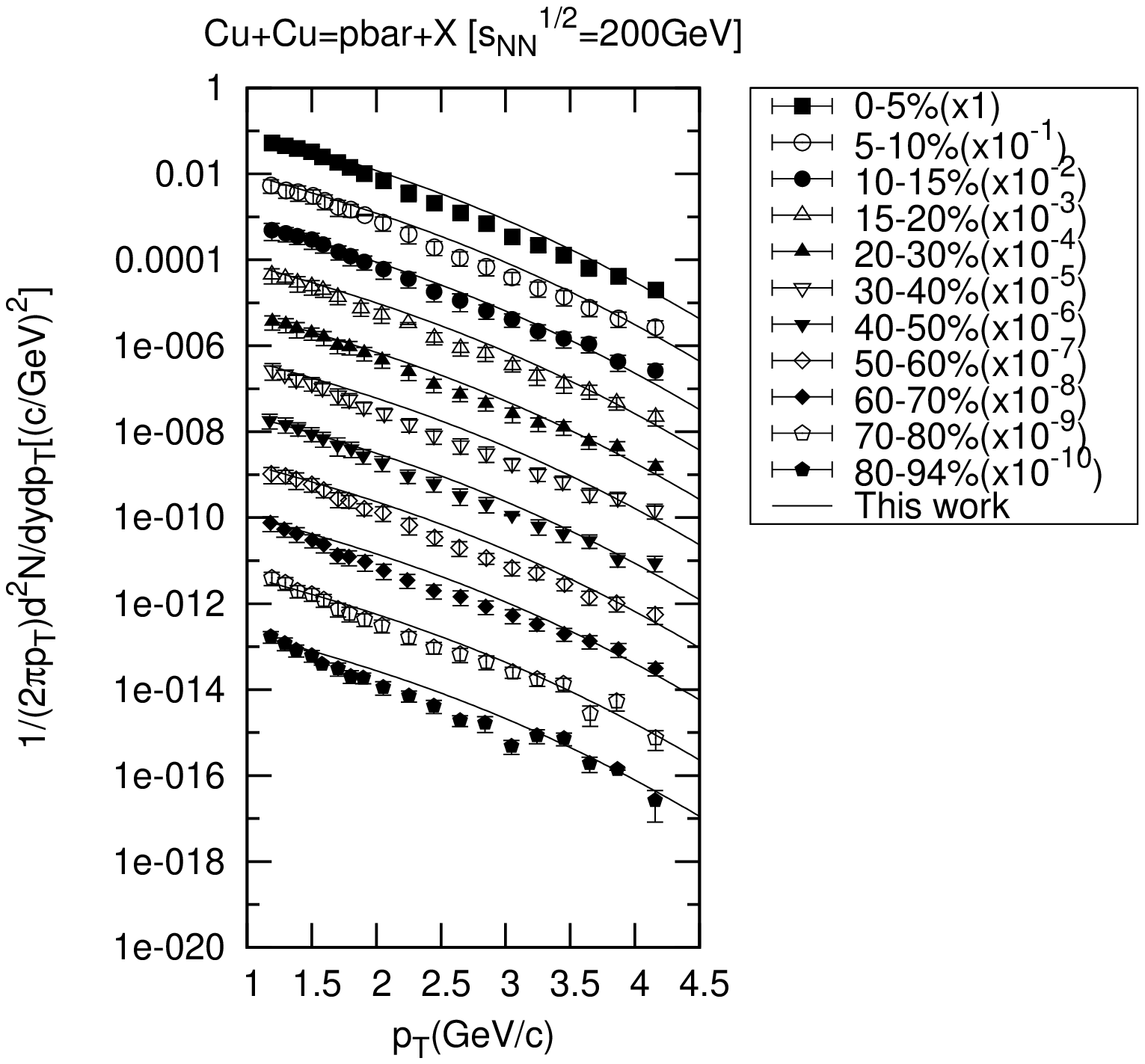}}
 \caption{{\small Centrality dependence of
the $p_T$ distribution for $\bar p$ for different centralities in
$Cu+Cu$ collisions at (a) $\sqrt {s_{NN}}$ =22.5 GeV \cite{phenix3},
(b)$\sqrt s_{NN}$ =62.4 GeV \cite{phenix3} and for (c) $\sqrt
s_{NN}$ =200 GeV \cite{phenix1}. The solid lines in the Figures show the
SCM calculations .}  }
\end{figure}
\begin{figure}
\subfigure[]{
\begin{minipage}{.5\textwidth}
\centering
\includegraphics[width=3.2in]{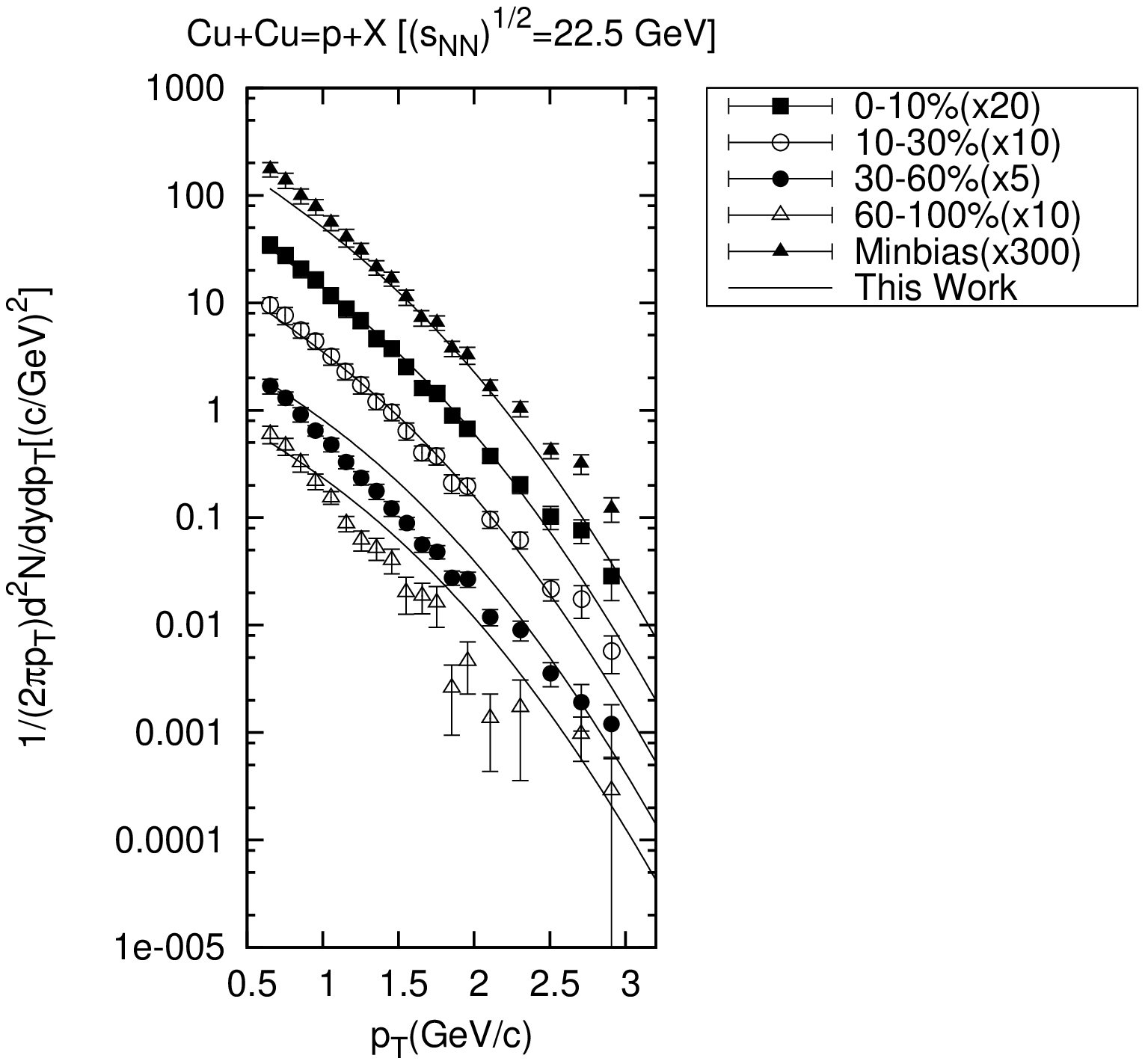}
\setcaptionwidth{2.6in}
\end{minipage}}%
\subfigure[]{
\begin{minipage}{0.5\textwidth}
\centering
 \includegraphics[width=3.2in]{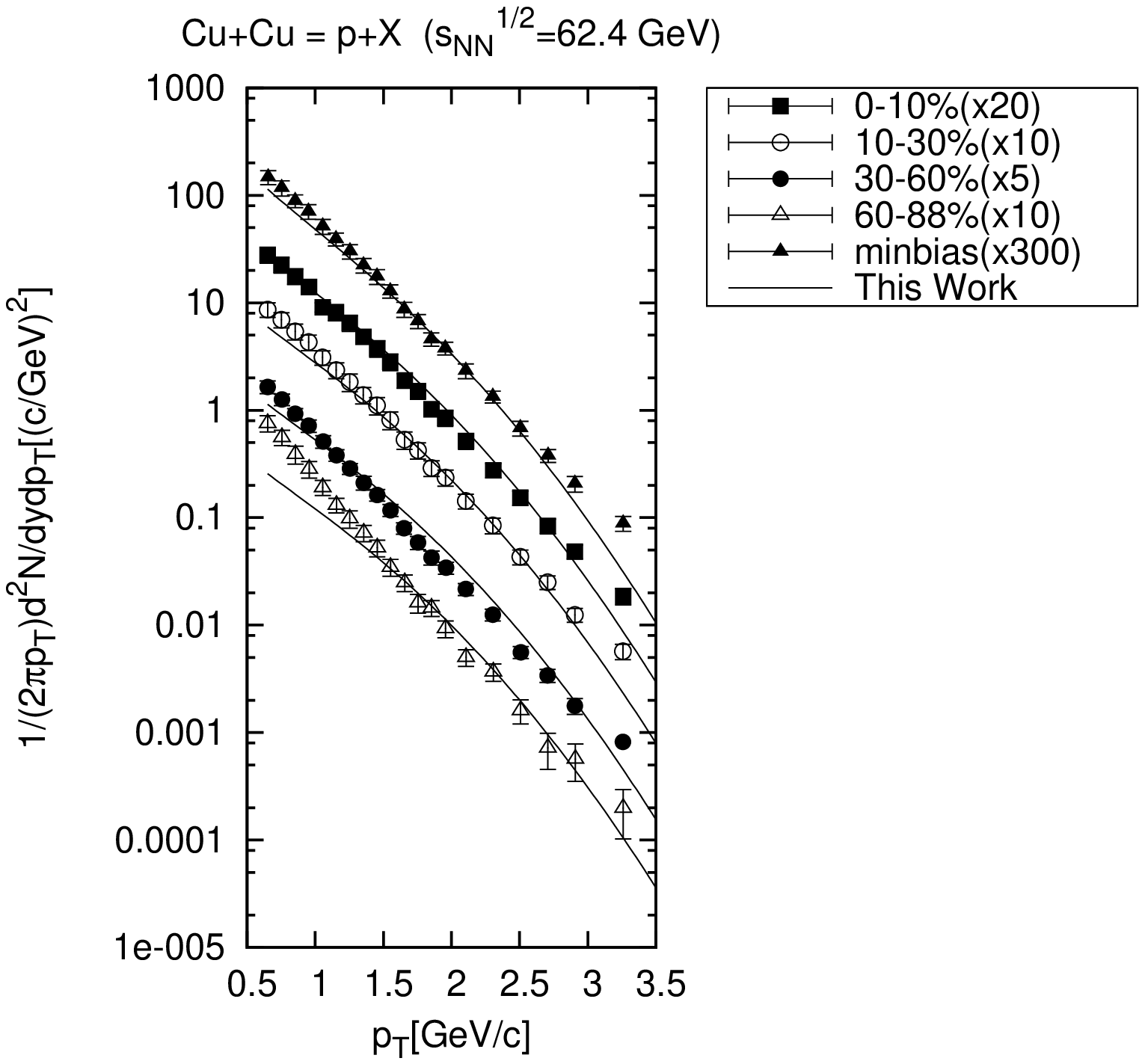}
  \end{minipage}}%
  \vspace{0.01in}
 \subfigure[]{
\centering
 \includegraphics[width=4.5in]{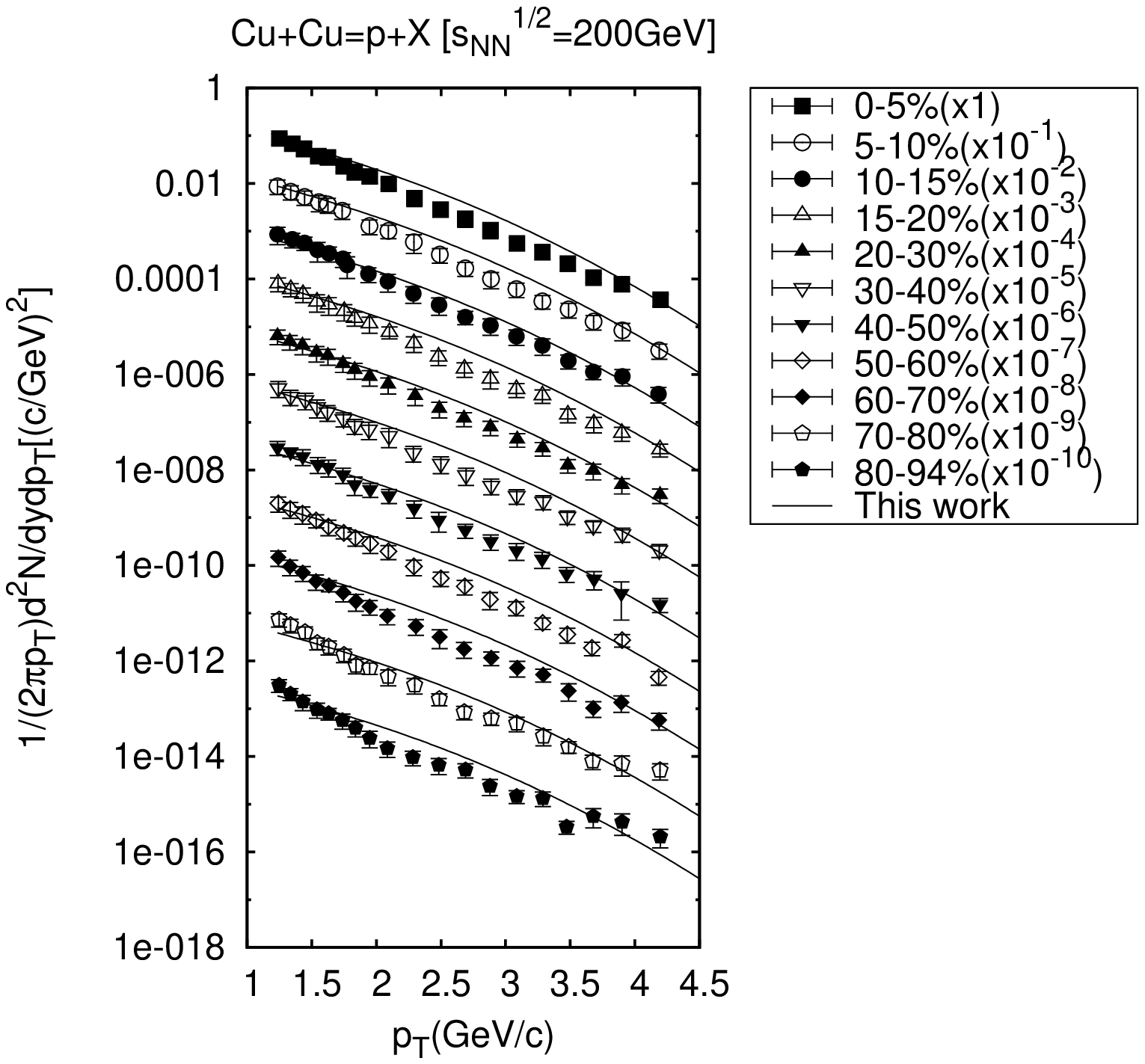}}
 \caption{{\small Centrality dependence of
the $p_T$ distribution for $p$ for different centralities in $Cu+Cu$
collisions at (a) $\sqrt {s_{NN}}$ =22.5 GeV \cite{phenix3},
(b)$\sqrt s_{NN}$ =62.4 GeV \cite{phenix3} and for (c) $\sqrt
s_{NN}$ =200 GeV \cite{phenix1}. The solid lines in the figures show the
SCM calculations .}  }
\end{figure}
\begin{figure}
\subfigure[]{
\begin{minipage}{.5\textwidth}
\centering
\includegraphics[width=2.5in]{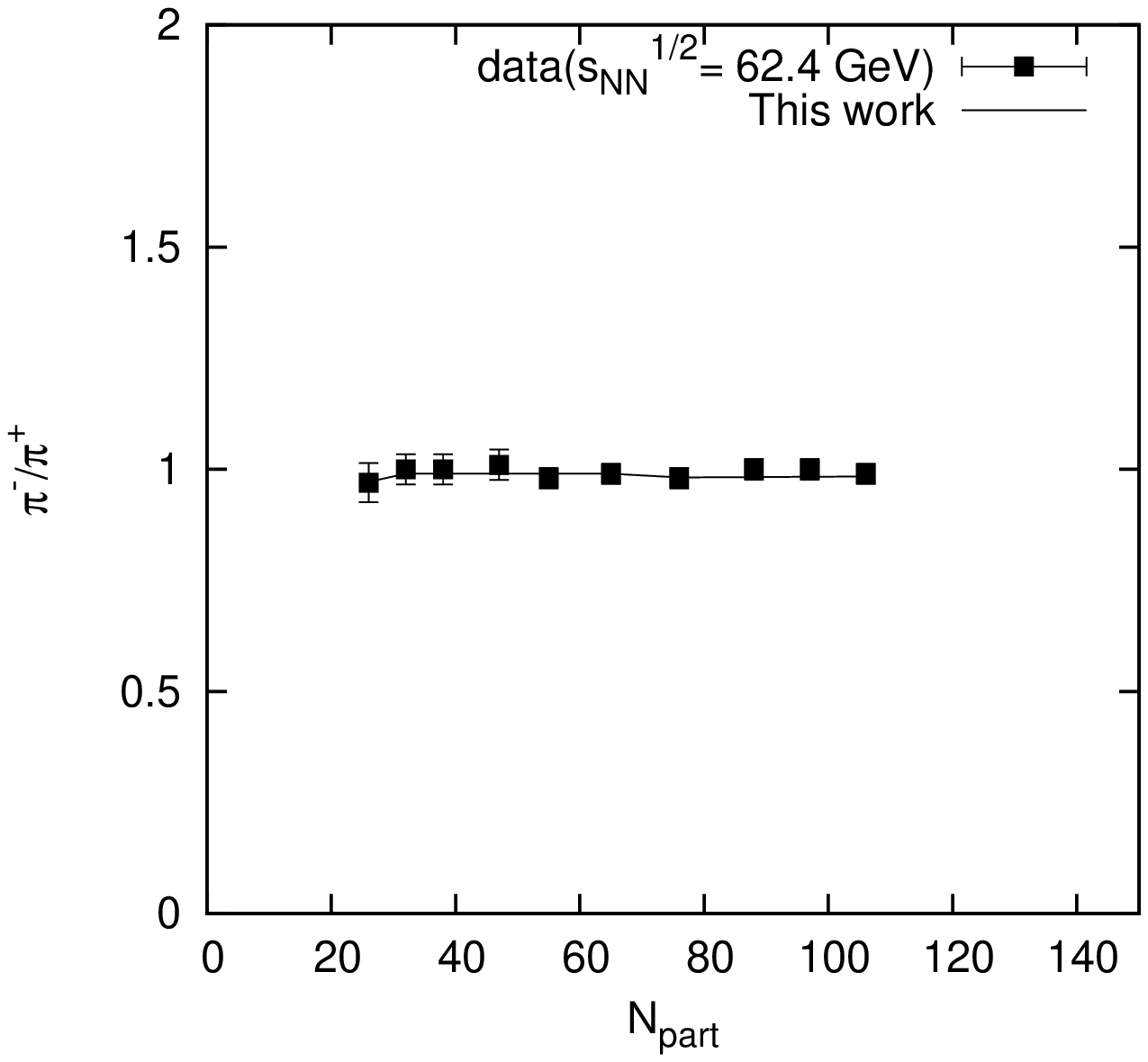}
\setcaptionwidth{2.6in}
\end{minipage}}%
\subfigure[]{
\begin{minipage}{0.5\textwidth}
\centering
 \includegraphics[width=2.5in]{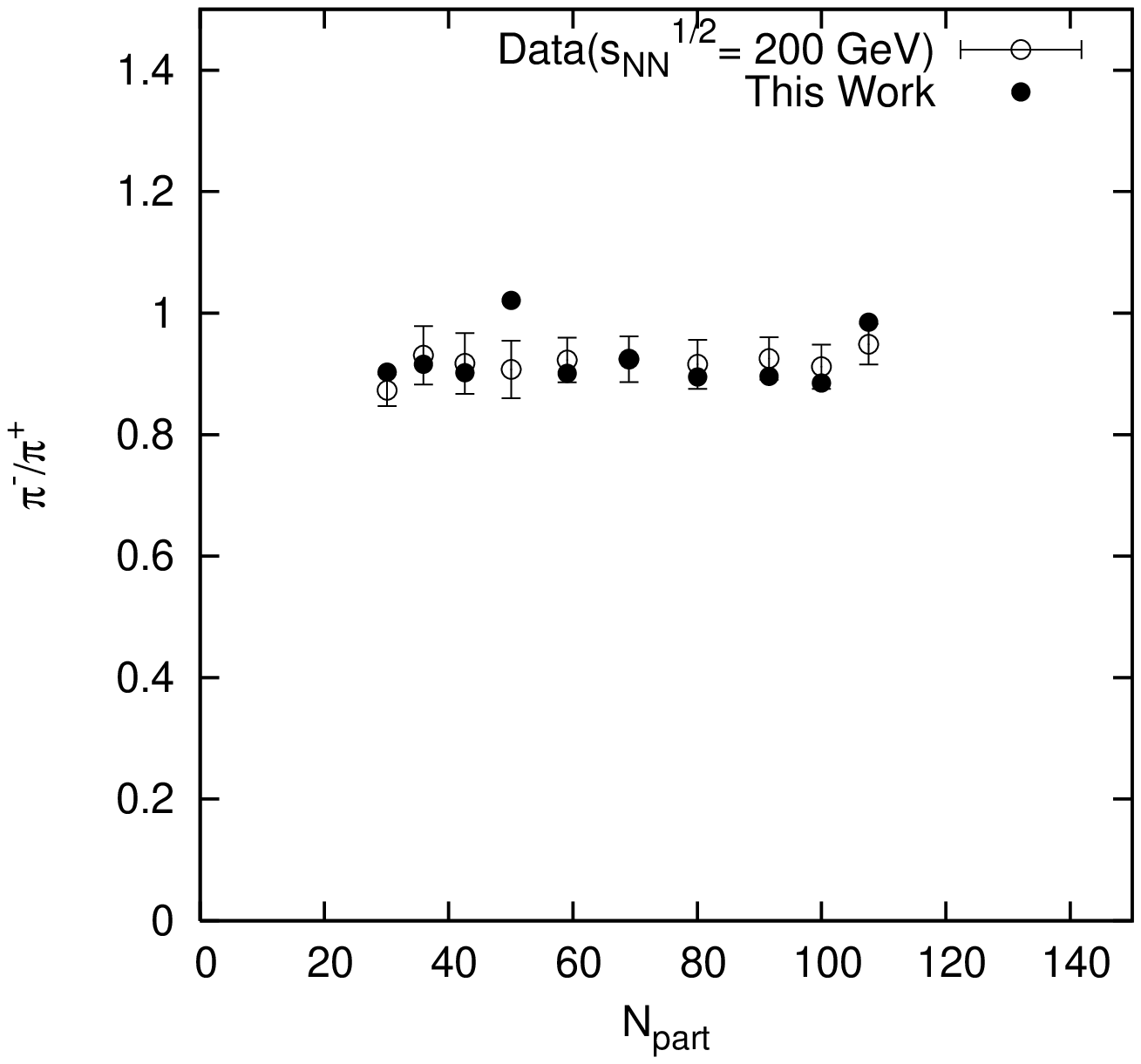}
  \end{minipage}}%
   \caption{{\small The $N_{part}$ versus $\pi^-/\pi^+$ ratio
behaviours at (a) $\sqrt {s_{NN}}$ =62.4 GeV and (b)$\sqrt s_{NN}$
=200 GeV. Data are taken from PHOBOS \cite{veres}, \cite{wosiek}.
The solid line in Fig. 8(a) and the filled circles in Fig. 8(b) show the
theoretically calculated values. } }
\end{figure}
\begin{figure}
\subfigure[]{
\begin{minipage}{.5\textwidth}
\centering
\includegraphics[width=2.5in]{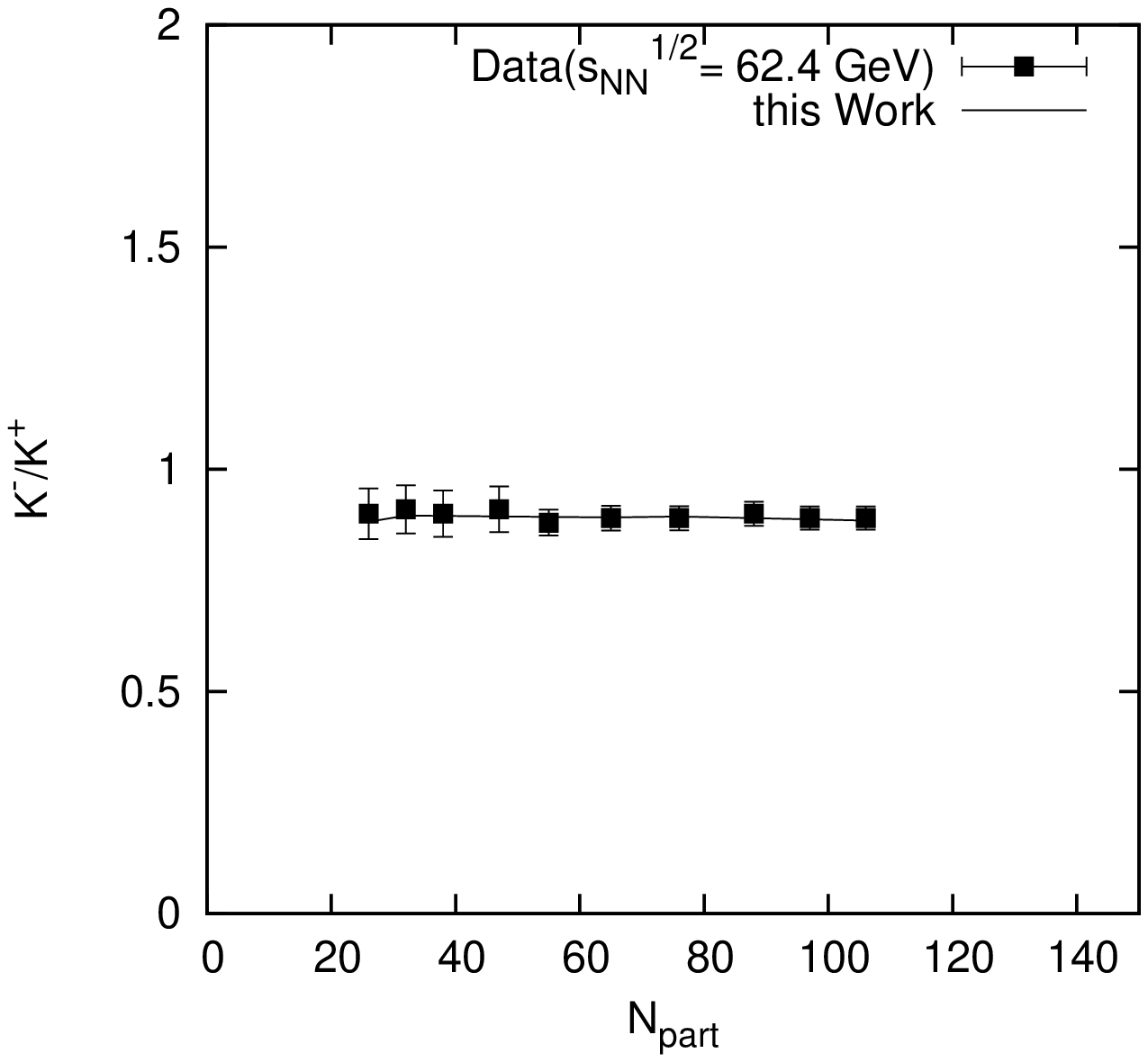}
\setcaptionwidth{2.6in}
\end{minipage}}%
\subfigure[]{
\begin{minipage}{0.5\textwidth}
\centering
 \includegraphics[width=2.5in]{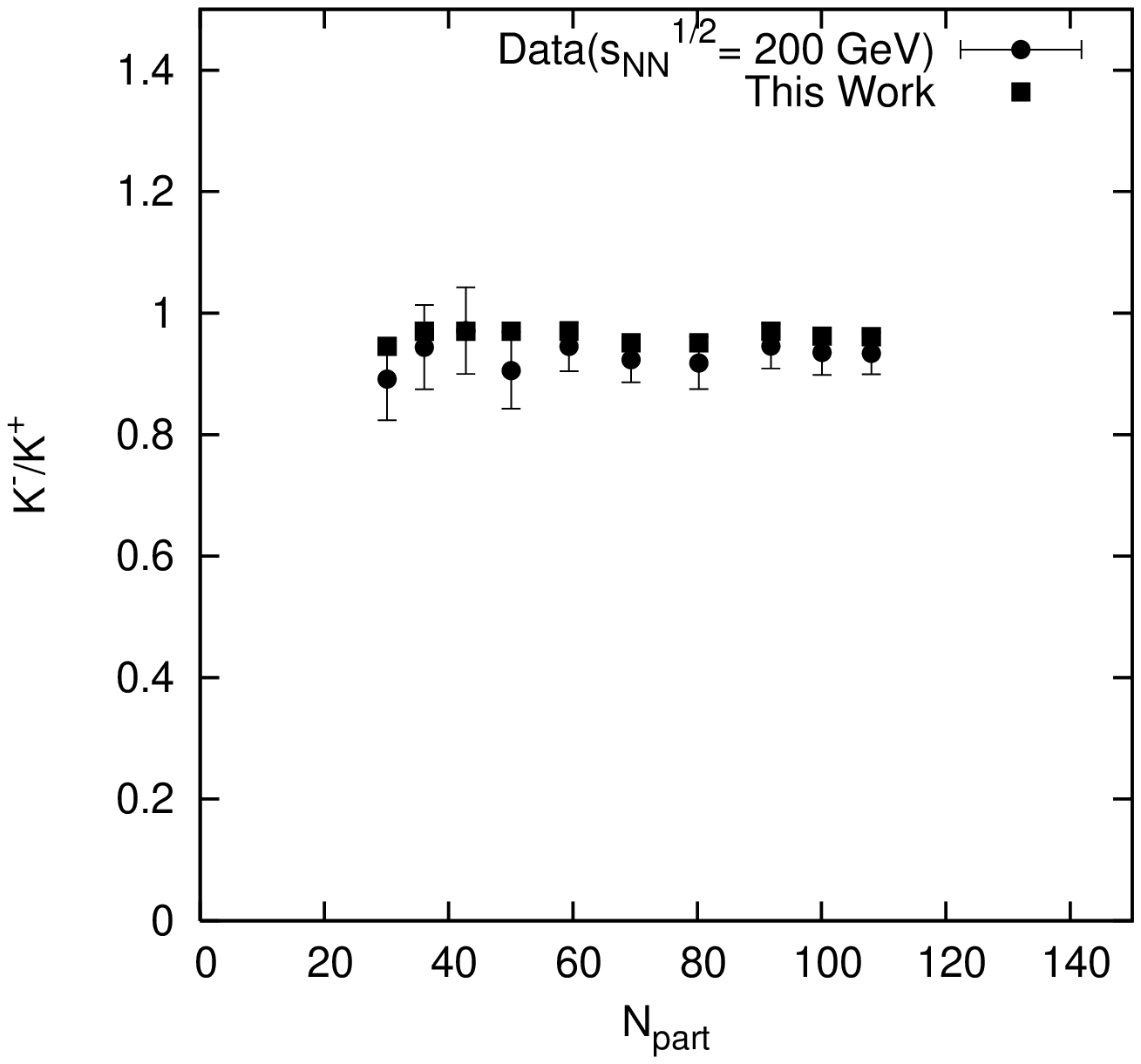}
  \end{minipage}}%
   \caption{{\small The $N_{part}$ versus $K^-/K^+$ ratio
behaviours at (a) $\sqrt {s_{NN}}$ =62.4 GeV and (b)$\sqrt s_{NN}$
=200 GeV. Data are taken from PHOBOS \cite{veres}, \cite{wosiek}.
The solid line in Fig. 9(a) and the filled squares in Fig. 9(b) show the
SCM-based calculated values.}  }
\end{figure}
\begin{figure}
\subfigure[]{
\begin{minipage}{.5\textwidth}
\centering
\includegraphics[width=2.5in]{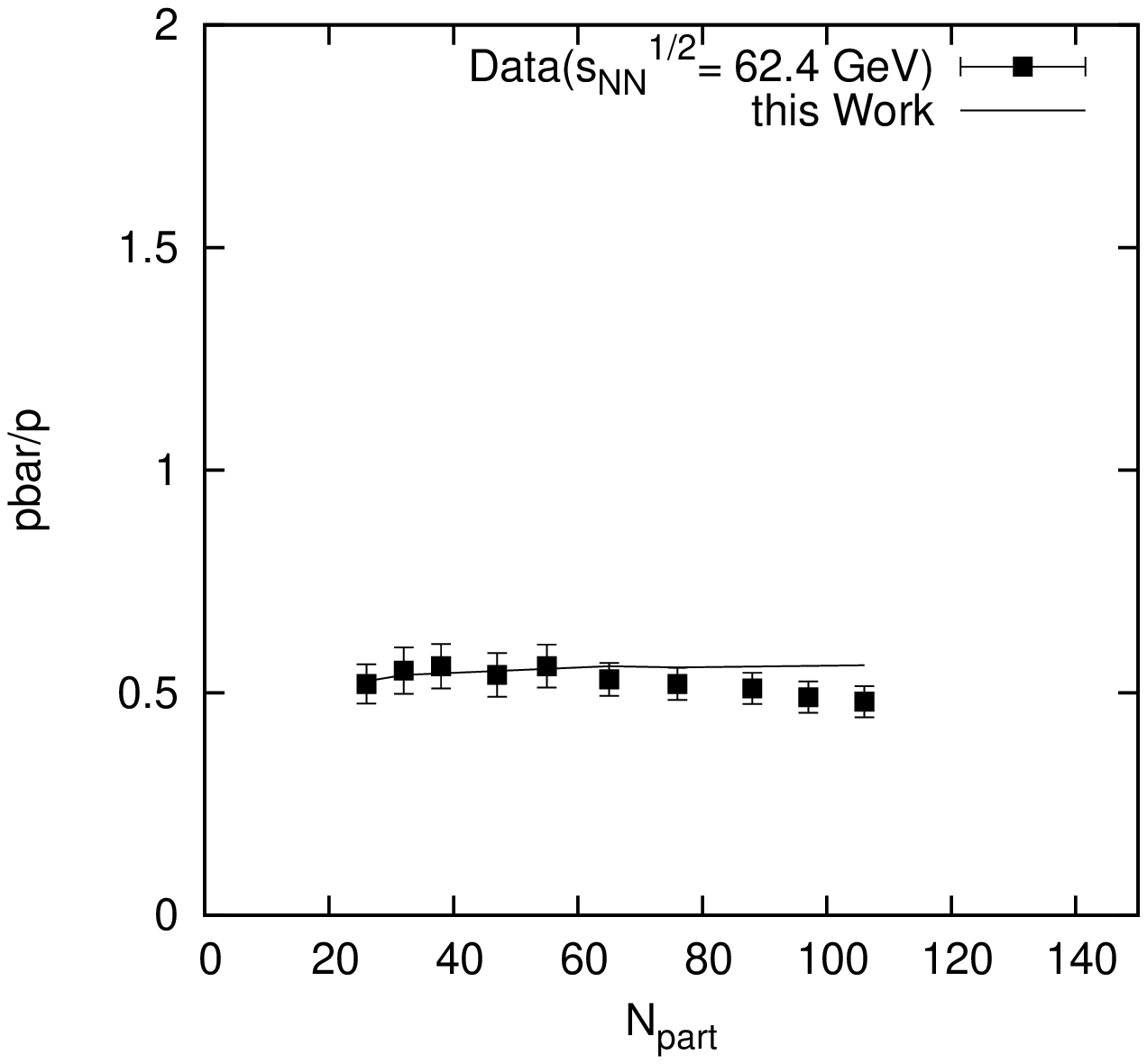}
\setcaptionwidth{2.6in}
\end{minipage}}%
\subfigure[]{
\begin{minipage}{0.5\textwidth}
\centering
 \includegraphics[width=2.5in]{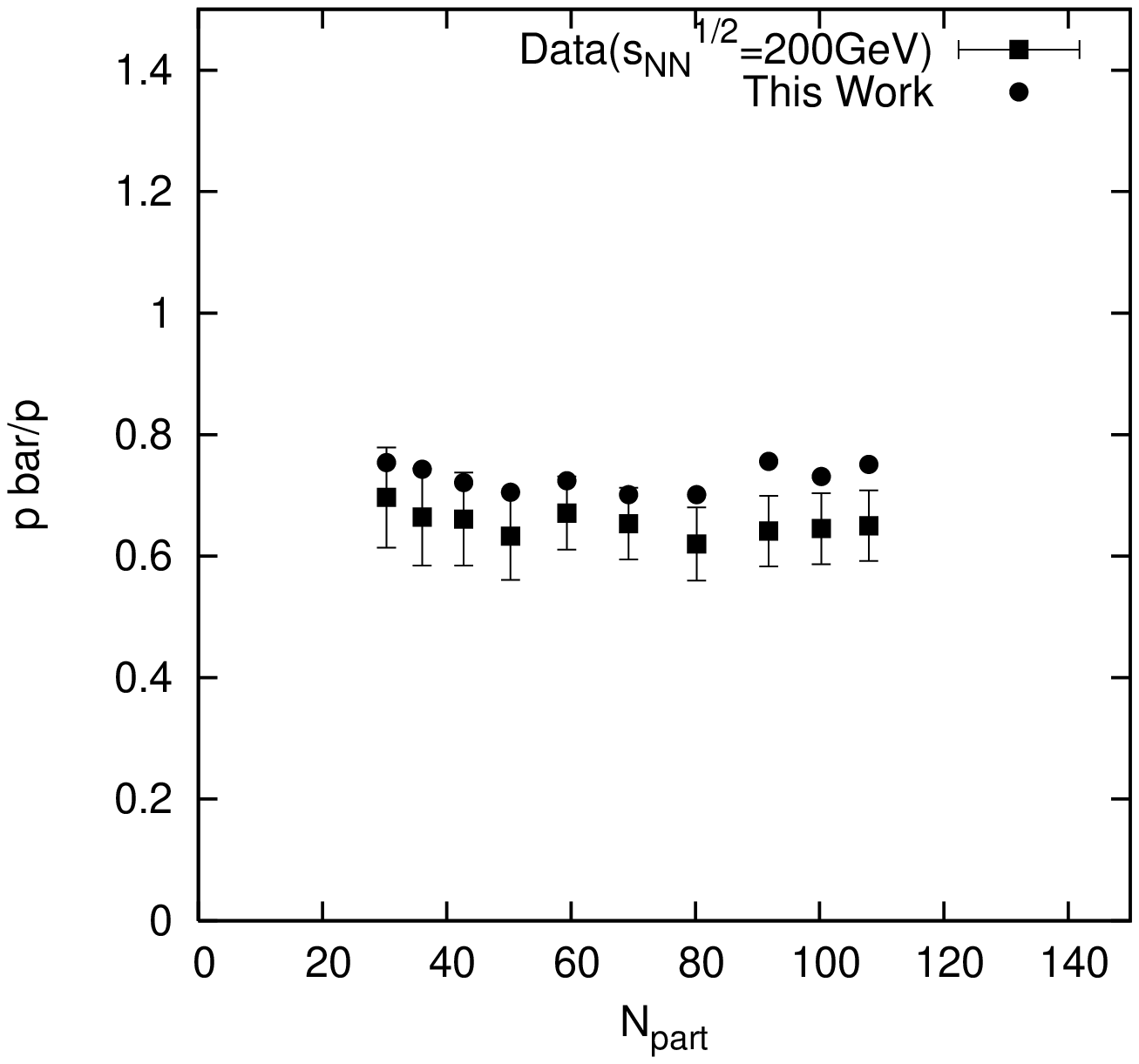}
  \end{minipage}}%
 \caption{{\small The $\bar p/p$ ratio
behaviours as a function of number of participants ($N_{part}$) in
$Cu+Cu$ reaction at (a) $\sqrt{s_{NN}}$ =62.4 GeV and (b) at
$\sqrt{s_{NN}}$ =200 GeV. Data are taken from PHOBOS \cite{veres},
\cite{wosiek}. The solid line in Fig. 10(a) and the filled circles in Fig.
10(b) show the SCM-based calculated values. } }
\end{figure}
\begin{figure}
\subfigure[]{
\begin{minipage}{.5\textwidth}
\centering
\includegraphics[width=2.5in]{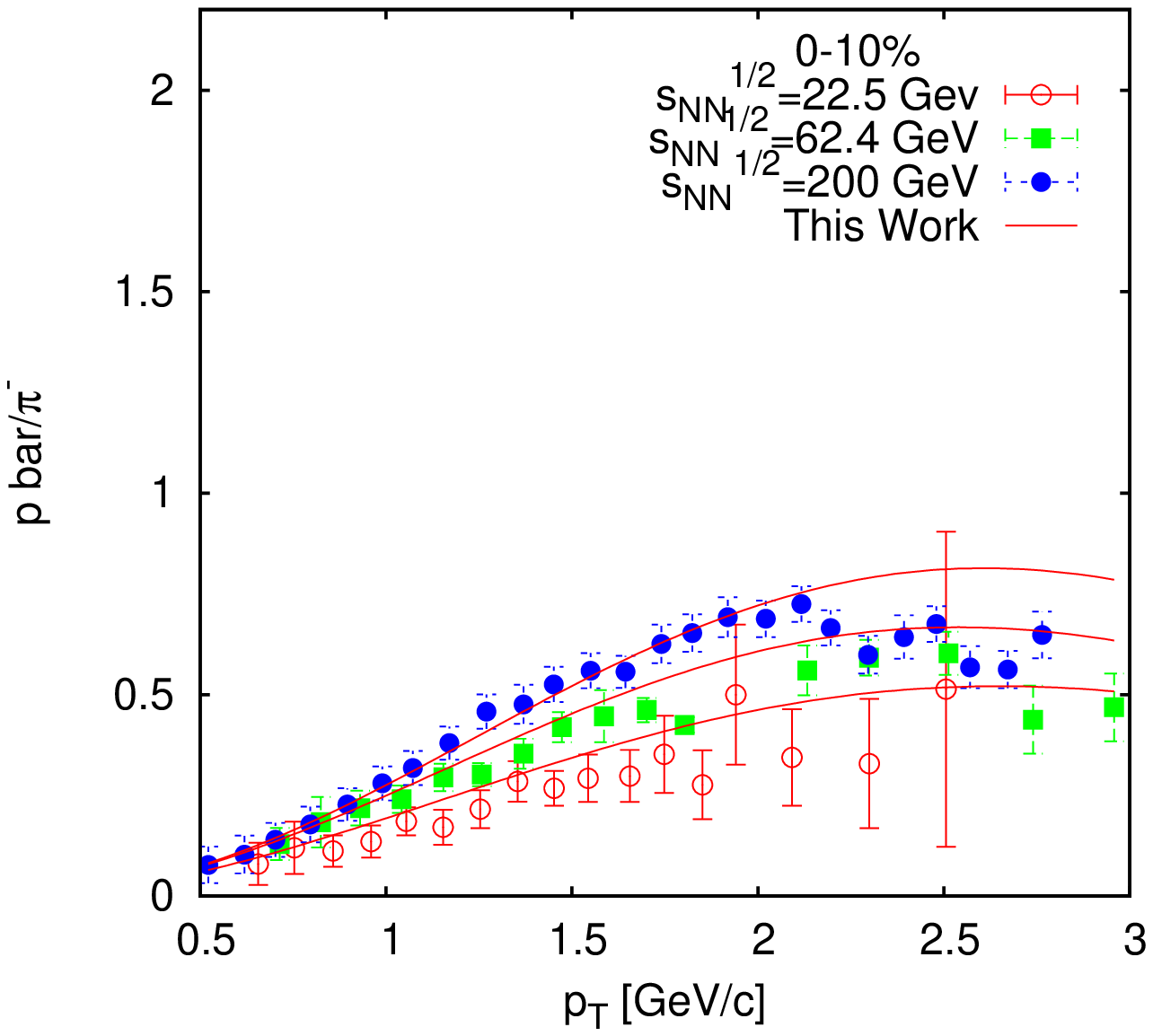}
\setcaptionwidth{2.6in}
\end{minipage}}%
\subfigure[]{
\begin{minipage}{0.5\textwidth}
\centering
 \includegraphics[width=2.5in]{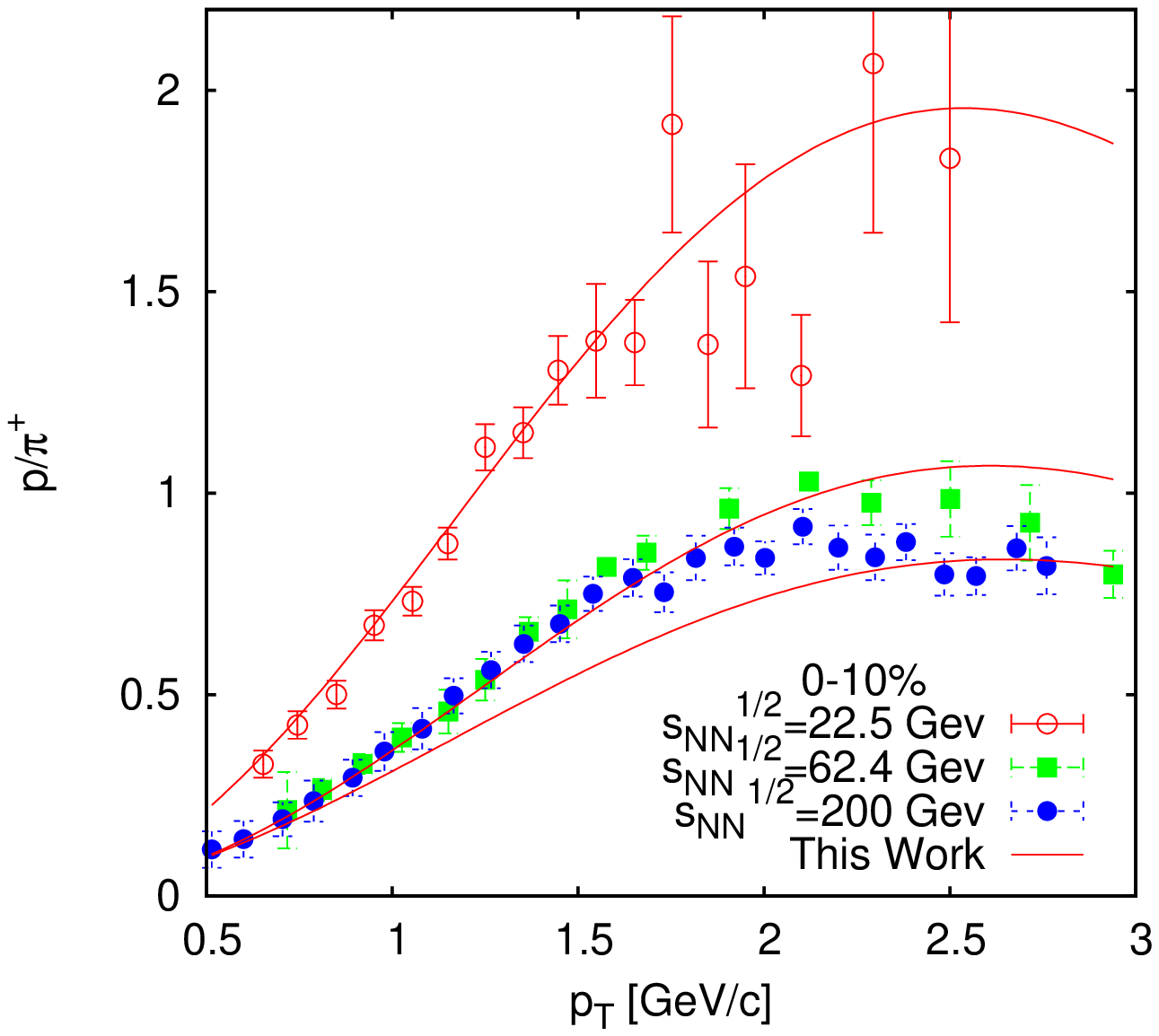}
  \end{minipage}}%
  \caption{{\small Ratios of (a) $\bar p/\pi^-$ and (b) $p/\pi^+$ as a
function $p_T$ for central $Cu+Cu$ reactions at $\sqrt {s_{NN}}$
=22.5, 62.4 and 200 GeV. Data in these Figures are taken from
\cite{chujo07}. The solid lines show the SCM-based results. } }
  \end{figure}
\begin{figure}
\subfigure[]{
\centering
\includegraphics[width=3.0in]{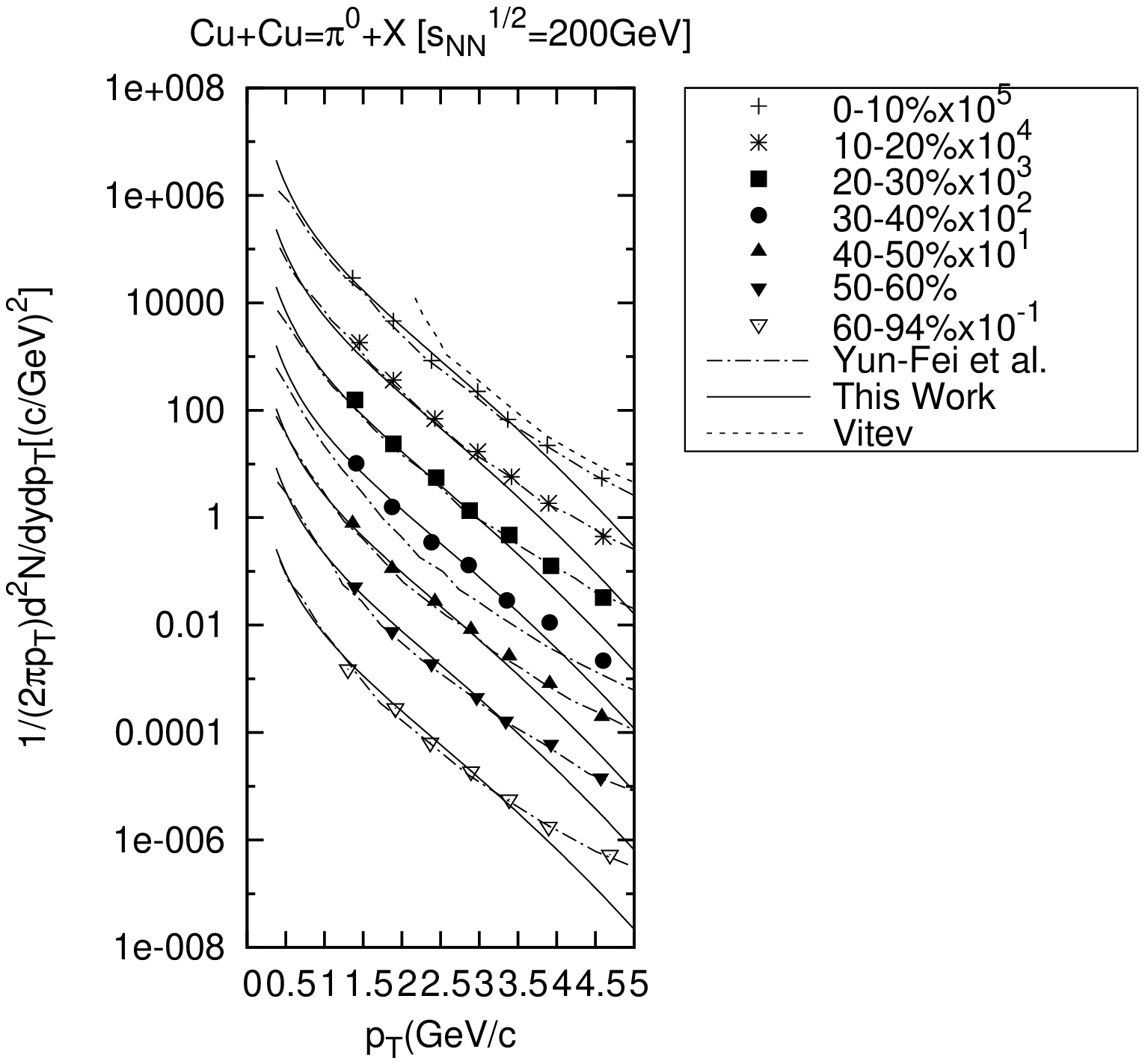}
\setcaptionwidth{2.6in}}
\vspace{0.01in}
\subfigure[]{
\centering
 \includegraphics[width=3.0in]{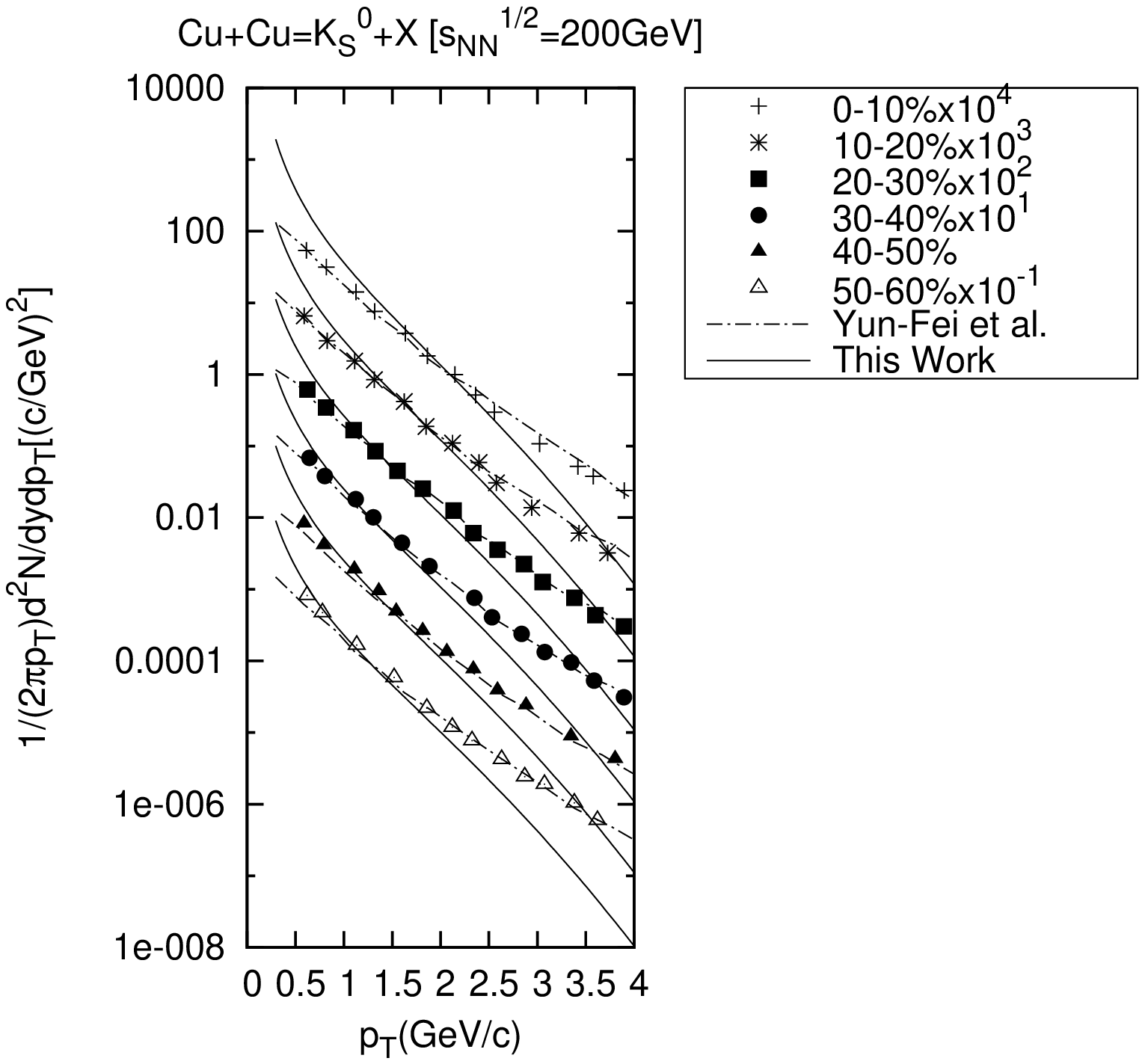}}
  \caption{{\small  Comparisons (a)
between the SCM, the Quark Combination Model \cite{fei} and the pQCD approach \cite{vitev} for the $\pi^0$ production at 200 GeV\cite{phenix1} and (b)between the SCM, the Quark Combination Model \cite{fei} for the $K_S^0$ production for $Cu+Cu$ reactions at $\sqrt {s_{NN}}$
=200 GeV. Data are taken from \cite{phenix07} and \cite{star07}. The solid lines in those figures represent SCM-based results, wherein the dashed and dotted lines represent the Quark Combination Model \cite{fei} and the pQCD approach \cite{vitev}.  } }
  \end{figure}
\begin{figure}
\subfigure[]{
\begin{minipage}{.5\textwidth}
\centering
\includegraphics[width=2.5in]{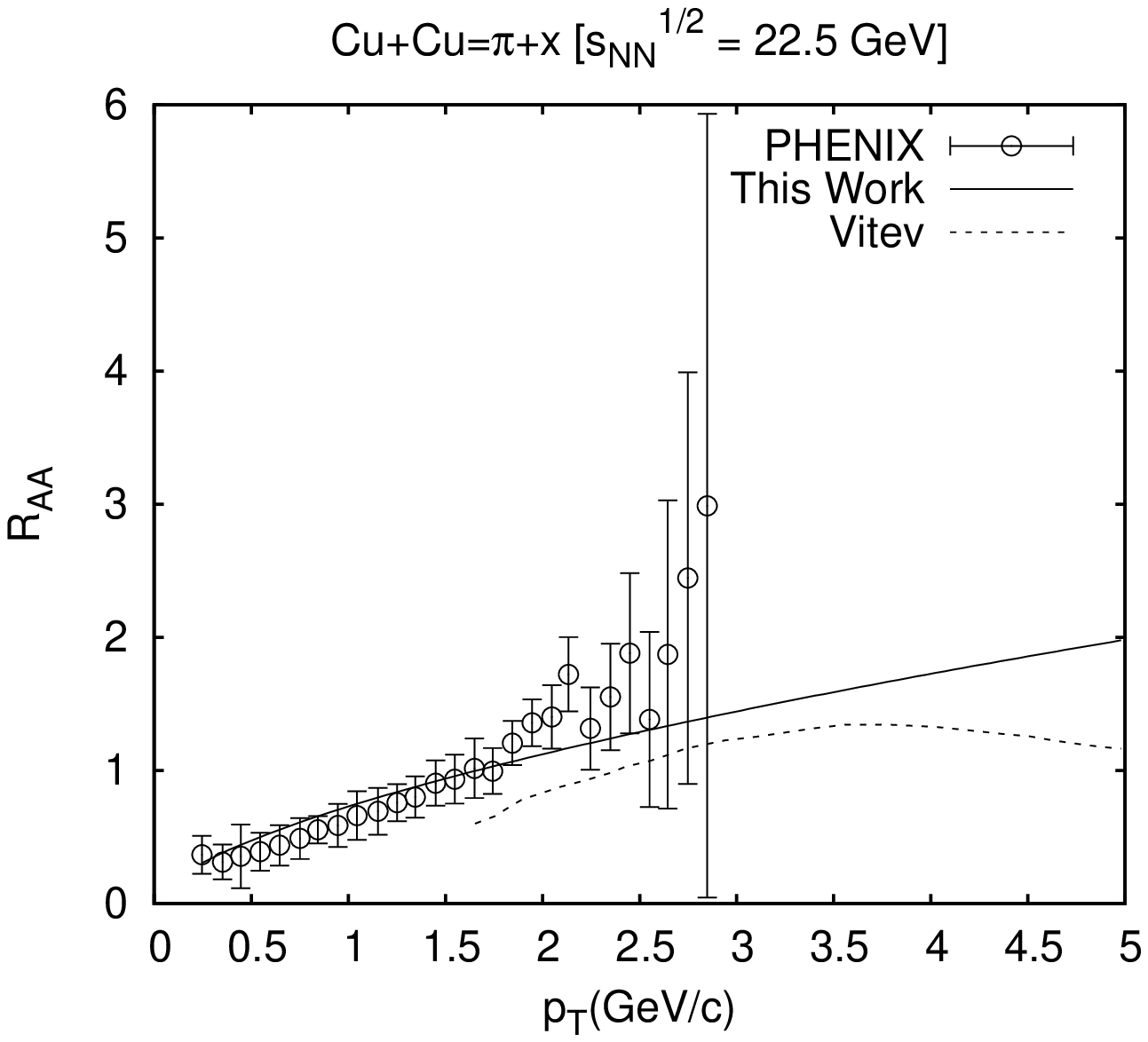}
\setcaptionwidth{2.6in}
\end{minipage}}%
\subfigure[]{
\begin{minipage}{0.5\textwidth}
\centering
 \includegraphics[width=2.5in]{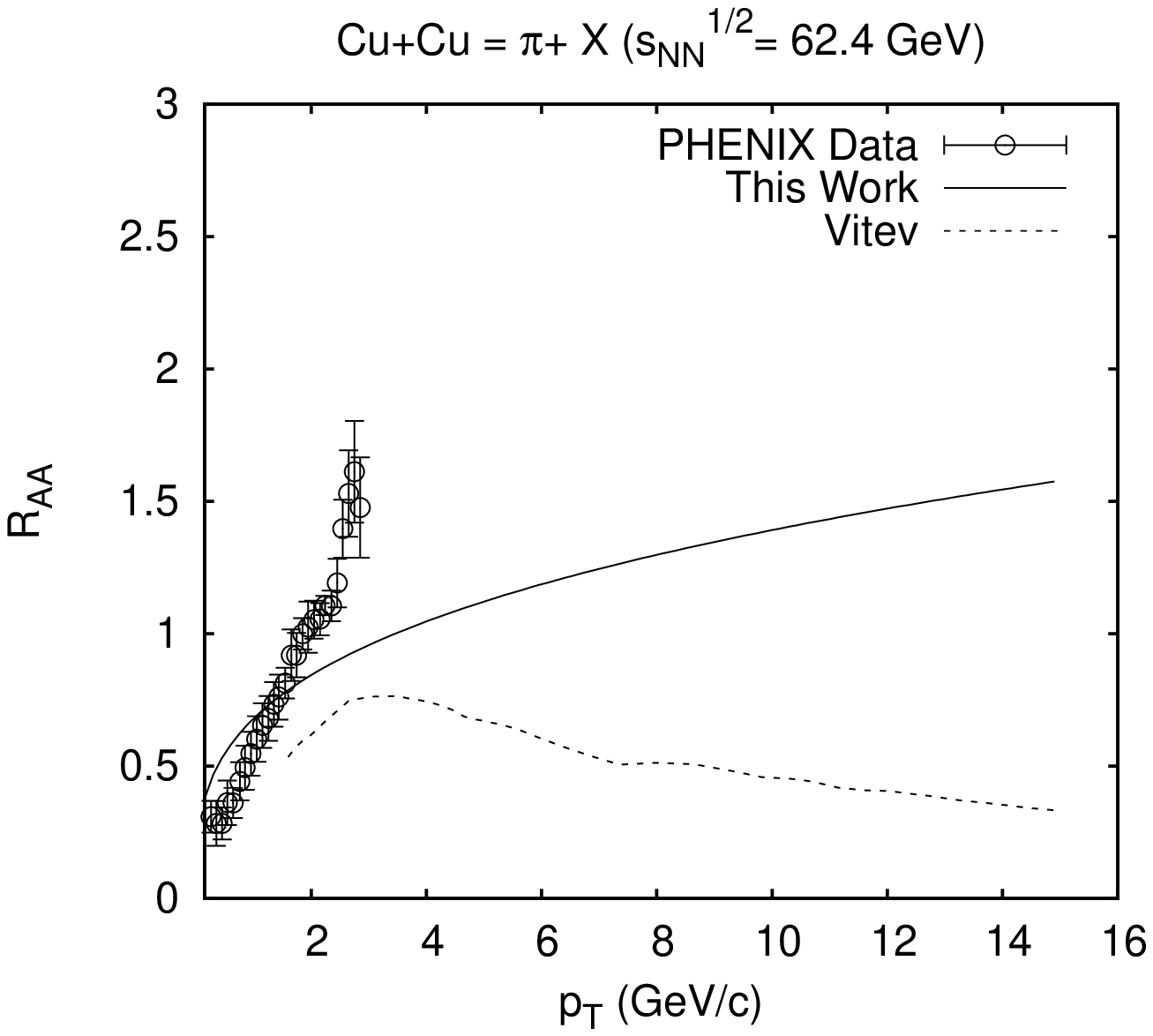}
  \end{minipage}}%
  \vspace{0.01in}
 \subfigure[]{
 \begin{minipage}{0.5\textwidth}
\centering
 \includegraphics[width=2.5in]{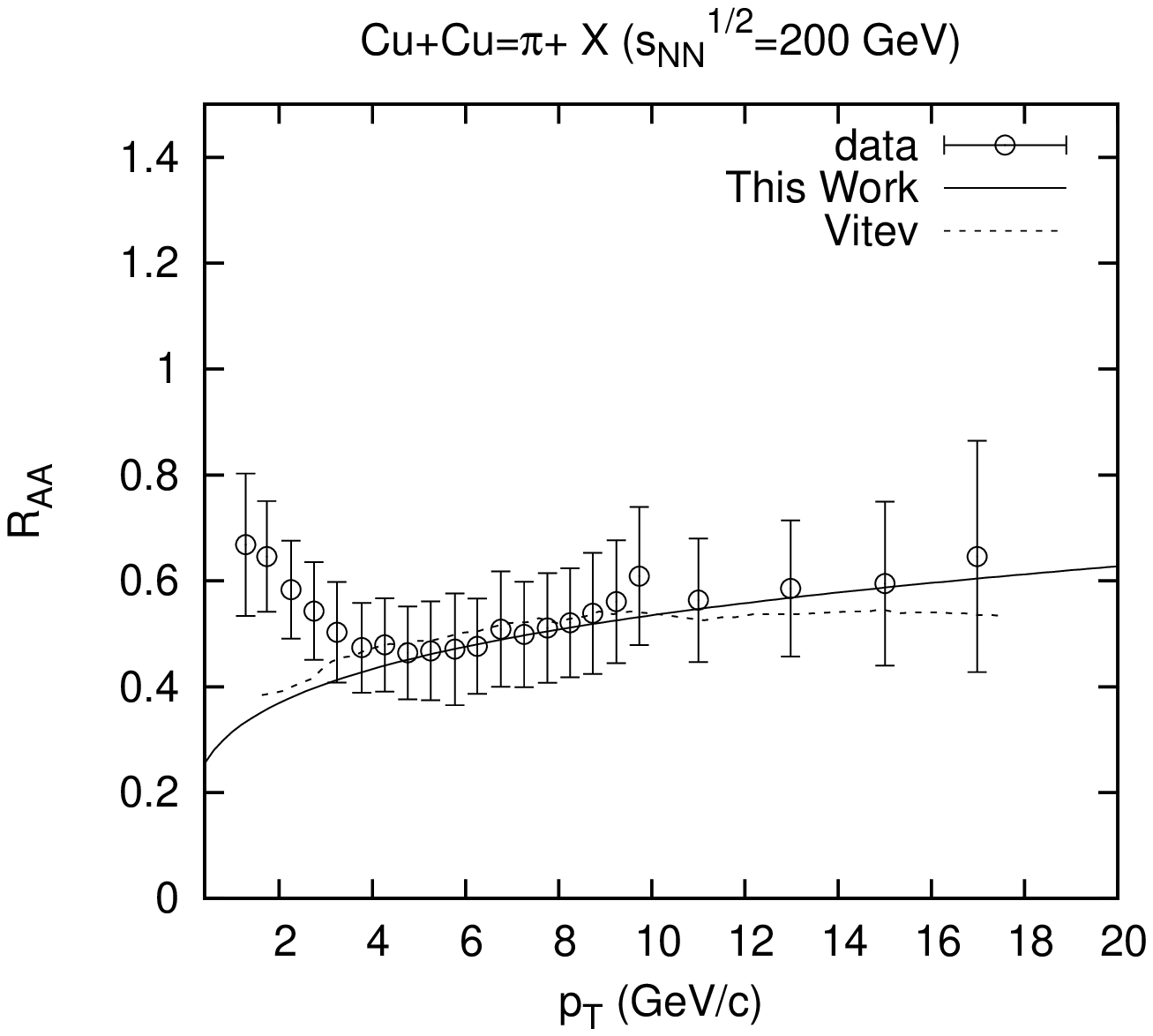}
 \end{minipage}}%
 \subfigure[]{
\begin{minipage}{0.5\textwidth}
\centering
 \includegraphics[width=2.5in]{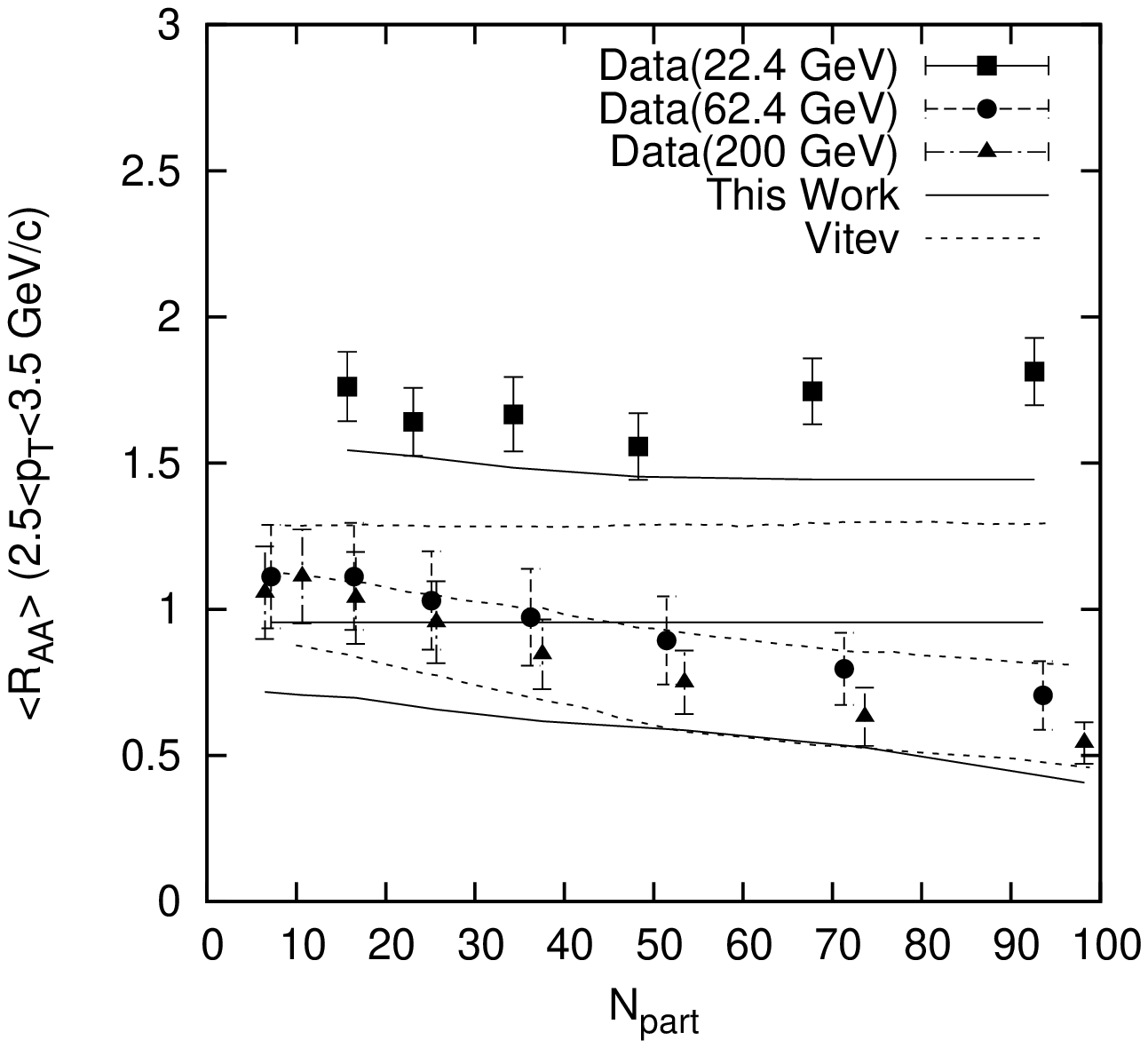}
  \end{minipage}}%
 \caption{{\small Plots of $p_T$ versus $R_{AA}$ (as defined in the text) at energies (a) $\sqrt {s_{NN}}$
=22.5 GeV, (b) $\sqrt{s_{NN}}$ = 62.4 and (c) $\sqrt{s_{NN}}$ = 200
GeV. Data in these Figures  are from \cite{chujo07}, \cite{awes} and
\cite{reygers}. The solid lines show the SCM-based results, wherein the dashed lines represent
the pQCD-oriented calculations \cite{phenix08},\cite{vitev}. (d) Plot of $<R_{AA}>$ vs. $N_{part}$ for $p_T$-ranges like
$2.5<p_T<3.5$ GeV/c. Comparisons of the nature of average
nuclear modification factors based on two sets of calculations, one done by SCM and represented by solid lines
and the other (dashed lines) obtained and shown by Vitev \cite{phenix08},\cite{vitev} are made here. } }
\end{figure}
\begin{figure}
\subfigure[]{
\begin{minipage}{.5\textwidth}
\centering
\includegraphics[width=2.5in]{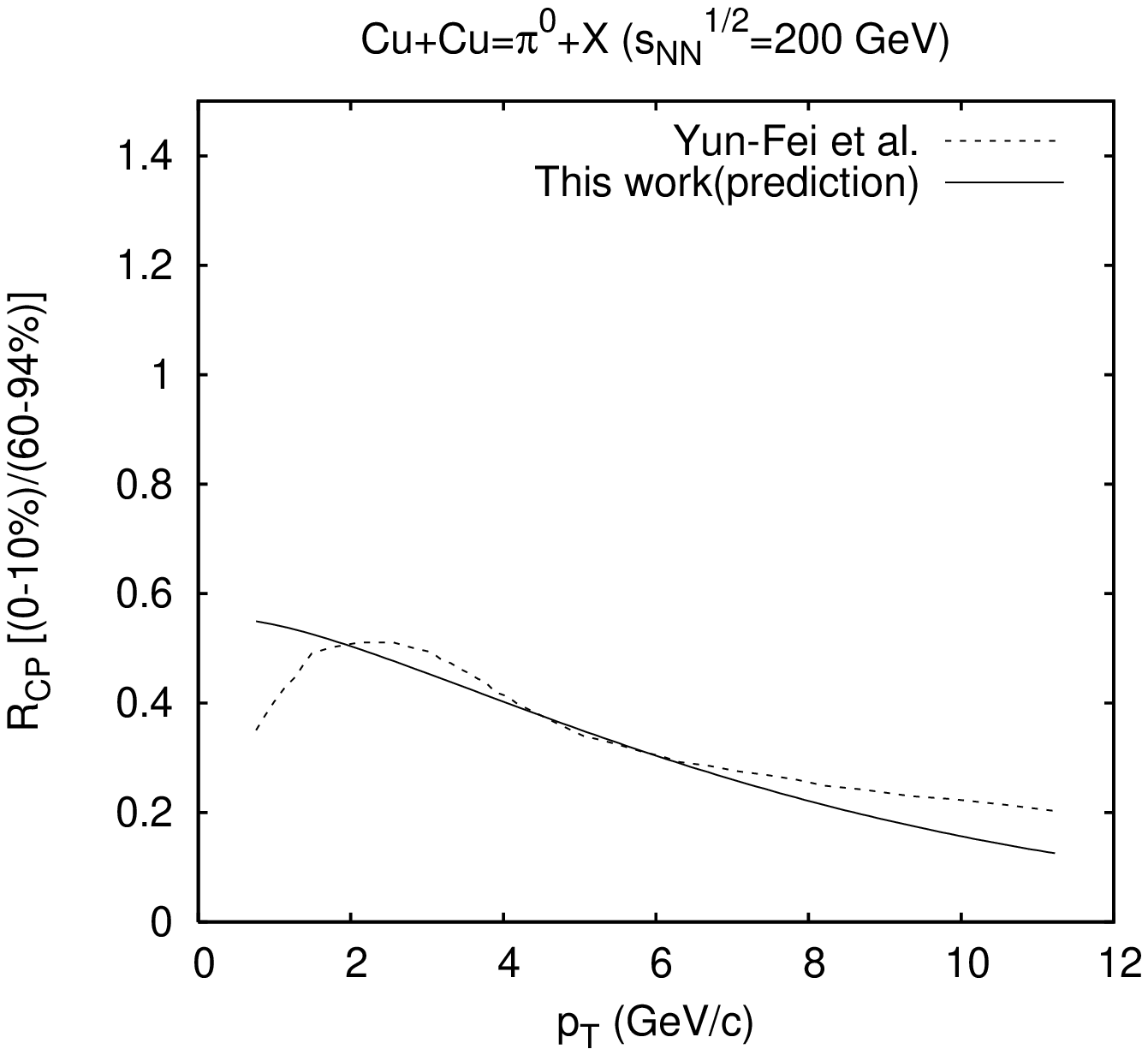}
\setcaptionwidth{2.6in}
\end{minipage}}%
\subfigure[]{
\begin{minipage}{0.5\textwidth}
\centering
 \includegraphics[width=2.5in]{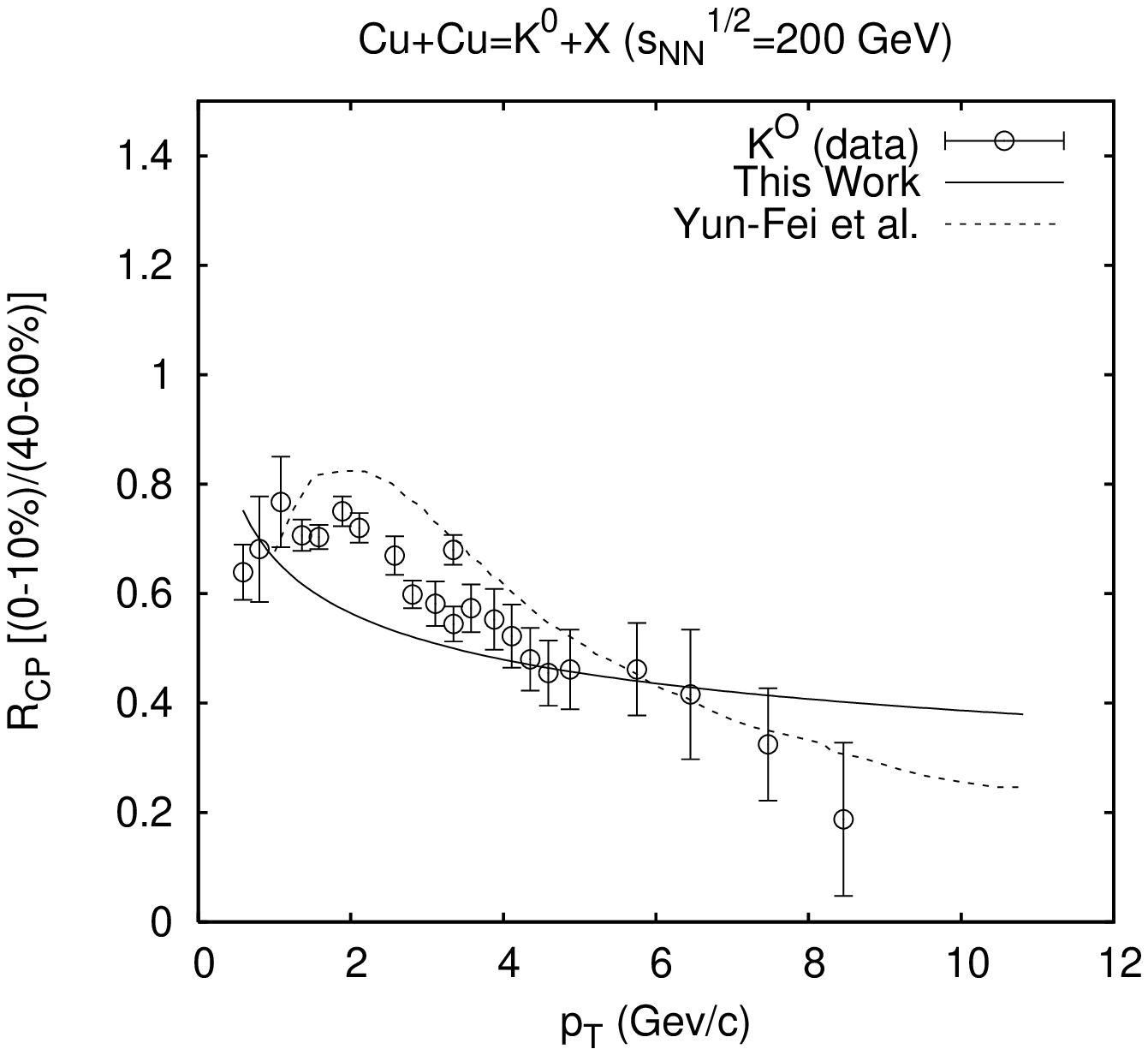}
  \end{minipage}}%
   \caption{{\small Plots for the $R_{CP}$
behaviours versus $p_T$ of (a) $\pi^0$ and (b)$K^0$. Pion data for
$Cu+Cu$ collisions has not yet available. Plots in Fig.14(a) are
predictive comparison between SCM-based result and result from Yun-Fei
et al.\cite{fei}. Data in Fig. (b) are taken from STAR
\cite{star07}. The solid line in Fig. 14(b) shows the SCM-based
results while the dotted line represents the results of Yun-Fei et
al.\cite{fei}.} }
\end{figure}
\end{document}